\documentclass[a4paper,11pt]{article}
\usepackage{jinstpub}
\usepackage{lineno}
\usepackage{graphicx}
\usepackage{subcaption}
\usepackage{float} 
\usepackage{amsmath}

\title{Charge Readout Electronics for the DUNE Horizontal Drift Far Detector: Design and Performance in ProtoDUNE-HD}

\usepackage{orcidlink}
\collaboration{The DUNE Collaboration}
\affiliation[0]{ Indian Institute of Science, Bengaluru, India, CV Raman Road, Bengaluru, Karnataka, 560012, India}
\affiliation[1]{University of Albany, SUNY, Albany, NY 12222, USA}
\affiliation[2]{Institute of Nuclear Physics at Almaty, Almaty 050032, Kazakhstan
}
\affiliation[3]{University of Amsterdam, NL-1098 XG Amsterdam, The Netherlands}
\affiliation[4]{Antalya Bilim University, 07190 D\"o{\c s}emealtı/Antalya, Turkey}
\affiliation[5]{University of Antananarivo, Antananarivo 101, Madagascar}
\affiliation[6]{University of Antioquia, Medell\'in, Colombia}
\affiliation[7]{Universidad Antonio Nari\~no, Bogot\'a, Colombia}
\affiliation[8]{Argonne National Laboratory, Argonne, IL 60439, USA}
\affiliation[9]{University of Arizona, Tucson, AZ 85721, USA}
\affiliation[10]{Universidad Nacional de Asunci\'on, San Lorenzo, Paraguay}
\affiliation[11]{University of Athens, Zografou GR 157 84, Greece}
\affiliation[12]{Universidad del Atl\'antico, Barranquilla, Atl\'antico, Colombia}
\affiliation[13]{Augustana University, Sioux Falls, SD 57197, USA}
\affiliation[14]{University of Bern, CH-3012 Bern, Switzerland}
\affiliation[15]{Beykent University, Istanbul, Turkey}
\affiliation[16]{University of Birmingham, Birmingham B15 2TT, United Kingdom}
\affiliation[17]{Universit\`a di Bologna, 40127 Bologna, Italy}
\affiliation[18]{Boston University, Boston, MA 02215, USA}
\affiliation[19]{University of Bristol, Bristol BS8 1TL, United Kingdom}
\affiliation[20]{Brookhaven National Laboratory, Upton, NY 11973, USA}
\affiliation[21]{University of Bucharest, Bucharest, Romania}
\affiliation[22]{University of California Berkeley, Berkeley, CA 94720, USA}
\affiliation[23]{University of California Davis, Davis, CA 95616, USA}
\affiliation[24]{University of California Irvine, Irvine, CA 92697, USA}
\affiliation[25]{University of California Riverside, Riverside CA 92521, USA}
\affiliation[26]{University of California Santa Barbara, Santa Barbara, CA 93106, USA}
\affiliation[27]{California Institute of Technology, Pasadena, CA 91125, USA}
\affiliation[28]{University of Cambridge, Cambridge CB3 0HE, United Kingdom}
\affiliation[29]{Universidade Estadual de Campinas, Campinas - SP, 13083-970, Brazil}
\affiliation[30]{Universit\`a di Catania, 2 - 95131 Catania, Italy}
\affiliation[31]{Universidad Cat\'olica del Norte, Antofagasta, Chile}
\affiliation[32]{Centro Brasileiro de Pesquisas F\'isicas, Rio de Janeiro, RJ 22290-180, Brazil}
\affiliation[33]{IRFU, CEA, Universit\'e Paris-Saclay, F-91191 Gif-sur-Yvette, France}
\affiliation[34]{CERN, The European Organization for Nuclear Research, 1211 Meyrin, Switzerland}
\affiliation[35]{Institute of Particle and Nuclear Physics of the Faculty of Mathematics and Physics of the Charles University, 180 00 Prague 8, Czech Republic }
\affiliation[36]{University of Chicago, Chicago, IL 60637, USA}
\affiliation[37]{Chung-Ang University, Seoul 06974, South Korea}
\affiliation[38]{CIEMAT, Centro de Investigaciones Energ\'eticas, Medioambientales y Tecnol\'ogicas, E-28040 Madrid, Spain}
\affiliation[39]{University of Cincinnati, Cincinnati, OH 45221, USA}
\affiliation[40]{Centro de Investigaci\'on y de Estudios Avanzados del Instituto Polit\'ecnico Nacional (Cinvestav), Mexico City, Mexico}
\affiliation[41]{Universidad de Colima, Colima, Mexico}
\affiliation[42]{University of Colorado Boulder, Boulder, CO 80309, USA}
\affiliation[43]{Colorado State University, Fort Collins, CO 80523, USA}
\affiliation[44]{Columbia University, New York, NY 10027, USA}
\affiliation[45]{Comisi\'on Nacional de Investigaci\'on y Desarrollo Aeroespacial, Lima, Peru}
\affiliation[46]{Centro de Tecnologia da Informacao Renato Archer, Amarais - Campinas, SP - CEP 13069-901, Brazil}
\affiliation[47]{Central University of South Bihar, Gaya, 824236, India
}
\affiliation[48]{Institute of Physics, Czech Academy of Sciences, 182 00 Prague 8, Czech Republic}
\affiliation[49]{Czech Technical University, 115 19 Prague 1, Czech Republic}
\affiliation[50]{Laboratoire d'Annecy de Physique des Particules, Universit\'e Savoie Mont Blanc, CNRS, LAPP-IN2P3, 74000 Annecy, France}
\affiliation[51]{Daresbury Laboratory, Cheshire WA4 4AD, United Kingdom}
\affiliation[52]{Dordt University, Sioux Center, IA 51250, USA}
\affiliation[53]{Drew University, Madison, NJ 07940, USA}
\affiliation[54]{Drexel University, Philadelphia, PA 19104, USA}
\affiliation[55]{Duke University, Durham, NC 27708, USA}
\affiliation[56]{Durham University, Durham DH1 3LE, United Kingdom}
\affiliation[57]{University of Edinburgh, Edinburgh EH8 9YL, United Kingdom}
\affiliation[58]{Universidad EIA, Envigado, Antioquia, Colombia}
\affiliation[59]{E\"otv\"os Lor\'and University, 1053 Budapest, Hungary}
\affiliation[60]{Erciyes University, Kayseri, Turkey}
\affiliation[61]{Faculdade de Ci\^encias da Universidade de Lisboa - FCUL, 1749-016 Lisboa, Portugal}
\affiliation[62]{Universidade Federal de Alfenas, Po{\c c}os de Caldas - MG, 37715-400, Brazil}
\affiliation[63]{Universidade Federal de Goias, Goiania, GO 74690-900, Brazil}
\affiliation[64]{Universidade Federal do ABC, Santo Andr\'e - SP, 09210-580, Brazil}
\affiliation[65]{Universidade Federal do Rio de Janeiro, Rio de Janeiro - RJ, 21941-901, Brazil}
\affiliation[66]{Fermi National Accelerator Laboratory, Batavia, IL 60510, USA}
\affiliation[67]{University of Ferrara, Ferrara, Italy}
\affiliation[68]{University of Florida, Gainesville, FL 32611-8440, USA}
\affiliation[69]{Florida State University, Tallahassee, FL, 32306 USA}
\affiliation[70]{Fluminense Federal University, 9 Icara\'i Niter\'oi - RJ, 24220-900, Brazil }
\affiliation[71]{Universit\`a degli Studi di Genova, Genova, Italy}
\affiliation[72]{Georgian Technical University, Tbilisi, Georgia}
\affiliation[73]{University of Granada \& CAFPE, 18002 Granada, Spain}
\affiliation[74]{Gran Sasso Science Institute, L'Aquila, Italy}
\affiliation[75]{Laboratori Nazionali del Gran Sasso, L'Aquila AQ, Italy}
\affiliation[76]{University Grenoble Alpes, CNRS, Grenoble INP, LPSC-IN2P3, 38000 Grenoble, France}
\affiliation[77]{Universidad de Guanajuato, Guanajuato, C.P. 37000, Mexico}
\affiliation[78]{Harish-Chandra Research Institute, Jhunsi, Allahabad 211 019, India}
\affiliation[79]{University of Hawaii, Honolulu, HI 96822, USA}
\affiliation[80]{Hong Kong University of Science and Technology, Kowloon, Hong Kong, China}
\affiliation[81]{University of Houston, Houston, TX 77204, USA}
\affiliation[82]{University of  Hyderabad, Gachibowli, Hyderabad - 500 046, India}
\affiliation[83]{Idaho State University, Pocatello, ID 83209, USA}
\affiliation[84]{Instituto de F\'isica Corpuscular, CSIC and Universitat de Val\`encia, 46980 Paterna, Valencia, Spain}
\affiliation[85]{Instituto Galego de F\'isica de Altas Enerx\'ias, University of Santiago de Compostela, Santiago de Compostela, 15782, Spain}
\affiliation[86]{Institute of High Energy Physics, Chinese Academy of Sciences, Beijing, China}
\affiliation[87]{Indian Institute of Technology Kanpur, Uttar Pradesh 208016, India}
\affiliation[88]{Jozef Stefan Institute, Jamova cesta 39, 1000 Ljubljana, Slovenia}
\affiliation[89]{Illinois Institute of Technology, Chicago, IL 60616, USA}
\affiliation[90]{Imperial College of Science, Technology and Medicine, London SW7 2BZ, United Kingdom}
\affiliation[91]{Indian Institute of Technology Guwahati, Guwahati, 781 039, India}
\affiliation[92]{Indian Institute of Technology Hyderabad, Hyderabad, 502285, India}
\affiliation[93]{Indiana University, Bloomington, IN 47405, USA}
\affiliation[94]{Istituto Nazionale di Fisica Nucleare Sezione di Bologna, 40127 Bologna BO, Italy}
\affiliation[95]{Istituto Nazionale di Fisica Nucleare Sezione di Catania, I-95123 Catania, Italy}
\affiliation[96]{Istituto Nazionale di Fisica Nucleare Sezione di Ferrara, I-44122 Ferrara, Italy}
\affiliation[97]{Istituto Nazionale di Fisica Nucleare Laboratori Nazionali di Frascati, Frascati, Roma, Italy}
\affiliation[98]{Istituto Nazionale di Fisica Nucleare Sezione di Genova, 16146 Genova GE, Italy}
\affiliation[99]{Istituto Nazionale di Fisica Nucleare Sezione di Lecce, 73100 - Lecce, Italy}
\affiliation[100]{Istituto Nazionale di Fisica Nucleare Sezione di Milano Bicocca, 3 - I-20126 Milano, Italy}
\affiliation[101]{Istituto Nazionale di Fisica Nucleare Sezione di Milano, 20133 Milano, Italy}
\affiliation[102]{Istituto Nazionale di Fisica Nucleare Sezione di Napoli, I-80126 Napoli, Italy}
\affiliation[103]{Istituto Nazionale di Fisica Nucleare Sezione di Padova, 35131 Padova, Italy}
\affiliation[104]{Istituto Nazionale di Fisica Nucleare Sezione di Pavia,  I-27100 Pavia, Italy}
\affiliation[105]{Istituto Nazionale di Fisica Nucleare Laboratori Nazionali di Pisa, Pisa PI, Italy}
\affiliation[106]{Istituto Nazionale di Fisica Nucleare Sezione di Roma, 00185 Roma RM, Italy}
\affiliation[107]{Istituto Nazionale di Fisica Nucleare Roma Tor Vergata , 00133 Roma RM, Italy}
\affiliation[108]{Istituto Nazionale di Fisica Nucleare Laboratori Nazionali del Sud, 95123 Catania, Italy}
\affiliation[109]{Istituto Nazionale di Fisica Nucleare, Sezione di Torino, Turin, Italy}
\affiliation[110]{Universidad Nacional de Ingenier\'ia, Lima 25, Per\'u}
\affiliation[111]{University of Insubria, Via Ravasi, 2, 21100 Varese VA, Italy}
\affiliation[112]{University of Iowa, Iowa City, IA 52242, USA}
\affiliation[113]{Iowa State University, Ames, Iowa 50011, USA}
\affiliation[114]{Institut de Physique des 2 Infinis de Lyon, 69622 Villeurbanne, France}
\affiliation[115]{Institute for Research in Fundamental Sciences, Tehran, Iran}
\affiliation[116]{Particle Physics and Cosmology International Research Laboratory	, Chicago IL,  60637 USA}
\affiliation[117]{Instituto Superior T\'ecnico - IST, Universidade de Lisboa, 1049-001 Lisboa, Portugal}
\affiliation[118]{Instituto Tecnol\'ogico de Aeron\'autica, Sao Jose dos Campos, Brazil}
\affiliation[119]{Institute for Theoretical Physics and Modeling, Yerevan 0036, Armenia}
\affiliation[120]{Iwate University, Morioka, Iwate 020-8551, Japan}
\affiliation[121]{Jackson State University, Jackson, MS 39217, USA}
\affiliation[122]{Jawaharlal Nehru University, New Delhi 110067, India}
\affiliation[123]{Jeonbuk National University, Jeonrabuk-do 54896, South Korea}
\affiliation[124]{Jyv\"askyl\"a University, FI-40014 Jyv\"askyl\"a, Finland}
\affiliation[125]{Kansas State University, Manhattan, KS 66506, USA}
\affiliation[126]{Kavli Institute for the Physics and Mathematics of the Universe, Kashiwa, Chiba 277-8583, Japan}
\affiliation[127]{High Energy Accelerator Research Organization (KEK), Ibaraki, 305-0801, Japan}
\affiliation[128]{Korea Institute of Science and Technology Information, Daejeon, 34141, South Korea}
\affiliation[129]{Taras Shevchenko National University of Kyiv, 01601 Kyiv, Ukraine}
\affiliation[130]{Lancaster University, Lancaster LA1 4YB, United Kingdom}
\affiliation[131]{Lawrence Berkeley National Laboratory, Berkeley, CA 94720, USA}
\affiliation[132]{Laborat\'orio de Instrumenta{\c c}\~ao e F\'isica Experimental de Part\'iculas, 1649-003 Lisboa and 3004-516 Coimbra, Portugal}
\affiliation[133]{University of Liverpool, L69 7ZE, Liverpool, United Kingdom}
\affiliation[134]{National Laboratory for Astrophysics, Rua dos Estados Unidos, 154
Bairro das Na{\c c}ões
Itajub\'a / MG - 37.504-364
Brasil}
\affiliation[135]{Los Alamos National Laboratory, Los Alamos, NM 87545, USA}
\affiliation[136]{Louisiana State University, Baton Rouge, LA 70803, USA}
\affiliation[137]{Laboratoire de Physique des Deux Infinis Bordeaux - IN2P3, F-33175 Gradignan, Bordeaux, France, }
\affiliation[138]{University of Lucknow, Uttar Pradesh 226007, India}
\affiliation[139]{Johannes Gutenberg-Universit\"at Mainz, 55122 Mainz, Germany}
\affiliation[140]{University of Manchester, Manchester M13 9PL, United Kingdom}
\affiliation[141]{Marmara University, Marmara Üniversitesi G\"oztepe Yerle{\c s}kesi 34722 Kadık\"oy - İstanbul, Turkey}
\affiliation[142]{Massachusetts Institute of Technology, Cambridge, MA 02139, USA}
\affiliation[143]{University of Medell\'in, Medell\'in, 050026 Colombia }
\affiliation[144]{University of Michigan, Ann Arbor, MI 48109, USA}
\affiliation[145]{Michigan State University, East Lansing, MI 48824, USA}
\affiliation[146]{Universit\`a di Milano Bicocca , 20126 Milano, Italy}
\affiliation[147]{Universit\`a degli Studi di Milano, I-20133 Milano, Italy}
\affiliation[148]{University of Minnesota Duluth, Duluth, MN 55812, USA}
\affiliation[149]{University of Minnesota Twin Cities, Minneapolis, MN 55455, USA}
\affiliation[150]{University of Mississippi, University, MS 38677 USA}
\affiliation[151]{Universit\`a degli Studi di Napoli Federico II , 80138 Napoli NA, Italy}
\affiliation[152]{Nikhef National Institute of Subatomic Physics, 1098 XG Amsterdam, Netherlands}
\affiliation[153]{National Institute of Science Education and Research, An OCC of Homi Bhabha National Institute, Bhubaneswar, Odisha, 752050, India}
\affiliation[154]{University of North Dakota, Grand Forks, ND 58202-8357, USA}
\affiliation[155]{Northern Illinois University, DeKalb, IL 60115, USA}
\affiliation[156]{Northwestern University, Evanston, Il 60208, USA}
\affiliation[157]{University of Notre Dame, Notre Dame, IN 46556, USA}
\affiliation[158]{University of Novi Sad, 21102 Novi Sad, Serbia}
\affiliation[159]{Ohio State University, Columbus, OH 43210, USA}
\affiliation[160]{Oregon State University, Corvallis, OR 97331, USA}
\affiliation[161]{University of Oxford, Oxford, OX1 3RH, United Kingdom}
\affiliation[162]{Pacific Northwest National Laboratory, Richland, WA 99352, USA}
\affiliation[163]{Universt\`a degli Studi di Padova, I-35131 Padova, Italy}
\affiliation[164]{Panjab University, Chandigarh, 160014, India}
\affiliation[165]{Universit\'e Paris-Saclay, CNRS/IN2P3, IJCLab, 91405 Orsay, France}
\affiliation[166]{Universit\'e Paris Cit\'e, CNRS, Astroparticule et Cosmologie, Paris, France}
\affiliation[167]{University of Parma,  43121 Parma PR, Italy}
\affiliation[168]{Universit\`a degli Studi di Pavia, 27100 Pavia PV, Italy}
\affiliation[169]{University of Pennsylvania, Philadelphia, PA 19104, USA}
\affiliation[170]{Pennsylvania State University, University Park, PA 16802, USA}
\affiliation[171]{Physical Research Laboratory, Ahmedabad 380 009, India}
\affiliation[172]{Universit\`a di Pisa, I-56127 Pisa, Italy}
\affiliation[173]{University of Pittsburgh, Pittsburgh, PA 15260, USA}
\affiliation[174]{Pontificia Universidad Cat\'olica del Per\'u, Lima, Per\'u}
\affiliation[175]{University of Puerto Rico, Mayaguez 00681, Puerto Rico, USA}
\affiliation[176]{Punjab Agricultural University, Ludhiana 141004, India}
\affiliation[177]{Queen Mary University of London, London E1 4NS, United Kingdom
}
\affiliation[178]{Radboud University, NL-6525 AJ Nijmegen, Netherlands}
\affiliation[179]{Rice University, Houston, TX 77005, USA}
\affiliation[180]{University of Rochester, Rochester, NY 14627, USA}
\affiliation[181]{Royal Holloway College London, London, TW20 0EX, United Kingdom}
\affiliation[182]{Rutgers University, Piscataway, NJ, 08854, USA}
\affiliation[183]{STFC Rutherford Appleton Laboratory, Didcot OX11 0QX, United Kingdom}
\affiliation[184]{Universit\`a del Salento, 73100 Lecce, Italy}
\affiliation[185]{Universidade do Estado de Santa Catarina , Santa Catarina, 89219-710, Brazil}
\affiliation[186]{Universidad del Magdalena, Santa Marta - Colombia}
\affiliation[187]{Sapienza University of Rome, 00185 Roma RM, Italy}
\affiliation[188]{Universidad Sergio Arboleda, 11022 Bogot\'a, Colombia}
\affiliation[189]{University of Sheffield, Sheffield S3 7RH, United Kingdom}
\affiliation[190]{SLAC National Accelerator Laboratory, Menlo Park, CA 94025, USA}
\affiliation[191]{University of South Carolina, Columbia, SC 29208, USA}
\affiliation[192]{South Dakota School of Mines and Technology, Rapid City, SD 57701, USA}
\affiliation[193]{South Dakota State University, Brookings, SD 57007, USA}
\affiliation[194]{Stony Brook University, SUNY, Stony Brook, NY 11794, USA}
\affiliation[195]{Sanford Underground Research Facility, Lead, SD, 57754, USA}
\affiliation[196]{University of Sussex, Brighton, BN1 9RH, United Kingdom}
\affiliation[197]{Syracuse University, Syracuse, NY 13244, USA}
\affiliation[198]{Universidade Tecnol\'ogica Federal do Paran\'a, Curitiba, Brazil}
\affiliation[199]{Tel Aviv University, Tel Aviv-Yafo, Israel}
\affiliation[200]{Texas A\&M University, College Station, TX 77840, USA}
\affiliation[201]{Texas A\&M University - Corpus Christi, Corpus Christi, TX 78412, USA}
\affiliation[202]{University of Texas at Arlington, Arlington, TX 76019, USA}
\affiliation[203]{University of Texas at Austin, Austin, TX 78712, USA}
\affiliation[204]{University of Toronto, Toronto, Ontario M5S 1A1, Canada}
\affiliation[205]{Tufts University, Medford, MA 02155, USA}
\affiliation[206]{Universidade Federal de S\~ao Paulo, 09913-030, S\~ao Paulo, Brazil}
\affiliation[207]{University College London, London, WC1E 6BT, United Kingdom}
\affiliation[208]{University of Kansas, Lawrence, KS 66045, USA}
\affiliation[209]{Universidad Nacional Mayor de San Marcos, Lima, Peru}
\affiliation[210]{Valley City State University, Valley City, ND 58072, USA}
\affiliation[211]{University of Vigo, E- 36310 Vigo Spain}
\affiliation[212]{Virginia Tech, Blacksburg, VA 24060, USA}
\affiliation[213]{University of Warsaw, 02-093 Warsaw, Poland}
\affiliation[214]{University of Warwick, Coventry CV4 7AL, United Kingdom}
\affiliation[215]{Wellesley College, Wellesley, MA 02481, USA}
\affiliation[216]{Wichita State University, Wichita, KS 67260, USA}
\affiliation[217]{William and Mary, Williamsburg, VA 23187, USA}
\affiliation[218]{University of Wisconsin Madison, Madison, WI 53706, USA}
\affiliation[219]{Yale University, New Haven, CT 06520, USA}
\affiliation[220]{York University, Toronto M3J 1P3, Canada}
\author[115]{S.~Abbaslu,}
\author[81]{F.~Abd Alrahman,}
\author[34]{A.~Abed Abud,}
\author[34]{R.~Acciarri,}
\author[198]{L.~P.~Accorsi,}
\author[12]{M.~A.~Acero,}
\author[198]{M.~R.~Adames,}
\author[72]{G.~Adamov,}
\author[66]{M.~Adamowski,}
\author[130]{K.~Adhikari,}
% ...

\abstract{DUNE (Deep Underground Neutrino Experiment) is a long-baseline neutrino oscillation experiment currently under construction, whose far detectors will be the largest liquid argon time projection chambers ever built. This detector design calls for custom-built cryogenic front-end electronics to meet its performance requirements. This paper describes the charge readout electronics that will be used in the DUNE horizontal drift (HD) far detector and presents performance results using data from the ProtoDUNE-HD detector, a 770 ton liquid argon time projection chamber operated at the CERN Neutrino Platform in 2024 that served as the final prototype of the DUNE HD design.}

\arxivnumber{2604.23966}

\begin{document}
\maketitle
\flushbottom

\section{Introduction}

The Deep Underground Neutrino Experiment (DUNE) is a long-baseline neutrino oscillation experiment currently under construction, designed to measure the neutrino mass ordering, determine the $\delta_{CP}$ phase in the neutrino sector, and test the limits of the standard three-flavor neutrino oscillation paradigm \cite{DUNE_LongBaseline}. To make these measurements, the experiment consists of a new high-intensity neutrino beam from Fermilab in Illinois, a near detector complex at the same location to measure the unoscillated neutrino flux \cite{ND_CDR}, and a set of far detectors 1300 km away and 1.5 km underground at the Sanford Underground Research Facility (SURF) in South Dakota to measure the oscillated neutrino spectrum \cite{DUNE_Intro}. In addition, DUNE will enable a number of important measurements beyond neutrino oscillations \cite{DUNE_physics}, including measuring the neutrino flux from any potential future nearby core-collapse supernovae during the lifetime of the detectors \cite{DUNE_Supernova} and searching for physics beyond the Standard Model \cite{DUNE_BSM}.

To accumulate the exposure needed for these measurements, the DUNE far detectors have been designed with correspondingly large volumes, with expected interior dimensions of 15.1\,m (w) × 14.0\,m (h) × 62.0\,m (l) each \cite{DUNE_Intro}. Their deep underground location in SURF will provide necessary shielding against cosmic ray muons \cite{SURF_Muon}, which would otherwise constitute an overwhelming background for non-beam events. Within these volumes, the detectors must maximize their target mass while retaining the ability to resolve neutrino events with sufficient precision for the aforementioned physics programs. Liquid argon time projection chambers (LArTPCs) meet these requirements by providing sub-cm spatial resolution of individual charged-particle tracks while remaining scalable to the required detector sizes \cite{RevModPhys.96.045001}. Due to these advantages, LArTPCs have now been employed in a variety of modern neutrino experiments \cite{app11062455}. The caverns at SURF have been excavated to allow for four far detector modules, with each one sized to hold 17 kilotonnes of liquid argon. The designs of the first two modules have been finalized as LArTPCs with the horizontal drift (HD) \cite{HD_TDR} and vertical drift (VD) designs \cite{VD_TDR}. The designs of the third and fourth modules remain under consideration \cite{DUNEPhaseII}.

A major challenge for the operation of large LArTPCs such as DUNE's far detectors is the design of their charge readout electronics, sometimes also referred to as the time projection chamber (TPC) electronics. In DUNE, these electronics must handle readout of hundreds of thousands of channels, and each channel must reliably resolve input charges of fewer than 10,000~electrons. In cryogenic detectors such as LArTPCs, cold electronics operating inside the detector offer significant advantages over warm electronics. These advantages become increasingly vital as detector sizes grow \cite{Radeka_2011}. When operated at 87\,K in liquid argon, complementary metal-oxide-semiconductor (CMOS) electronics benefit from a number of improvements in their fundamental properties that reduce the overall electronics noise \cite{CryoCMOS}. In addition, cold electronics can be placed closer to the detector's sensing components, greatly reducing their input capacitance compared to warm electronics, which must bring signals out of the cryostat for readout. This contributes further to their lower noise levels. Most importantly, cold electronics can substantially reduce the required amount of cold cables by providing multiplexing capabilities, which simplifies the mechanical design of the detector. For these reasons, several LArTPC-based neutrino experiments have incorporated cold front-end electronics in their charge readout systems to varying degrees \cite{CHEN2023167571}. However, cold electronics also face challenges that warm electronics do not, stemming from the need to operate them in a cryogenic environment without the possibility of replacement for extended periods of time. In the case of DUNE, the intended operational lifetime is several decades. Correspondingly, it is essential that these electronics be tested under realistic detector conditions and at the scales required for DUNE.

A central part of the DUNE program has been the CERN Neutrino Platform \cite{Pietropaolo_2017}, which has provided the infrastructure for testing large-scale LArTPC prototypes. The first DUNE far detector prototype tested at the Neutrino Platform was the single-phase (SP) design, in the form of the ProtoDUNE-SP demonstrator \cite{PDSP_TDR,PDSP_design}. ProtoDUNE-SP operated as a LArTPC with an active volume of $(7.2 \times 6.1 \times 7.0)$\,m$^3$ from 2018 to 2020 \cite{PDSP_results}, recording interactions from particles from the H4-VLE tertiary beamline \cite{PhysRevAccelBeams.20.111001,PhysRevAccelBeams.22.061003} and producing several new measurements of hadron-argon interactions \cite{PDSP_kaon,PDSP_lowkaon,PDSP_pionproton,PDSP_pion}. While ProtoDUNE-SP showed that its detector design was fundamentally sound and could meet the requirements of a DUNE far detector, it also showed there was room for improvement in a number of the detector's subsystems, including the high voltage systems, charge readout electronics, and photon detectors.

The current HD design evolved from the SP design after updating these various detector components, incorporating lessons learned from the ProtoDUNE-SP experience. The Neutrino Platform has once again been used to test this updated design, in the form of the \mbox{ProtoDUNE-HD} detector. ProtoDUNE-HD reused the ProtoDUNE-SP cryostat, but it replaced all of the interior components of the TPC with newly assembled final versions of hardware expected to be used in DUNE's HD far detector. ProtoDUNE-HD ran for 7 months from May 2024 to December 2024. During that time, it collected 10 weeks of test beam data from the same H4-VLE beamline that ProtoDUNE-SP used.

This paper presents the design and performance of the TPC electronics used for charge readout in ProtoDUNE-HD. These electronics will also be used in DUNE's HD far detector and the bottom half of DUNE's VD far detector. Other subsystems in ProtoDUNE-HD, as well as analysis of the test beam data it collected, will be described in future work. Section \ref{sec:overview} summarizes the principles of charge signal formation in a LArTPC and provides an overview of the TPC electronics system, including how it interfaces with the rest of the detector. This overview section incorporates references to sections \ref{sec:ASICs} and \ref{sec:FEMB}, which contain detailed descriptions of the cryogenic components of the electronics chain, and to sections \ref{sec:wib} and \ref{sec:ptc}, which contain detailed descriptions of the warm interface electronics. Section \ref{sec:performance} presents analysis of the TPC electronics performance in ProtoDUNE-HD, and we conclude in section \ref{sec:conclusion}.

\section{TPC and Electronics Chain Overview}
\label{sec:overview}

In the HD LArTPC design employed by ProtoDUNE-HD, the TPC's drift volume is filled with liquid argon and subjected to an electric field of ${\sim}500$\,V/cm. When charged particles pass through this region, they ionize the argon atoms. The electrons freed by this ionization are then drifted by the applied electric field toward anode plane assemblies (APAs) at one end of the volume, where the signals are collected.

The APAs are instrumented with conductive wires arranged in four layers spaced 4.75\,mm away from each other and with 4.7--4.8\,mm pitch. Starting from the layer closest to the detector's drift volume, these four layers are biased at voltages of $-665, -370, 0,$ and \mbox{$+820$\,V} respectively. The wire layer set to $-665$\,V, directly facing the drift volume, serves as a shield layer to block long-range induction effects; signals from this layer are not recorded. The inner three layers are connected to the TPC electronics, which are responsible for reading out the electron-induced signals on these wires. Signals from these three layers of wires provide the bulk of the detector's tracking and calorimetric information. The wires biased at $-370$\,V and $0$\,V are called induction layers, as they only pick up inductive signals from ionization electrons drifting past them. The final layer at \mbox{$+820$\,V}, farthest from the detector's drift volume, is referred to as the collection layer. Ionization electrons are directed by the electric field towards this layer, where they are collected. Further details of the DUNE HD LArTPC design can be found in ref. \cite{HD_TDR}.

\begin{figure}
    \centering
    \includegraphics[width=\linewidth,trim=1.5cm 2cm 3.5cm 2cm,clip]{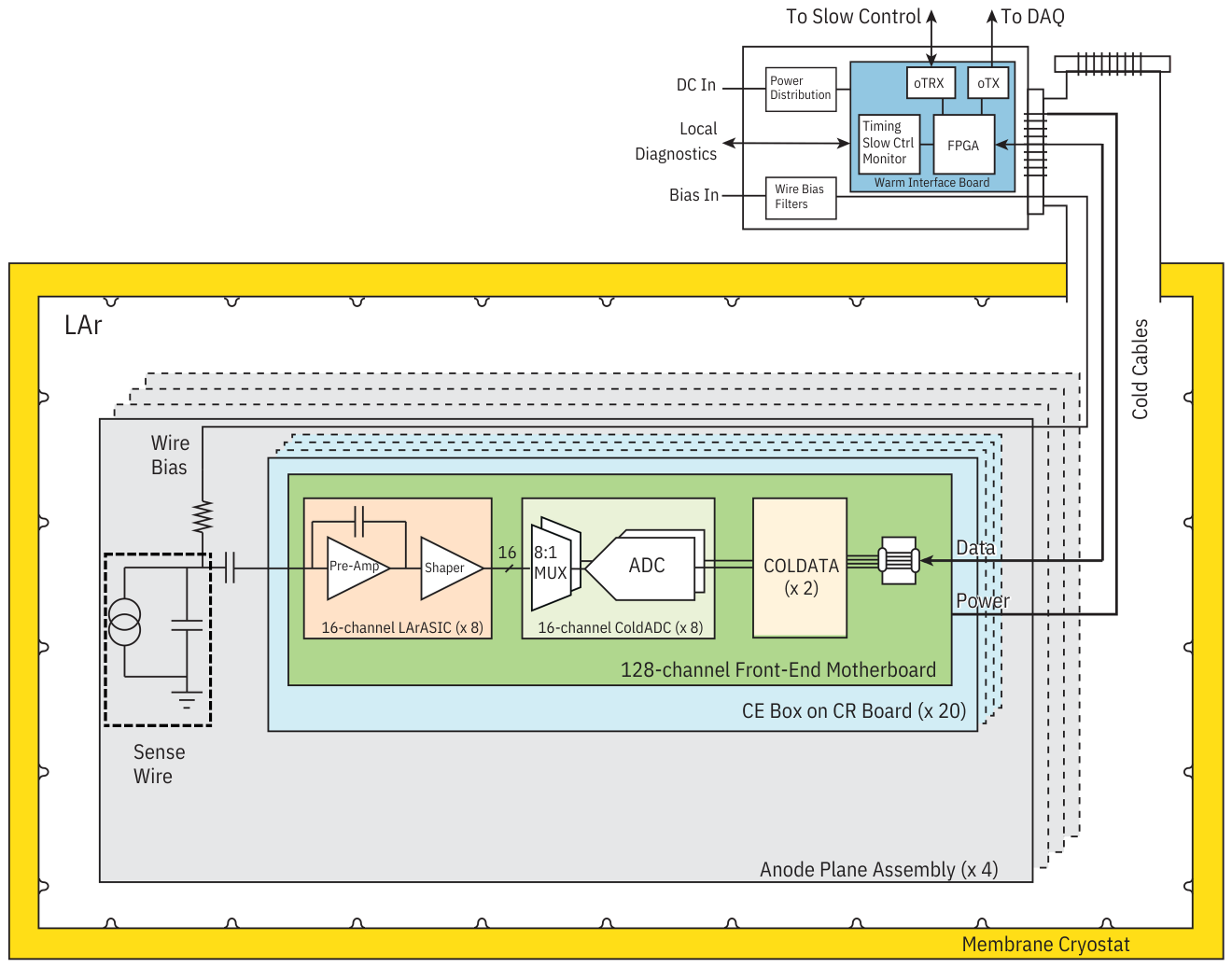}
    \caption{\label{fig:ce-chain} Schematic of the TPC electronics readout chain for ProtoDUNE-HD. The DUNE HD far detector uses the same design, scaled up to 150 APAs. The APA sense wires are represented on the left as an equivalent current source for electron signals from the TPC and equivalent capacitance. Their signals are connected to the front end of LArASICs through a physical capacitor. The LArASIC, ColdADC, and COLDATA ASICs are described in further detail in section \ref{sec:ASICs}. The ASICs are arranged onto front-end motherboards, which are described in section \ref{sec:FEMB}. These motherboards are contained within cold electronics (CE) boxes, which are plugged into capacitive-resistive (CR) boards on the APA \cite{HD_TDR}. The interface to all external systems is managed by warm interface boards shown on top of the cryostat, which are described in section \ref{sec:wib}.}
\end{figure}

 A schematic of the full ProtoDUNE-HD TPC electronics readout chain is shown in figure~\ref{fig:ce-chain}. Each induction and collection wire is connected to one channel of an application-specific integrated circuit (ASIC) called LArASIC \cite{LArASIC_CHARMS} that serves as a charge amplifier. These signals are then fed to the ColdADC ASIC \cite{ColdADC}, which performs digitization. Finally, the digital signals are encoded and transmitted to the warm electronics by the COLDATA ASIC. These three ASICs are all \mbox{custom-designed} to operate in liquid argon and fulfill DUNE's far detector performance requirements. They are described in further detail in section \ref{sec:ASICs}. Eight LArASIC chips, eight ColdADC chips, and two COLDATA chips are assembled onto a single front-end motherboard (FEMB), which is responsible for the readout of 128 channels. FEMBs are installed directly on the APAs inside the liquid argon volume of the TPC, with each APA holding 20 FEMBs to read out its 2,560 wires (800 from each of the two induction layers, and 960 from the collection layer). The layout and design of the FEMBs is described further in section \ref{sec:FEMB}.

Each FEMB is connected to a warm interface board (WIB) by a power cable and a data cable. The WIB uses the power cable to supply voltages to the power regulators on the FEMB. The data cable allows the WIB to transmit timing signals and configuration commands to the FEMB, and it allows the FEMB to transmit data from the ASICs up to the WIB. The WIBs are responsible for continuously transmitting that data to the data acquisition (DAQ) system in appropriate formats. Each WIB sits in a warm interface electronics crate (WIEC) on a penetration on top of the cryostat. Each APA is connected to a single WIEC with five WIBs, which is sufficient to manage the APA's 20 FEMBs. The WIBs receive power and a global timing signal from a power and timing card (PTC) located in the WIEC, with the PTC receiving external power and timing. The WIBs are described in further detail in section \ref{sec:wib}, and the PTCs are described more in section \ref{sec:ptc}.

\section{Cryogenic ASICs}
\label{sec:ASICs}

The core of DUNE's cryogenic TPC electronics design is the 3-ASIC solution, which uses three types of ASICs operating entirely in liquid argon to divide the tasks of charge amplification (LArASIC), digitization (ColdADC), and transmission to warm electronics (COLDATA). All three ASICs are fabricated using CMOS processes from Taiwan Semiconductor Manufacturing Company (TSMC). The LArASIC is produced using their 180 nm CMOS process, and the ColdADC and COLDATA are produced using their 65 nm CMOS process. This section provides details on the design and operation of each of these ASICs, as well as their interfaces to each other. Particular attention is given to COLDATA, which, unlike the LArASIC and ColdADC, has not been fully described in previous work.

\subsection{LArASIC}
\label{sec:larasic}

LArASIC is the front-end charge amplifier in the readout chain, responsible for amplifying and shaping the raw charge signals coming from the LArTPC. Earlier versions of this ASIC have been used in other LArTPC-based neutrino experiments \cite{CHEN20121287,8533137}, including in ProtoDUNE-SP, where it was referred to as the ``FE ASIC'' \cite{Adams_2020}. Since details of this ASIC's overall architecture have been previously described in other work \cite{5874057,LArASIC_CHARMS}, this section summarizes only some of the high-level features of the LArASIC and discusses the configuration choices relevant to operation in ProtoDUNE-HD.

\begin{figure}
    \includegraphics[width=\linewidth]{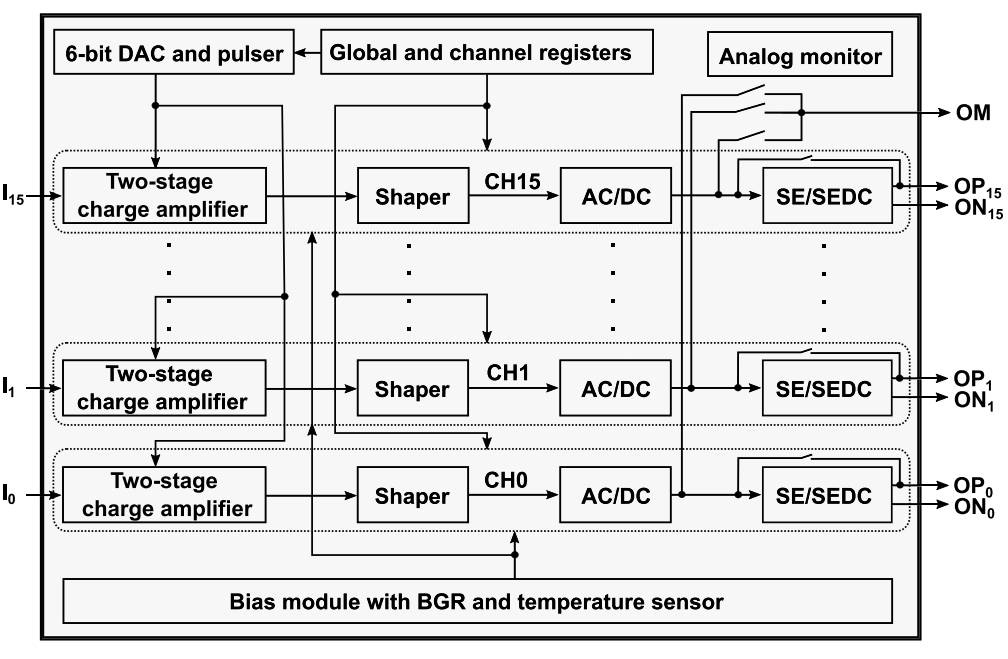}
    \caption{\label{fig:LArASICBlock} Block diagram for the LArASIC, reproduced from ref. \cite{LArASIC_CHARMS}. The chip has both global and per-channel configuration registers, which are managed through a serial peripheral interface. Each channel input $I_n$ is connected to an APA sense wire. Each channel consists of its own charge amplifier and shaping amplifier, whose output can be AC-coupled or DC-coupled and passed through a SE or SEDC buffer. The outputs $OP_n$ (and $ON_n$ if the SEDC buffer is used) are routed to the 16 inputs of a single ColdADC. Biasing circuits all use an on-chip bandgap reference (BGR). The 6-bit DAC is used to control an on-chip pulser that injects charge into the charge amplifiers. As part of the analog monitor system, the amplifier outputs can be directed to the output monitor (OM) line for diagnostic purposes, described further in section \ref{sec:femb_calibmonitor}.}
\end{figure}

The high-level block diagram in figure \ref{fig:LArASICBlock} shows the components of the LArASIC described in this section. The LArASIC consists of 16 channels that can be configured independently through a serial peripheral interface (SPI) with the COLDATA, with a range of settings available to suit different operating conditions. Incoming charges first go through a two-stage charge amplifier, with the second stage having an adjustable gain. The amplified charges next pass through a 5th-order shaping amplifier, which converts the fast current pulses into semi-Gaussian voltage pulses with well-defined widths that are suitable for digitization. These shaped pulses can be configured with a peaking time $t_p$ (defined as time from $1\%$ to peak amplitude) of 0.5, 1, 2, or 3\,$\mu$s. The total gain can be configured to 4.7, 7.8, 14, or 25\,mV/fC, corresponding to input charge ranges of 300, 180, 100, or 56\,fC respectively. The amplifier's performance is also impacted by the choice of adaptive reset current (also referred to as the reset quiescent current), which can be set to 0.1, 0.5, 1, or 5\,nA. All results in this paper have this current set to 0.5\,nA, which optimizes for high enough current to prevent saturation effects under normal circumstances while avoiding the increase in noise associated with larger reset currents \cite{9875569}. The gain and peaking time are chosen to maximize dynamic range while minimizing noise; ProtoDUNE-HD operated with 7.8\,mV/fC gain and 2\,$\mu$s peaking time. This choice is discussed further in section \ref{sec:performance}.

The output of the amplifier can be AC-coupled or DC-coupled, and it can be optionally sent through a single-ended (SE) or single-ended-to-differential converter (SEDC) buffer to drive larger capacitive loads. This can be useful when the LArASIC output is physically located far from digitization electronics, as can be the case when a cryogenic analog-to-digital converter (ADC) is not available. However, on typical FEMBs, the immediate proximity of the ColdADC to the LArASIC eliminates this need. Correspondingly, in all results in this paper, the LArASICs were operated with DC-coupled outputs and with the SE and SEDC buffers both bypassed.

Each channel output can be configured to a pedestal of either 200\,mV or 900\,mV, which are respectively near the bottom and middle of the LArASIC's 0--1.8\,V output range. The 200\,mV pedestal is used for channels connected to collection wires in the TPC, which expect to see primarily unipolar signals. The 900 mV pedestal is used for channels connected to induction wires, which see primarily bipolar signals. This maximizes the available dynamic range for both types of channels.

For calibration and test purposes, the LArASIC is equipped with an on-chip pulser that can inject known amounts of charge to each channel by using a series of fixed output voltages through a test capacitor. The full charge injection range of this pulser varies with the chosen gain setting. It spans 238, 180, 100, or 56\,fC, corresponding respectively to gain settings of 4.7, 7.8, 14, or 25\,mV/fC. An on-chip 6-bit digital-to-analog converter (DAC) controls the voltages used to supply these injected pulses, providing uniform steps from zero to the maximum permitted pulse amplitude. The voltage output of each step of the DAC is measured during individual quality control (QC) tests and has been found to be linear to within $0.3\%$. The test capacitor through which the voltage output is injected is a 185\,fF metal-insulator-metal (MIM) capacitor built into each channel on the chip. With the CMOS process used for LArASICs, the MIM capacitance values are expected to vary by $<2\%$ between chips and $<2\%$ between 300\,K and 87\,K. While the standard LArASIC QC process does not measure each MIM capacitor to this level of precision, separate dedicated measurements of a random sample of LArASICs have confirmed that the variation in capacitance among the MIM capacitors is within these bounds. Through the known voltage outputs of the DAC and the known capacitance of the MIM capacitors, controlled quantities of input charge are used to provide initial calibrations of the gain of each channel. This charge injection scheme is described further in section~\ref{sec:femb_calibmonitor}, and the corresponding analysis procedure is described in more detail in section \ref{sec:pulsers}.

\subsection{ColdADC}

The ProtoDUNE-SP ADC ASIC had notable performance issues under cryogenic conditions, despite acceptable warm performance. Among its primary limitations were the appearance of missing codes\footnote{Some output codes would never appear in the data as a result of being skipped over in the ADC response function.}, sticky codes\footnote{The ADC would get stuck on certain output bits, giving the same output code for a wide range of input voltages.}, and non-monotonic sections of its transfer curve\footnote{Within certain input ranges, an increase in the input voltage could result in a smaller output code from the ADC.}~\cite{HD_TDR}. While extensive quality control screening was conducted to select the fraction of ADC ASICs least affected by these issues \cite{Adams_2020}, the selected ASICs still required special treatment during analysis of ProtoDUNE-SP data to correct for the remaining non-idealities \cite{PDSP_results}. This problem is believed to have arisen from the ADC architecture's particular susceptibility to mismatches in transistor characteristics, which worsened considerably under cryogenic conditions. As a result, the collaboration decided to overhaul the ADC ASIC with a completely new design.

The ColdADC ASIC is the result of this overhaul. Like the ProtoDUNE-SP ADC ASIC, it is responsible for digitizing the amplified analog signals coming from the LArASIC. It is placed in close proximity to the LArASIC outputs to minimize noise pickup during transmission between the two chips. The ColdADC has eliminated all major issues seen with the ProtoDUNE-SP ADC, and so special screening for the least poorly behaved chips is no longer required. Details of the ColdADC's design and performance have been previously reported \cite{ColdADC}, but this section summarizes its features that are most pertinent to the general FEMB layout.

\begin{figure}
    \includegraphics[width=\linewidth,trim=0cm 4.5cm 4.5cm 3.5cm,clip]{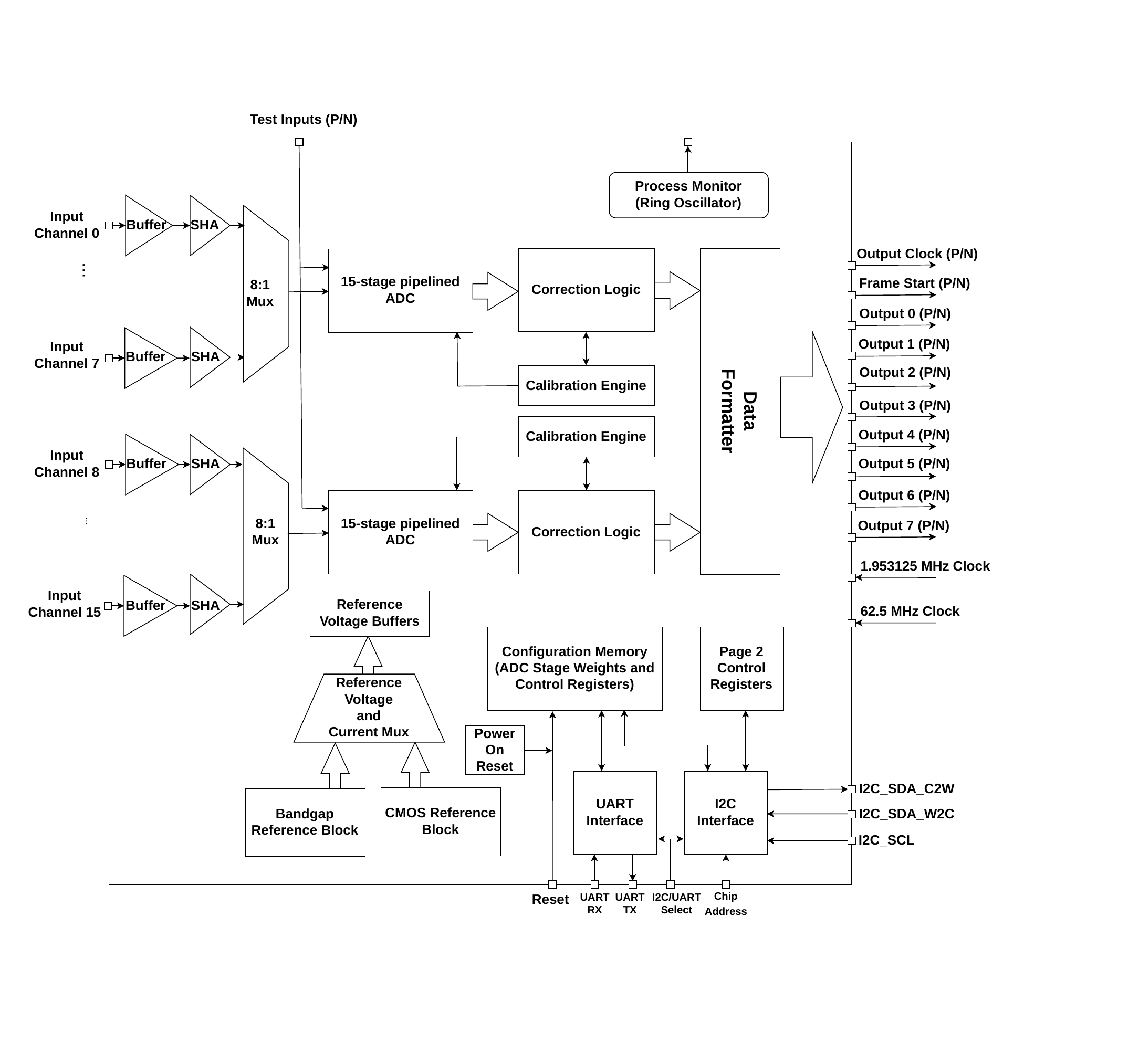}
    \caption{\label{fig:ColdADCDiagram} Block diagram for the ColdADC, adapted from ref.~\cite{ColdADC}. The 16 input channels on the left are connected to the 16 outputs of one LArASIC, and incoming data flows through the digitization pipeline described in the text. Dedicated test input pins allow for signals to be injected directly into the pipeline ADCs, bypassing the SHAs. Digitized data from the \mbox{16 channels} is transmitted along eight digital outputs, which are sent to the COLDATA. The 62.5 MHz and 1.953125 MHz input clocks are provided by a controlling COLDATA. The I2C interface is used by the COLDATA to configure the ColdADC, but a UART interface is also provided for testing purposes. A ring oscillator is included to allow tests to check for variations in the manufacturing process.}
\end{figure}

The block diagram for the ColdADC is shown in figure \ref{fig:ColdADCDiagram}. Each ColdADC ASIC handles \mbox{16 channels}, which are internally processed by two 8-channel ADCs. The ColdADC channel inputs are directly connected to the outputs of one LArASIC. The channels can operate with either single-ended or differential inputs to match the LArASIC output mode. For all results presented in this paper, they were operated in single-ended mode, for the reasons discussed in section \ref{sec:larasic}.

Each input feeds into a sample-and-hold amplifier (SHA). The 16 SHAs are divided into two blocks of eight. Each block of eight SHAs is processed by its own pipeline ADC, which samples the SHAs in turn through an 8:1 multiplexer (MUX). The pipeline ADCs use a 1.5 bit/stage architecture with 15 stages, digitizing each channel into 16-bit values at a rate of one sample per 512\,ns. This digitization rate is set by the clock that the COLDATA feeds to the ColdADC. The pipeline ADCs output 16-bit two's complement integers, which are converted to offset binary format (so that the smallest output code is 0x0000 and the largest output code is 0xFFFF) before being sent to the COLDATA. The COLDATA is responsible for truncating the ColdADC outputs back down to the 14-bit values that are used in the final data.

In order to achieve the desired linearity without relying on precise capacitor matching, which can become unreliable under cryogenic conditions, the pipeline ADCs include self-calibration engines \cite{KaranicolasCalibration}. The self-calibration procedure takes advantage of the fact that the desired level of precision is easily satisfied by the last stages of the pipeline ADC. It thus targets the first seven stages, starting from the seventh stage and then proceeding through the more significant stages. For each stage it calibrates, it uses the known decision threshold voltages as inputs to the stage and checks the output of the rest of the pipeline ADC after forcing the normal comparator outputs to 1 or 0. The differences in results are used to automatically adjust the weights of that stage in the ADC to correct for any gain and offset errors. This procedure requires approximately 250\,ms to run, and the ADC ignores input during that time. As a result, the self-calibration procedure cannot be used during active data collection. However, the calibration constants have been found to be stable over the course of months, and so it is not necessary to frequently re-run the calibration procedure. Instead, it is opportunistically run whenever the electronics are reconfigured for a new detector data acquisition period. Since data collection must be paused during reconfiguration anyway, this causes negligible loss in livetime.

\subsection{COLDATA}
\label{sec:coldata}

COLDATA is a control and communication ASIC designed to control four LArASICs and four ColdADCs, as well as to concentrate data from four ColdADCs. Each COLDATA can receive commands from either a WIB or another COLDATA. This flexibility removes the need for dedicated communication lines between every COLDATA and the WIBs, reducing the required number of cold cable connections. Commands carry information identifying their intended recipient, allowing each COLDATA to respond to those addressed to itself and relay all others to the appropriate ASIC, including passing responses from the destination back to the sender.

COLDATA uses low-voltage differential signaling (LVDS) over a twinax cable to communicate with the WIB controlling it, following an I2C-like protocol.  The key distinction between this protocol and standard I2C \cite{I2C} arises from the use of LVDS lines instead of the typical single-ended I2C lines. This is necessary in order to meet the requirement that these communications be able to traverse long cables between the WIB and COLDATA; the lengths of these cables reach 22\,m in the HD design and can reach up to 35\,m in alternative DUNE detector designs. The serial clock (SCL) line can use LVDS without issue, but since LVDS is not compatible with bi-directional signaling, separate serial data (SDA) lines are used for data sent from the WIB to COLDATA (W2C = warm to cold) and for data sent from COLDATA to the WIB (C2W = cold to warm). For simplicity, this modified protocol will be referred to as I2C.

Commands intended for the other COLDATA or for one of the eight ColdADCs are relayed to the target ASIC using single-ended CMOS signals, which are less power-intensive than LVDS. The I2C protocol is used to read and write control and configuration registers in both COLDATAs and ColdADCs, as well as to download LArASIC configuration data to COLDATA. COLDATA uses a custom serial programming interface to write and read back LArASIC configuration data.

\begin{figure}
    \includegraphics[width=\linewidth]{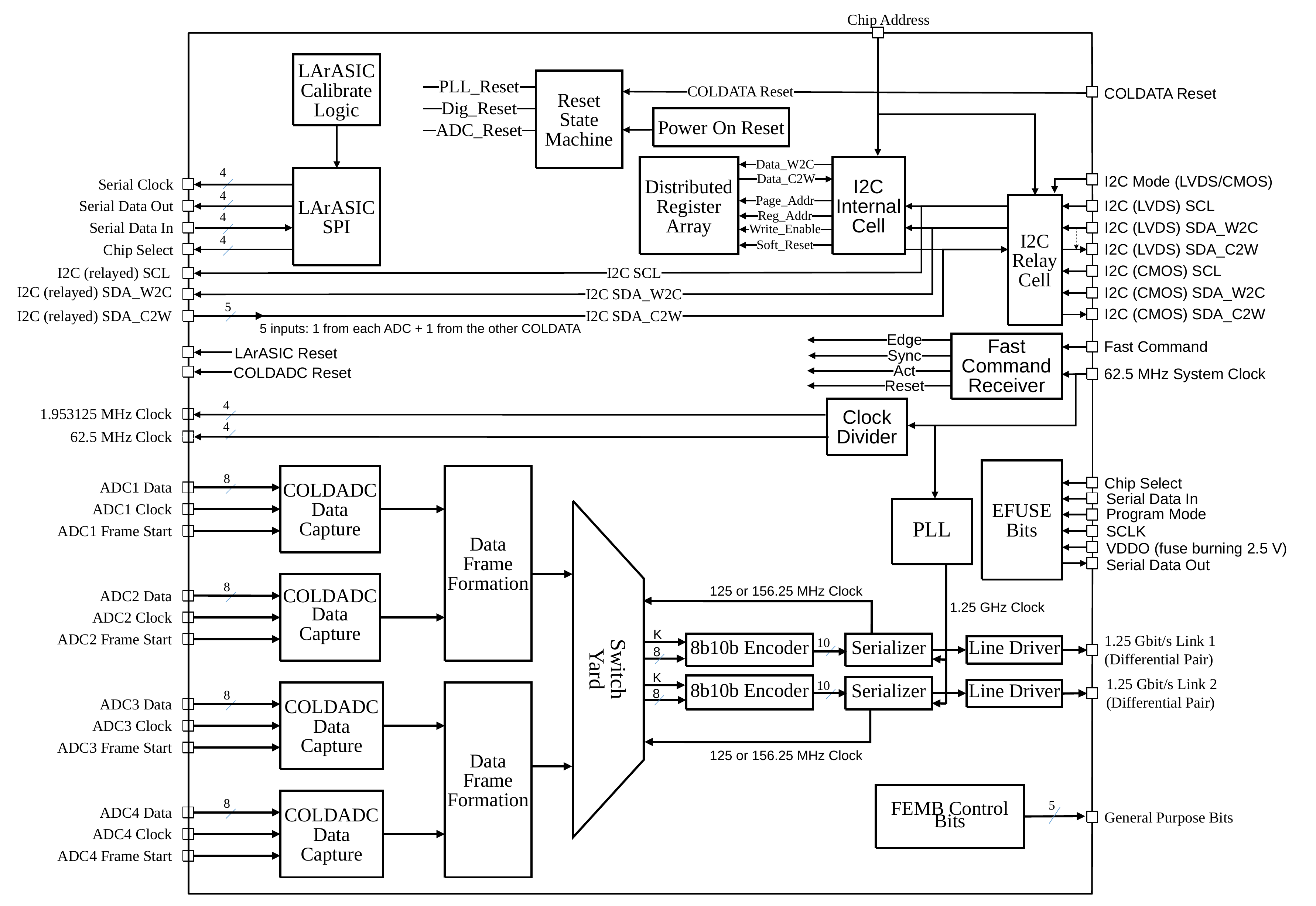}
    \caption{\label{fig:ColdDATA_BlockDiagram} COLDATA block diagram. Connections to the LArASICs and ColdADCs are shown on the left, and connections to the WIB or other pieces of the FEMB are shown on the right. Communication blocks are shown in the upper section. I2C communications are mediated by the I2C relay cell, which forwards all messages to their designated recipients. The input 62.5\,MHz system clock is used to generate the 62.5\,MHz and 1.953125\,MHz clocks that the ColdADC uses, as well as to feed the COLDATA's phase-locked loop (PLL). The bottom section shows the flow of data from the ColdADCs on the left to the 1.25 Gbit/s output links on the right. Five general purpose inputs/outputs can be configured with the general purpose bits shown in the bottom right.}
\end{figure}

Besides handling commands, COLDATA is also responsible for cold to warm data transmission. It merges the data streams it receives from four ColdADCs, provides standard 8b10b encoding \cite{8b10b}, serializes the data, and sends the data over two 1.25 Gigabit per second (Gbit/s) links to its controlling WIB. The speed of these links provides enough bandwidth to continuously transmit the 14-bit values being received from each ColdADC every 512\,ns\footnote{With two serial links, each one must carry the data of 32 ColdADC channels, which amounts to a raw data rate of 0.875 Gbit/s. The overhead from 8b10b encoding and frame headers and trailers brings this up to 1.25 Gbit/s.}. These links are driven by line drivers with programmable transmitter equalization and pre-emphasis.

The block diagram of the entire COLDATA is shown in figure \ref{fig:ColdDATA_BlockDiagram}. All of the functions of COLDATA are synchronized to a global 62.5\,MHz DUNE system clock, which it receives directly from its controlling WIB through a LVDS pair. COLDATA derives and provides the ColdADC sample clock by dividing the 62.5\,MHz clock by 32, resulting in the ColdADC's one sample per 512\,ns.

\begin{figure}
    \includegraphics[width=\linewidth]{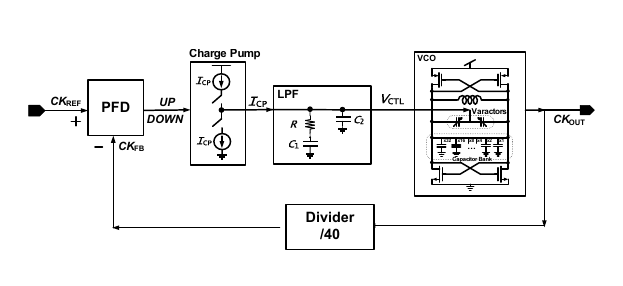}
    \caption{\label{fig:PLL_circuit_diagram} Block diagram of the COLDATA PLL, which generates a 2.5\,GHz clock for use by the data serializers. The phase and frequency detector (PFD) takes the 62.5\,MHz reference clock and the PLL's feedback clock as input. The differences are used to produce $UP$ and $DOWN$ signals used as input to the charge pump, which uses a tunable bias current $I_{CP}$. The charge pump output goes through a low-pass filter (LPF) before being used to drive the voltage-controlled oscillator (VCO), which produces the output clock. The feedback clock is extracted from this output with a divide-by-40 circuit, which cuts the 2.5\,GHz output back down to match the 62.5\,MHz input clock.}
\end{figure}

\paragraph{{Phase-Locked Loop}}

The COLDATA Phase-Locked Loop (PLL), shown in figure \ref{fig:PLL_circuit_diagram}, receives the 62.5\,MHz clock from the WIB and produces a phase-locked 2.5\,GHz output clock. The PLL output is then passed through a divide-by-2 circuit to create a 1.25\,GHz output clock\footnote{The PLL does not natively output the required clock frequency because it was designed before the speed of the COLDATA data links was finalized.}. This clock is used to feed the two Serializer blocks, so that their data streams are synchronized with the system clock. The PLL consists of a phase and frequency detector (PFD), a charge pump, a low-pass filter (LPF), and a digital divide-by-40 circuit. The PFD calculates phase and frequency differences between the 62.5\,MHz reference clock ($CK_{REF}$) and the feedback clock that the PLL generates by dividing its own output clock by 40 ($CK_{FB}$). These differences are used to generate the $UP$ and $DOWN$ signals that drive the charge pump to charge or discharge the low-pass filter. The control voltage ($V_{CTL}$) developed at the output of the low-pass filter tunes the varactors used as voltage-controlled variable capacitors in the Voltage-Controlled Oscillator (VCO), thereby adjusting the phase and frequency of the PLL's output clock $CK_{out}$.  The charge pump current and the capacitance in the LC-based VCO are programmable to compensate for process and temperature-dependent variations. The bandwidth of the PLL is designed to be 1.5\,MHz.  The inductor in the VCO has an inductance of 1.4\,nH and a quality factor of 14, and the VCO's gain $K_{VCO}$ is set to be $\sim$100\,MHz/Volt.

\paragraph{Data Flow}

COLDATA receives data in 16-bit words from each of four ColdADC ASICs over LVDS pairs. The data from a single ColdADC is handled by a single ColdADC Data Capture block. The input to this block includes eight LVDS pairs carrying data from the ColdADC at a rate of \mbox{62.5 Megabits} per second (Mbit/s) each, synchronous to the 62.5\,MHz system clock. This block also receives a copy of the ColdADC's 62.5\,MHz clock as input, phased so that it can be used to acquire the serial data. Lastly, there is a dedicated line for a frame start signal from the ColdADC, which the block uses to recognize the 4-bit nibbles of data in each 16-bit word.

In the next step, each Data Frame Formation block receives data from two ColdADCs (\mbox{32 channels}), formatted by their respective ColdADC Data Capture blocks. The Data Frame Formation blocks can be configured to truncate the 16-bit ADC data to either 12 or 14 bits and then pack the data into bytes. To assist with testing, they can also be configured to ignore input ColdADC data and instead pack 12-bit dummy data into bytes. While the 12-bit frame format was included to allow for a possible scenario where the ADC only had 12-bit precision, the ColdADC possesses sufficient precision for 14-bit ADC values to be useful. Consequently, the 14-bit frame format is used for all normal data taking. In anticipation of the data being 8b10b encoded, as described below, the Data Frame Formation block creates frames that include two 8b10b K character header bytes, a 15-bit timestamp pulled internally from the COLDATA, an 8-bit checksum for the data from each ColdADC, and a trailer K character byte that indicates the end of the data frame.

The Switch Yard block repeatedly switches between its input data frames to pass them sequentially to the 8b10b Encoder blocks, which apply 8b10b encoding to the data. These receive entire 8-bit bytes at a time as input. A dedicated K line input flags bytes that should be interpreted as K characters. In normal operation, the 8b10b Encoder blocks use a lookup table to produce 10-bit words, which are transmitted in their entirety to the Serializer blocks at a rate of 125\,MHz. The 8b10b Encoder block can also be configured to ignore input from the Switch Yard and send a PRBS7 pattern to the Serializer blocks at 156.25\,MHz. This can be useful for debugging the COLDATA output, as PRBS7 patterns can be used by many oscilloscopes and field programmable gate arrays (FPGAs) to generate eye diagrams to quantify the signal quality\footnote{PRBS7 is often used to validate serial links that use 8b10b encoded data due to its similarity in DC balance and numbers of sequential 1s or 0s.}.

\begin{figure}
    \begin{center}
        \includegraphics[width=\linewidth]{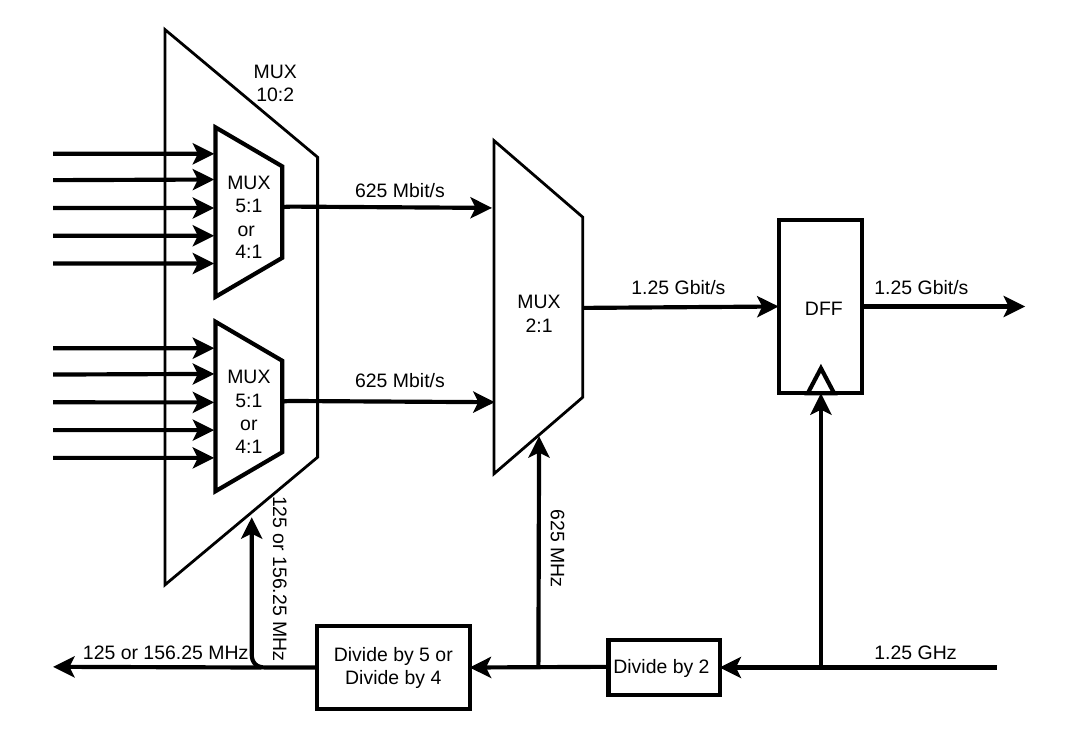}
    \end{center}
    \caption{\label{fig:coldata_serializer} Block diagram for the COLDATA Serializer. 10-bit words are received as input from the 8b10b Encoder block on the left. These go through two 5:1 multiplexers (MUX), resulting in two 625 Mbit/s outputs. These outputs are then combined by a 2:1 MUX to result in the final \mbox{1.25 Gbit/s} output, which goes through a D flip-flop (DFF) before being passed on to the line driver. The Serializer receives a 1.25\,GHz clock from the PLL as input, which it internally divides to generate the clocks needed for its multiplexers. In debugging mode, the Serializer uses a \mbox{156.25\,MHz} clock instead of 125\,MHz and converts the 5:1 multiplexers into 4:1 ones.}
\end{figure}

The Serializer block is responsible for serializing the bytes coming out of the 8b10b Encoder block so that they can be transmitted along a single line. A diagram of this block is shown in figure~\ref{fig:coldata_serializer}.  In the normal 10-bit mode, the 10 bits coming in at 125\,MHz are input to two 5:1 multiplexers. The outputs of these multiplexers are then sent to a 2:1 multiplexer to create a single output at the desired 1.25 Gbit/s rate. This output is registered by a D flip-flop to align it to the 1.25\,GHz clock and then finally sent to a line driver. The 625\,MHz and 125\,MHz clocks required by the multiplexers are produced by dividers within each Serializer out of the 1.25\,GHz clock supplied to it by the PLL. The 125\,MHz clock is also sent to the Switch Yard, which it uses to transmit its data to the Serializer.  An 8-bit mode is included for debugging purposes.  In this mode, no 8b10b encoding is done and the serial output is not DC balanced.  In 8-bit mode, parallel data is instead input at 156.25\,MHz and the first multiplexer stage operates as two 4:1 multiplexers.

Lastly, the Serializer's output 1.25 Gbit/s data stream is handed to a hybrid-mode line driver, which transmits the data to the warm electronics. These line drivers are designed with current-mode transmitter equalization and voltage-mode pre-emphasis to be able to drive 25--35\,m twinax cables \cite{COLDATA_LD}, which are the upper limit of expected possible cable lengths for DUNE.  The current-mode transmitter equalization circuit uses a finite impulse response filter to distort the data pulse to compensate for the large frequency-dependent signal loss over a long twinax cable.  The voltage-mode main driver and pre-emphasis circuit use source-series-terminated output stages to provide a large output voltage swing and low static power consumption.  The line driver is highly programmable and can also be operated in a pure current-mode or a pure voltage-mode, with the pure current-mode with no equalization or pre-emphasis being appropriate for use with short cables. This configurability allows COLDATA to be used with a variety of detectors with different cable lengths, under both warm and liquid argon conditions. 



\paragraph{EFUSE Registers}

Each COLDATA contains a 32-bit Electrical Fuse (EFUSE) register, which can be permanently programmed using dedicated I/O pins. A unique value is programmed into each COLDATA's EFUSE register after it passes quality control tests.  The contents of the EFUSE register can be transferred to four read-only I2C registers and read out by the WIB, allowing different COLDATA chips to be distinguished after they are installed.  This feature is used to verify that data cables and WIB data links are connected as intended.

\paragraph{Fast Commands}

Some commands, referred to as Fast Commands, are time-sensitive and must be executed by COLDATA at specific times. These commands are transmitted on a dedicated LVDS pair and are formatted as DC-balanced 8-bit words. The rising edges of Fast Command bits are synchronized with the rising edges of the 62.5\,MHz clock. The valid Fast Commands are listed in table \ref{tab:fast_commands}. The Act Fast Command is specially used to execute a number of Fast:Act Commands. For each of these, a dedicated I2C register is first loaded with the desired command. When an Act Fast Command is received, the Fast:Act Command loaded into that register is started on the next rising edge of the 62.5\,MHz system clock. The 10 valid Fast:Act Commands are summarized in table \ref{tab:fastact}, a few of which are described in more detail in the rest of this section. 

\begin{table}[h]
\centering
\caption{\label{tab:fast_commands} List of available Fast Commands and their functions. A received Fast Command is executed on the next rising edge of the 62.5\,MHz clock.}
\begin{tabular}{|l|p{9cm}|}
\hline
\textbf{Fast Command} & \textbf{Function} \\
\hline
Alert & Synchronize the Fast Command receiver with the sender \\ \hline
Idle  & No action \\ \hline
Edge  & Move the edge of the ADC sample clock to the next rising edge of the 62.5\,MHz clock \\ \hline
Sync  & Zero out the COLDATA timestamp \\ \hline
Act   & Perform the Fast:Act Command stored in the Act Command register \\ \hline
Reset & Reset the COLDATA \\
\hline
\end{tabular}
\end{table}

\begin{table}[h]
\centering
\caption{\label{tab:fastact} List of available Fast:Act Commands and their functions. A Fast:Act Command must be preloaded into the appropriate COLDATA register before activation, and it is executed immediately following the reception of the Act Fast Command.}
\begin{tabular}{|l|p{10cm}|}
\hline
\textbf{Fast:Act Command} & \textbf{Function} \\
\hline
Idle & No action \\ \hline
LArASIC-Pulse & Initiate the test pulse sequence defined in the relevant I2C registers. Issuing this command a second time stops the test pulse sequence \\ \hline
Save Timestamp & Save the current 15-bit COLDATA timestamp value into two read-only I2C registers \\ \hline
Save Status & Save various COLDATA status bits into the other two read-only I2C registers \\ \hline
Clear Saves & Clear all four read-only registers referenced in the previous two commands \\ \hline
Reset ColdADCs & Reset all four ColdADCs \\ \hline
Reset LArASICs & Reset all four LArASICs using their reset pads \\ \hline
SPI Reset & Reset all four LArASICs using SPI \\ \hline
Program LArASICs & Download the stored daisy-chain control bits from the \mbox{COLDATA} I2C registers to the LArASICs using SPI \\ \hline
Relay I2C-SDA & Echo incoming I2C SDA W2C signals back to the WIB along the I2C SDA C2W line\\
\hline
\end{tabular}
\end{table}

The Program LArASICs Fast:Act Command configures the LArASICs by reading a number of designated I2C registers on the COLDATA, which can be written beforehand in a non-time-sensitive manner. Upon execution of this command, the COLDATA SPI shifts the daisy-chain control bits stored in those I2C registers into the LArASICs twice.  Since SPI shifts a bit out each time a bit is shifted in, this allows the interface to check that the LArASICs were programmed correctly by verifying that the bits shifted out while the control bits are being shifted in the second time match the bits stored in the COLDATA's I2C registers.  The result of these comparisons is loaded into a read-only I2C register when a Save Status Fast:Act Command is issued.

The Relay I2C-SDA Fast:Act Command allows the WIB to measure the cable delay associated with the long signal cables connecting the COLDATA to the WIB. When this Fast:Act Command is issued, the COLDATA connects its incoming I2C SDA W2C line with a very small time delay to its outgoing I2C SDA C2W line for approximately 64~$\mu$s.  The WIB firmware uses this to measure the signal cable delay and then issues Edge and Sync Fast Commands at appropriate offsets. This ensures that all ColdADCs sample at the same time and that timestamps are synchronized for all COLDATAs, regardless of the length of the signal cables between each COLDATA and its corresponding WIB. This synchronization is achieved to a precision of one 62.5\,MHz clock tick (16\,ns).

\paragraph{Power}

The COLDATA bias voltages and typical current consumptions are shown in table \ref{tab:COLDATA_Power}. The current drawn from VDDIO, which sees the largest power consumption, is dominated by the current used by the LVDS drivers. While standard LVDS uses 3--4\,mA, LVDS drivers on the COLDATA can be programmed to use one of four different current settings: 2, 4, 6, or 8\,mA. A COLDATA that uses LVDS for I2C communication with the WIB is typically configured to use the 8\,mA setting, resulting in a total current draw from VDDIO of approximately 70\,mA. A COLDATA that receives I2C communications from another COLDATA has no termination resistors connected to most of its LVDS outputs and draws only around 30\,mA from VDDIO.

\begin{table}
    \centering
    \caption{COLDATA Power Domains. The current draw for VDDIO depends on the current setting for LVDS outputs. VDD-LArASIC current is very close to zero, except when the LArASICs are being programmed.}
    \label{tab:COLDATA_Power}
    \begin{tabular}{|c|c|c|c|}
    \hline
        \textbf{Name} & \textbf{Voltage} & \textbf{Usage} & \textbf{Typical Current}\\ \hline
        VDDIO & 2.25\,V &  All I/O except to/from LArASICs& 30-70\,mA\\ \hline
        VDD-LArASIC & 1.8\,V & I/O to/from LArASICs & 0\,mA\\ \hline
        VDDCORE & 1.2\,V & Core digital logic & 11\,mA\\ \hline
        VDDD & 1.2\,V & \begin{tabular}{@{}c@{}} Digital logic in PLL, serializers,\\and 1.25 Gbit/s line drivers\end{tabular} & 22\,mA\\ \hline
       VDDA  & 1.2\,V & PLL analog circuits & 9\,mA\\
    \hline
    \end{tabular}
    
\end{table}

\paragraph{Resets}

The COLDATA's Reset State Machine resets the entire chip in a prescribed order starting with the PLL, setting all control registers to default values in the process. This Reset State Machine can be started by issuing a Reset Fast Command, pulling the COLDATA Reset pad to ground, or issuing a reset signal through the Power On Reset circuit. The Power On Reset circuit is activated when VDDCORE is turned on, so that the COLDATA is brought up in its predefined default state. This circuit includes a Schmitt trigger for hysteresis. The length of time that the reset signal is asserted for when the power is turned on is temperature-dependent.  Simulations indicate that the duration of the reset signal is at least 8\,$\mu$s at 300\,K and 800\,$\mu$s at 87\,K.

\section{Front-End Motherboard}
\label{sec:FEMB}

The Front-End Motherboard (FEMB) integrates the ASICs described in section \ref{sec:ASICs} in order to perform digitized readout of 128 APA sense wires while immersed in liquid argon. Each FEMB holds eight LArASICs, eight ColdADCs, and two COLDATAs. This design is referred to as the monolithic FEMB design, in contrast with the ProtoDUNE-SP FEMB design that featured a mezzanine card with an FPGA \cite{Adams_2020}; functions handled by the FPGA in the ProtoDUNE-SP FEMB design are now handled by COLDATA. This section describes the layout and design of the FEMB outside the internal workings of the component ASICs, which are described in section \ref{sec:ASICs}.

\subsection{Layout}

\begin{figure}
    \centering
    \includegraphics[width=0.95\linewidth]{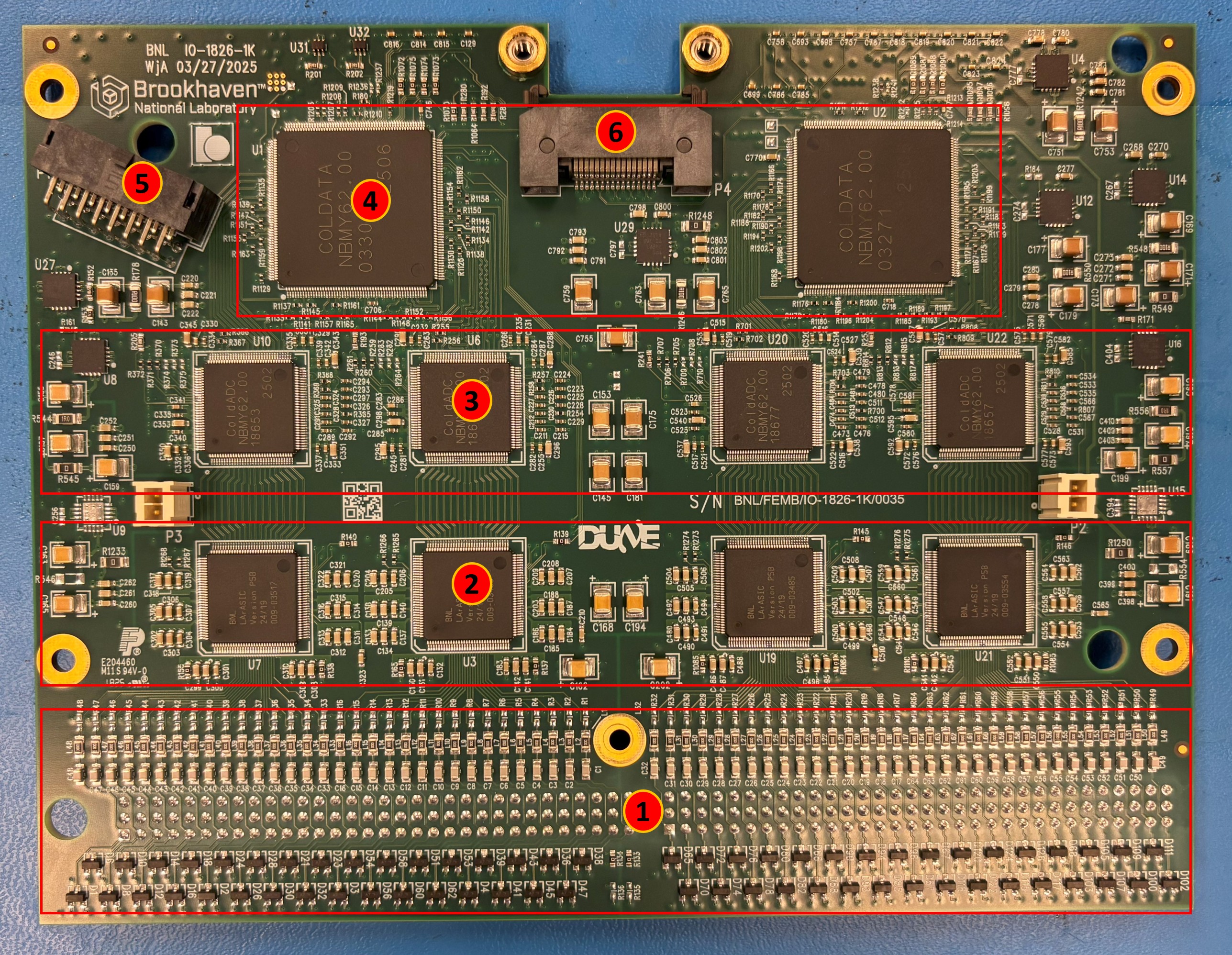}
    \caption{\label{fig:femb_layout} Picture of the top layer of a FEMB. The locations of input connectors and protection diodes (1), four LArASICs (2), four ColdADCs (3), two COLDATAs (4), the Samtec IPL1 connector for power (5), and the Samtec ASP 36-pin data connector (6) are indicated. Note that another four LArASICs and four ColdADCs are placed on the other side of the PCB, directly opposite the ones that are shown.}
\end{figure}

The FEMB hosts both analog and digital components within a double-sided printed circuit board (PCB). The organization of these components on the PCB is designed to minimize interference between them and maximize signal integrity. Figure \ref{fig:femb_layout} shows the physical separation of the different functional blocks. The bottom section provides the interface to the detector's sense wires with two 3-row, 96-pin through-hole connectors. Protection diodes are installed in the section immediately following these connectors in order to suppress any transient high-voltage discharges and safeguard the front-end amplifiers. Above this, the front-end section contains the LArASICs, which handle all analog amplification and shaping. This is followed by the middle section with ColdADCs, performing all analog-to-digital conversion. Lastly, COLDATAs occupy the top section, operating entirely in the digital domain.

\begin{figure}
    \centering
    \includegraphics[width=0.95\linewidth]{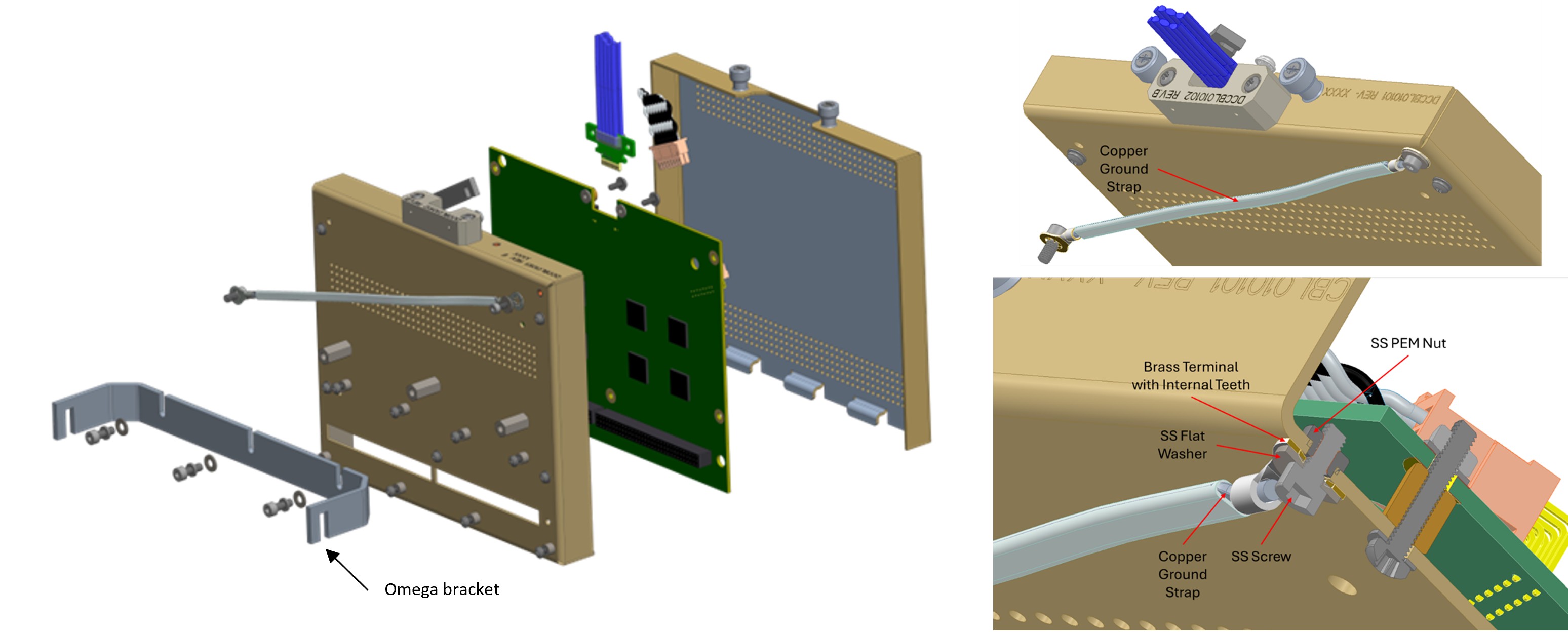}
    \caption{\label{fig:femb_assembly} Drawing of how a FEMB is assembled into its enclosing CE box, including a ground strap and omega bracket used for mounting onto an APA. Details of the ground strap connection to the CE box are shown on the right.}
\end{figure}

The PCB consists of 14 layers, including two whole copper planes that provide low-resistance common return paths. The FEMB is housed within a metallic enclosure, referred to as the cold electronics (CE) box, which serves three primary functions: mechanical support for the FEMB, shielding against electromagnetic interference, and strain relief for both power and data cables. A FEMB is mounted within its CE box using five brass standoffs, secured with stainless-steel screws tightened against conical spring lock washers or tooth washers. The CE box is mounted on the APA with an omega bracket, electrically grounding the FEMB and CE box to the APA frame. This connection is supplemented with a ground strap between the CE box and APA. By minimizing contact resistance, this design ensures a consistent common return path across all 20 FEMBs on a single APA. This assembly is depicted in figure \ref{fig:femb_assembly}.

\subsection{Signal and Control Interface}
\label{sec:femb_signal}

\begin{figure}
    \centering
    \includegraphics[width=0.95\linewidth]{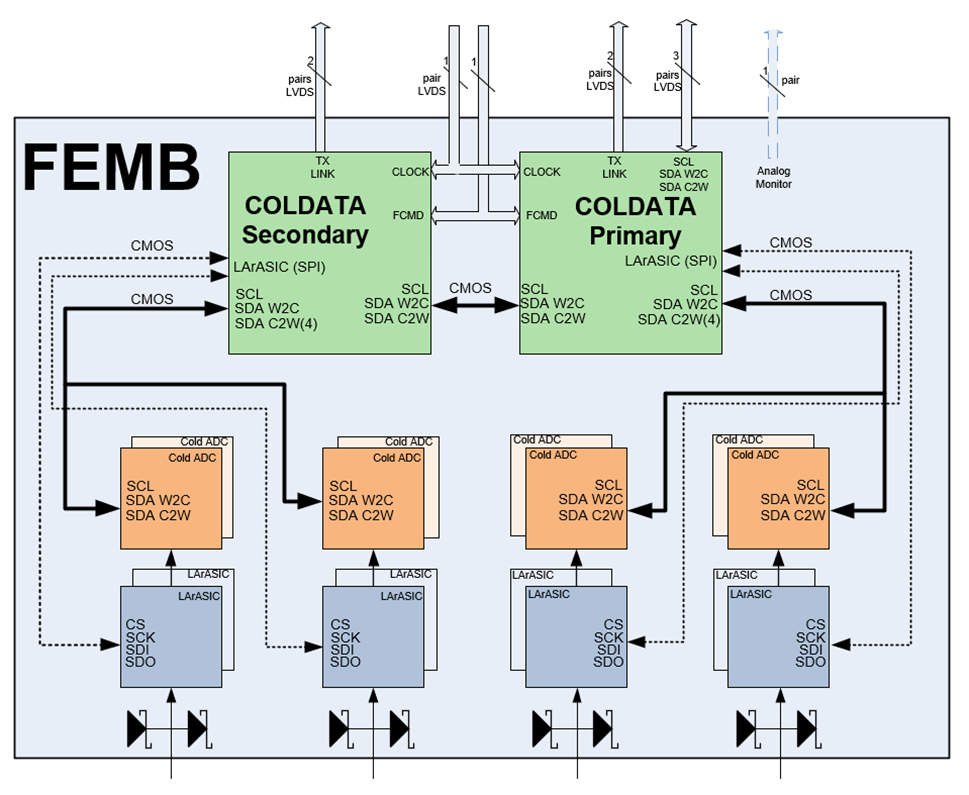}
    \caption{\label{fig:femb_block_diagram} Block diagram for the monolithic FEMB design used in ProtoDUNE-HD, with connections to the WIB shown at the top. The three lines for I2C communication (SCL, SDA W2C, SDA C2W), four lines for data transmission (TX LINK), one line for system clock (CLOCK), and one line for Fast Commands (FCMD) are shown. I2C and SPI communications within the FEMB are relayed with single-ended CMOS signals. Sense wires come in at the bottom to the front end of each LArASIC channel, where external diodes are installed for protection against potential discharges.}
\end{figure}

A custom Samtec 10-pair twinax cable is used for signal transmission between each FEMB and its corresponding WIB. The allocation of the 10 differential pairs within one of these data cables is summarized in table \ref{tab:cable_assignments}. A schematic block diagram of the FEMB showing these connections is provided in figure \ref{fig:femb_block_diagram}. As described in section \ref{sec:coldata}, each COLDATA is able to receive commands from either a WIB or another COLDATA. In order to minimize the number of needed cable connections, the FEMB thus operates the two COLDATAs in a primary-secondary topology. Configuration and control of ASICs on the FEMB are done through an I2C-like protocol between the primary COLDATA and the WIB. Three pairs of LVDS lines are dedicated to this purpose: one for the usual SCL signal, one unidirectional SDA line from warm to cold (SDA W2C), and one unidirectional SDA line from cold to warm (SDA C2W). To minimize power consumption on the FEMBs, I2C communication within the FEMB itself employs single-ended CMOS rather than LVDS. The primary COLDATA communicates with the WIB through the cable's I2C lines and extends the I2C link to the secondary COLDATA. Each COLDATA interfaces with its four ColdADCs through I2C and controls its four LArASICs via SPI. Each LArASIC is provided with an independent SPI link to its COLDATA for additional robustness in control.

\begin{table}[ht]
\centering
\caption{\label{tab:cable_assignments} Signal type assignments for each of the 10 twinax pairs within a Samtec data cable connecting a FEMB to a WIB.}
\begin{tabular}{|p{2.5cm}|l|c|l|p{5cm}|}
\hline
\textbf{Signal Name} & \textbf{Type} & \textbf{\begin{tabular}[x]{@{}c@{}}Number\\of Pairs\end{tabular}} & \textbf{I/O Standard} & \textbf{Function} \\
\hline
Data Links & Differential & 4 & LVDS & 1.25 Gbit/s links carrying data from the ASICs\\ \hline
I2C SCL & Differential & 1 & LVDS & I2C serial clock line\\ \hline
I2C SDA C2W & Differential & 1 & LVDS & I2C serial data from \mbox{COLDATA} to WIB\\ \hline
I2C SDA W2C & Differential & 1 & LVDS & I2C serial data from WIB to COLDATA\\ \hline
Fast Command & Differential & 1 & LVDS & Fast Command \mbox{transmission}\\ \hline
62.5\,MHz Clock & Differential & 1 & LVDS & Global clock from DTS\\ \hline
Analog Monitor or Calibration & Single-ended & \textonehalf & Analog & Monitoring of FEMB \mbox{devices} or injection of calibration pulses \\ \hline
WIB Control or Ground & Single-ended & \textonehalf & Analog & 1.8\,V CMOS signal or ground return \\ 
\hline
\end{tabular}
\end{table}

Each COLDATA is allocated two LVDS pairs for data transmission to the WIB, with each pair capable of transmitting at 1.25 Gbit/s. The resulting four 1.25 Gbit/s links from each FEMB are connected to a transceiver equalizer on the WIB. These data links have undergone benchtop tests in both room temperature and liquid argon conditions for combined cable lengths of up to 35\,m. No bit errors were observed after continuous testing for over 24 hours, corresponding to error rates of less than $1\times10^{-14}$.

The two on-board COLDATAs receive their 62.5\,MHz system clocks from a shared clock line coming through one LVDS pair from the WIB. This clock is derived from the external DUNE Timing System (DTS) \cite{Barcock_2023}, which is distributed through the PTC to the WIBs. This timing distribution scheme provides a single unified clock domain for all FEMBs, which is used to synchronize charge readout of the entire TPC. Another LVDS pair is used to transmit Fast Commands to both COLDATA at once, allowing separate FEMBs to receive globally coordinated instructions from either the DTS or the WIB. For instance, a Sync Fast Command issued by the DTS would simultaneously reset the 15-bit timestamp in every COLDATA in the detector, aligning data across all transmission links.

Lastly, one line on the cable is allocated as a bidirectional analog path. This line can be driven from the FEMB side in order to monitor various voltages on the FEMB, and it can be driven from the WIB side in order to inject external signals for calibrating the front-end electronics. The details of this feature are described in section \ref{sec:femb_calibmonitor}.

\subsection{Power Distribution and Consumption}
\label{sec:femb_power}

\begin{figure}
    \centering
    \includegraphics[width=\linewidth,trim=1cm 1.5cm 1cm 1.5cm,clip]{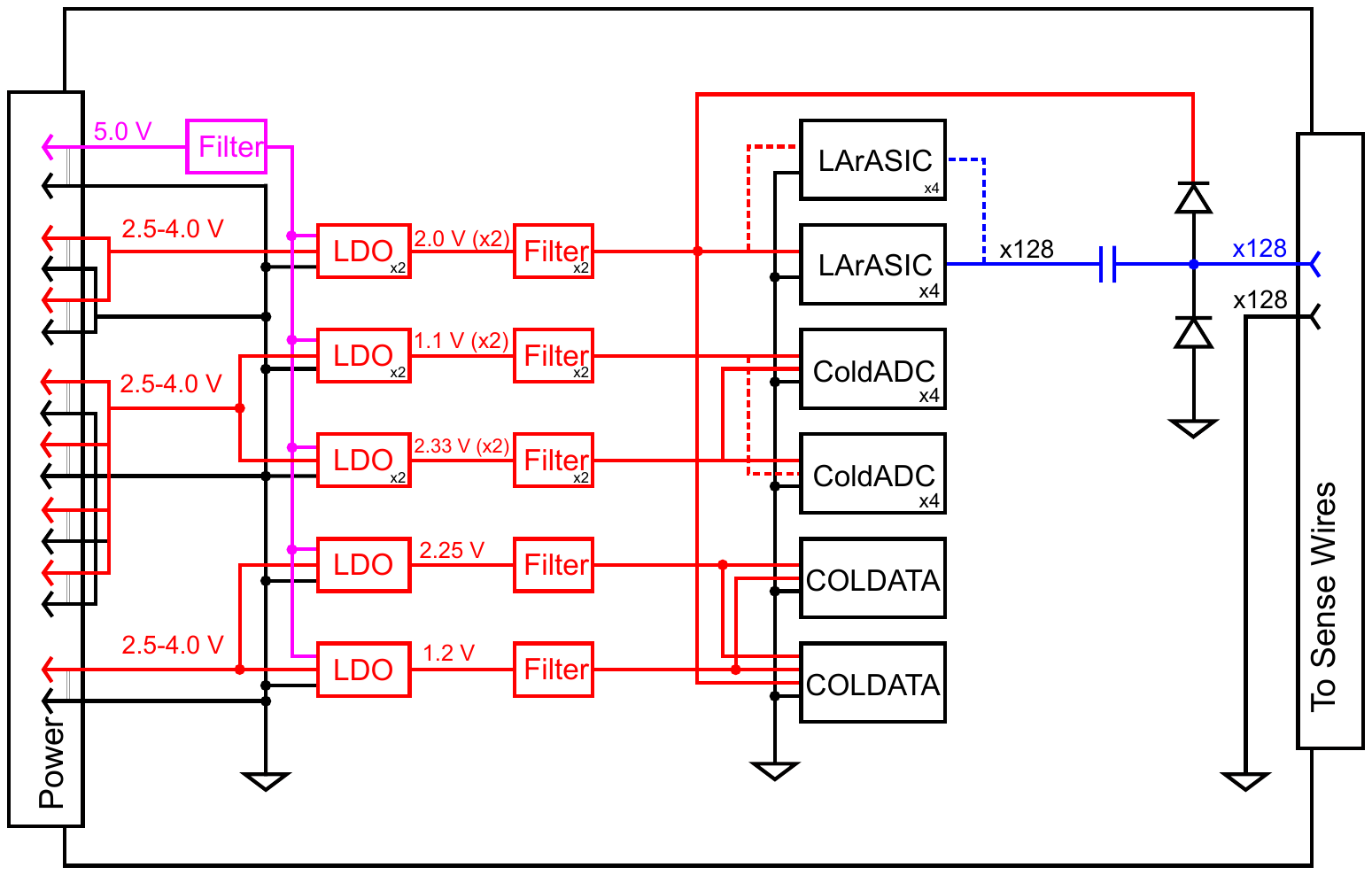}
    \caption{\label{fig:femb_power} Power distribution scheme for the ASICs on a FEMB, with power lines coming along the power cable from the WIB on the left. These voltages serve as inputs to LDOs, whose outputs are filtered and then used to supply the ASICs. Connections on the right go to the sense wires that are being read out.}
\end{figure}

The primary goal of the FEMB's power distribution scheme is to supply reliable, low-noise DC voltages to the on-board ASICs. Each FEMB receives three independent low-voltage power rails (adjustable from 2.5--4.0\,V, but all configured to \mbox{4.0\,V} for normal detector operations) and a \mbox{5\,V} bias supply from the WIB through a Samtec IPL1 series connector, with a current rating of \mbox{2.1\,A} per pin. The allocation of these power rails across connector pins is illustrated in figure \ref{fig:femb_power}. After receiving these power rails, the FEMB passes them through a low-noise power distribution architecture to mitigate any electrical ripple noise. Each incoming rail is stepped down by an on-board cryogenic-compatible low-dropout (LDO) regulator, with each LDO having noise levels of approximately \mbox{20\,$\mu$V} root mean square (RMS). The LDO outputs are then passed through RC filters in order to create clean, stable DC voltages for the LArASICs, ColdADCs, and COLDATAs. The input power lines are divided into:
\begin{itemize}
    \item \textbf{LDO Bias}: A single pin delivers a 5\,V bias voltage for all of the LDOs on the FEMB. The LDOs require this bias input for the operation of their error amplifiers, voltage references, and internal control circuitry. The total bias current needed to operate all eight LDOs on a board is less than 10\,mA.
    \item \textbf{LArASIC Power}: Two pins provide an adjustable 2.5--4.0\,V supply, which is stepped down by LDOs to 2.0\,V for the eight LArASIC chips. Each LArASIC consumes approximately 170\,mW under the configuration with maximum power requirements, corresponding to a total current draw of about 540\,mA through the power cable's LArASIC lines.
    \item \textbf{ColdADC Power}: Four pins provide an adjustable 2.5--4.0\,V supply, which is stepped down by LDOs to 2.33\,V and 1.1\,V for the eight ColdADC chips. Each ColdADC has a nominal power consumption of 325\,mW, yielding a combined current draw of approximately \mbox{1180\,mA} through the power cable's ColdADC lines.
    \item \textbf{COLDATA Power}: One pin supplies an adjustable 2.5--4.0\,V supply, stepped down by LDOs to 2.25\,V and 1.2\,V for the two COLDATA chips. Each COLDATA consumes approximately 170\,mW, corresponding to a total current draw of about 140\,mA through the power cable's COLDATA line.
\end{itemize}

To prevent single-point failures from open connections, cross-coupling diodes are installed between the LDO bias and ColdADC power lines. Under normal operating conditions these diodes are reverse-biased, as the LDO bias is maintained at a higher voltage than the ColdADC power line. In the event that the LDO bias line is disconnected for any reason, raising the ColdADC power line to 4.0\,V keeps the LDO bias line above 3.7\,V, allowing it to sustain proper LDO biasing. Cross-coupling diodes are installed between the COLDATA power and LArASIC power lines as well, operating on similar redundancy principle.

The overall power consumption of the FEMB depends on the configuration of the ASICs. The LArASIC's optional single-ended output buffer and the ColdADC's optional single-ended-to-differential input buffer were included in their designs for flexibility in case they had to handle long transmission lines between the amplifier and ADC. However, the LArASIC and ColdADC are arranged in close physical proximity in the FEMB design. As a result, these buffers are unnecessary and can be bypassed to reduce power consumption. The nominal FEMB configuration instead uses a simple single-ended interface between the LArASICs and ColdADCs. Under this configuration, the average measured power consumption is approximately 29\,mW per channel. The maximum envisioned power consumption arises from activation of the LArASIC's SEDC buffer, which can help mitigate crosstalk between channels by creating a differential interface to the ColdADC. This configuration increases power consumption to approximately 34\,mW per channel. Including the additional overhead from voltage regulation, with an estimated 300\,mV dropout tolerance, this places the maximum power consumption of a monolithic FEMB at less than ${\sim}45$\,mW per channel, corresponding to ${\sim}5.8$\,W per board. This value is considered safely below the 50\,mW per-channel limit that was established to prevent local argon boiling during cryogenic operations \cite{HD_TDR}.

\subsection{Charge Calibration and Monitoring Scheme}
\label{sec:femb_calibmonitor}

\begin{figure}
    \centering
    \includegraphics[width=0.95\linewidth,trim=1cm 1cm 1cm .8cm,clip]{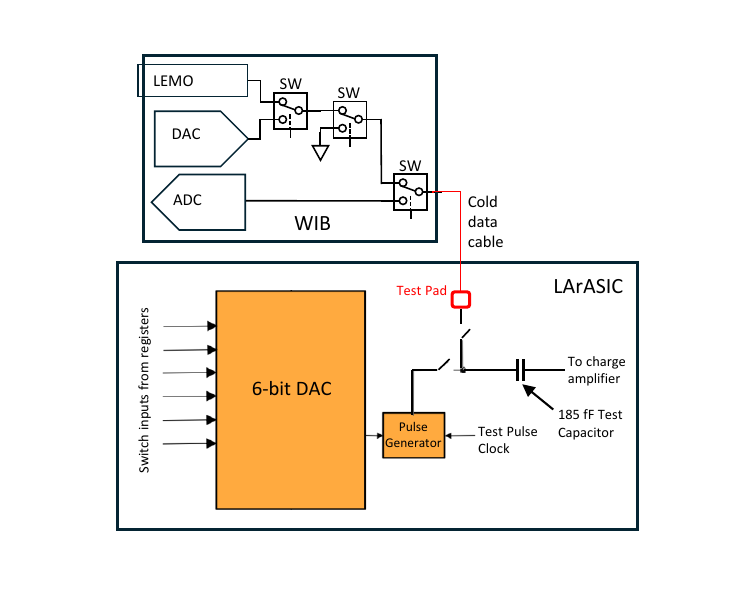}
    \caption{\label{fig:femb_calibration} Schematic for avenues of charge injection into the front end of the LArASICs on a FEMB, which can be used for charge calibration. The internal calibration path and the depicted switches are controlled by internal registers on the LArASIC. The external calibration path can use either voltages generated by the WIB's on-board DAC or pulses injected externally through the WIB's LEMO connector. This calibration path between the WIB and FEMB can also be used as a monitoring path to transmit quantities on the FEMB up to the WIB. From there, these quantities can be read through an ADC on the WIB or externally through a LEMO connector.}
\end{figure}

Calibration of the response of the on-board ASICs requires the ability to inject known amounts of charge to each channel. The FEMB provides multiple options for this, shown in figure \ref{fig:femb_calibration}. The 185\,fF MIM capacitor in each LArASIC channel can be individually enabled through LArASIC registers, allowing voltage pulses to be injected through it to provide calibration charges. The nominal charge calibration procedure, described further in section \ref{sec:pulsers}, uses input pulses generated through the LArASIC's internal 6-bit DAC described in section \ref{sec:larasic}.

The LArASIC's built-in DAC has imperfect linearity that can vary slightly from chip to chip. Although this nonlinearity is small and can be measured and corrected for, external calibration signals of known magnitude provide a useful independent cross-check. To enable this external fine-tuning, each COLDATA has five general-purpose input/outputs (GPIOs) that the FEMB makes available to the WIB. These can be used to generate pulses using a 16-bit DAC on the WIB or to directly inject external pulses through LEMO connectors on the front panel of the WIB. The WIB Control signal mentioned in table \ref{tab:cable_assignments} is used in combination with these GPIOs to control the injection of calibration pulses with the WIB, allowing the pulse timing to be synchronized with external systems. 

The GPIOs on the COLDATA can also be used to provide various analog monitoring paths. Experience from the ProtoDUNE-SP and SBND experiments \cite{8533137} has shown the importance of being able to directly probe FEMBs even when they are being operated in liquid argon, particularly during QC tests, installation, and detector commissioning. The monolithic FEMB design includes diagnostic features that allow key parameters to be transmitted through the data cable's analog monitoring line to the WIB. On the WIB, they can be either digitized by the on-board ADC or accessed through a LEMO connector on the WIB's front panel. This monitoring path is shared with the line used for external charge calibration.

The FEMB's analog monitor feature can be used to inspect a number of quantities from the LArASICs and ColdADCs. The LArASIC's monitoring output line can be connected to its bandgap reference voltage, its temperature sensor, its 6-bit DAC output, or an individual channel's amplifier response. The ColdADC can be configured to connect any of its internal reference voltages to its monitoring output line. The FEMB can select one of these values from one ASIC at a time to send to the WIB for monitoring. In addition, the FEMB can be configured to monitor the voltage output of any of the on-board LDOs. This allows QC tests to screen out any defective regulators. Using this procedure, a small fraction ($<1\%$) of LDOs were found to have output voltages more than 50\,mV below their nominal values under cryogenic conditions and had to be replaced.

\subsection{Performance Characterization}
\label{sec:femb_performance}

\begin{figure}
    \centering
    \includegraphics[width=0.47\linewidth]{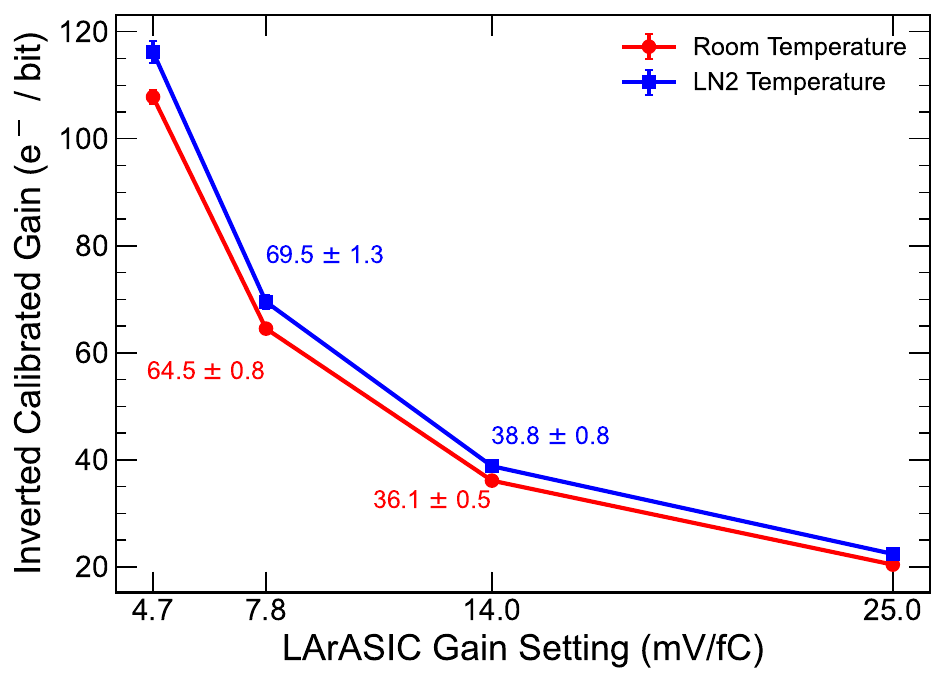}
    \includegraphics[width=0.49\linewidth]{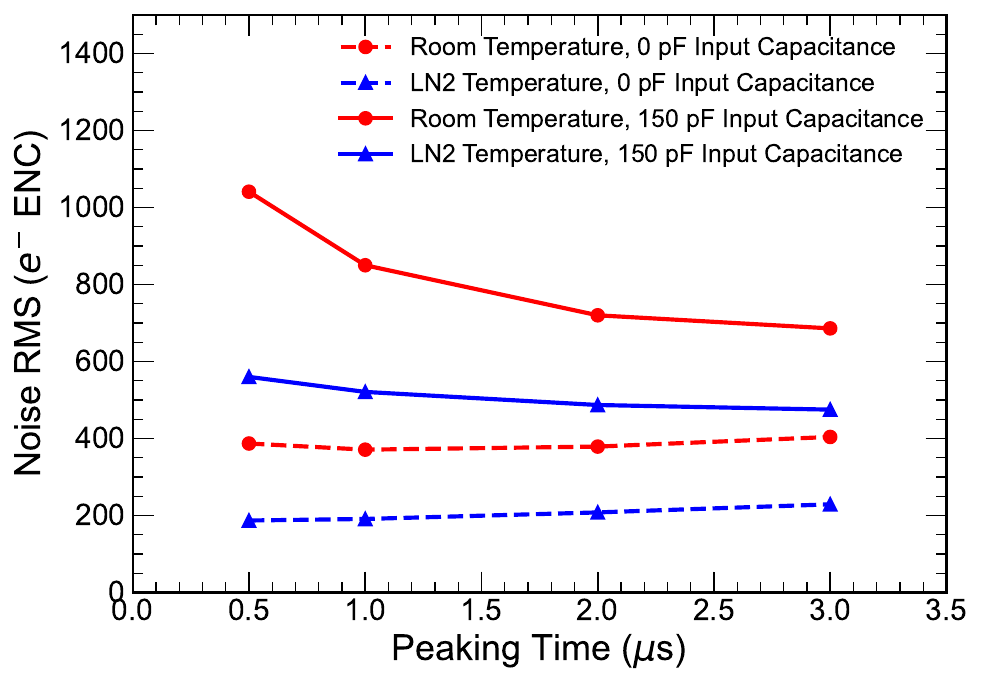}
    \caption{\label{fig:femb_noise} \textbf{Left:} average calibrated gains for each gain setting from pulser scans of a typical FEMB. The values at room temperature and in liquid nitrogen are annotated for the two primary gain settings under consideration for use in DUNE. Error bars represent the standard deviation of results from the 128 channels of a single FEMB. \textbf{Right:} noise levels for a typical FEMB measured in benchtop tests at 14\,mV/fC gain, under both room temperature and liquid nitrogen conditions, and with and without an input capacitor simulating the effect of an actual detector.}
\end{figure}

The performance of each FEMB is characterized during benchtop QC tests before it is sent for installation in its intended detector. Key metrics include the electronic noise, gain, linearity, and cross-talk under both room temperature and cryogenic conditions. In these tests, liquid nitrogen is typically used instead of liquid argon, since the difference between their respective temperatures of 77\,K and 87\,K is negligible for evaluating FEMB performance. In addition, rather than connecting the FEMB to long sense wires, a 150\,pF mica capacitor is connected to the front end of each channel to emulate the effect of a typical wire plane's input capacitance. This capacitor choice provides minimal variation between warm and cryogenic conditions, and the capacitance value is similar to the capacitance expected from typical wire lengths in DUNE detectors\footnote{APA wires in the DUNE HD design carry capacitances of ${\sim}20$\,pF/m and lengths of 6 to 7.6\,m.}. Typical calibration values and the corresponding noise levels are shown in figure \ref{fig:femb_noise}. Noise levels quantify the magnitude of fluctuations in the electronics response in the absence of any signal, measured as a number of electrons in equivalent noise charge (ENC). This quantity is defined as the number of electrons whose collection would produce a signal amplitude equivalent to the RMS noise amplitude. Benchtop noise levels with the 150\,pF input capacitor have been found to be much better than the 1000 electrons ENC required by DUNE \cite{HD_TDR,VD_TDR}. These tests have measured the percentage integral non-linearity to be less than $0.3\%$ and cross-talk to be much less than $1\%$ in all channels, satisfying the DUNE requirements for those quantities as well.

\section{Warm Interface Board}
\label{sec:wib}

The Warm Interface Board (WIB) serves as the bridge between FEMBs located inside the cryostat and the back-end DAQ and slow control systems located outside the cryostat. To minimize cryogenic penetrations, WIBs are housed in WIECs, which sit on top of the cryostat's signal feedthroughs. Each WIEC contains slots for one PTC and up to six WIBs, with the slots connected to each other through a power and timing backplane (PTB). The WIB is powered with a 12\,V input, which is normally supplied by a PTC through the PTB. The WIB steps down this power into low-noise power rails for both its own on-board components and for up to four FEMBs that it controls. It must also receive the system clock and any timing commands from the global DTS and distribute these to the FEMBs.

The WIB's data management responsibilities include receiving a total of 16 continuous 1.25~Gbit/s data streams from four FEMBs through their corresponding cold cables, formatting the raw data, and multiplexing the data for transmission over two 10 Gbit/s optical links to the DAQ using UDP/IP. On the control and diagnostics side, the WIB must provide a Gigabit Ethernet (GbE) interface using TCP/IP. This is used by the DAQ Control, Configuration, and Monitoring (CCM) framework to configure and monitor both the WIB itself and its FEMBs. It is also used by the detector slow control system to perform diagnostics on the WIB without interrupting the data flow to the DAQ. This section describes the hardware, firmware, and software used on the WIB to enable all of these functions.

\subsection{Hardware}

\begin{figure}
    \centering
    \includegraphics[width=0.67\linewidth,angle=90]{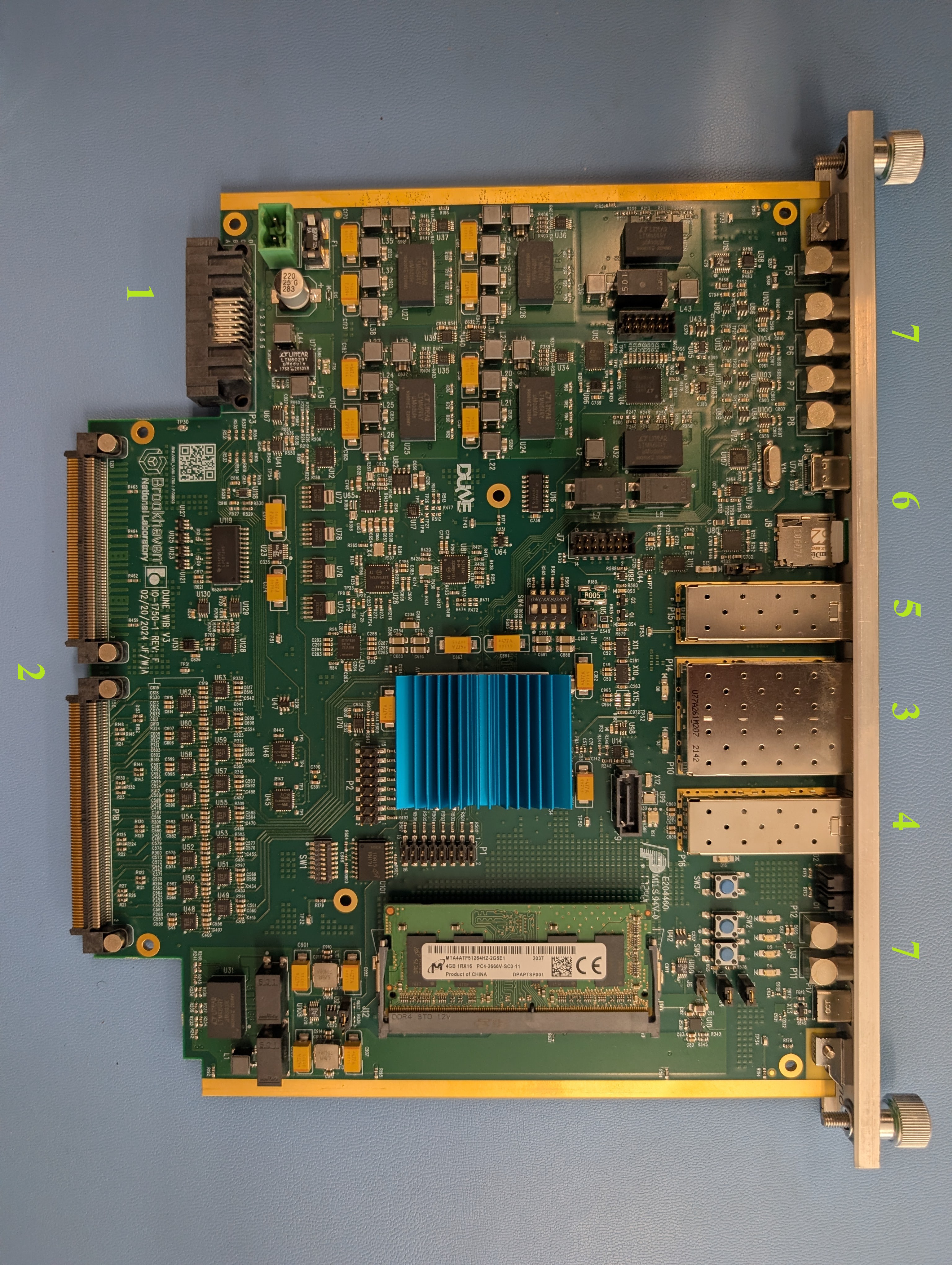}
    \caption{\label{fig:wib_photo} Photograph of one of the WIBs used in ProtoDUNE-HD, with the front panel at the top and the connections to the WIEC backplane on the bottom. The PTC/PTB interface (1), flange board interface (2), optical links (3), GbE interface (4), spare SFP port (5), debug and storage (6), and front-panel LEMO connectors (7) are labeled and described further in the text.}
\end{figure}

To fulfill its needed functions, the WIB incorporates several electronic interfaces to the other parts of the TPC electronics system and to other detector subsystems, which are marked in figure \ref{fig:wib_photo}:
\begin{enumerate}
\item \textbf{PTC/PTB Interface}: A Samtec MPTC series connector at the back of the WIEC connects the PTC and all of the WIBs in the WIEC. This connector carries 12\,V power, timing information, I2C communications, and crate/slot addressing to the WIB.
\item \textbf{Flange Board Interfaces}: Two Samtec ERM8 series connectors connect the WIB to its four FEMBs. These connectors distribute independent power rails, synchronized clocks, \mbox{COLDATA} Fast Commands, 16 high-speed data links, custom LVDS-based I2C communications, and analog paths for calibration and monitoring. The FEMB sides of these connections are described in sections \ref{sec:femb_signal} and \ref{sec:femb_power}.
\item \textbf{Optical Links}: Two 10 Gbit/s enhanced small form-factor pluggable (SFP+) modules \mbox{transmit} digitized charge signals from a total of 512 front-end channels (four FEMBs with \mbox{128 channels} each) to the DAQ.
\item \textbf{Configuration and Slow Control}: A 1000Base-LX small form-factor pluggable (SFP) module supports GbE communication for configuring and monitoring the WIB.
\item \textbf{Spare SFP Port}: This port is not used during normal detector operations, but it can be used to receive timing signals directly through the WIB's front panel instead of through the PTB. It can also be used as an additional GbE connection.
\item \textbf{Debug and Storage}: A USB Type-C connector offers UART access for debugging purposes, while a removable micro-SD card stores the Zynq FPGA firmware and software.
\item \textbf{Front-Panel Connectors}: Two GPIO LEMO connectors and five additional LEMO ports can be used for calibration pulse injection and analog monitoring.
\end{enumerate}

\begin{figure}
    \centering
    \includegraphics[width=\linewidth,trim=1cm .1cm .6cm .2cm,clip]{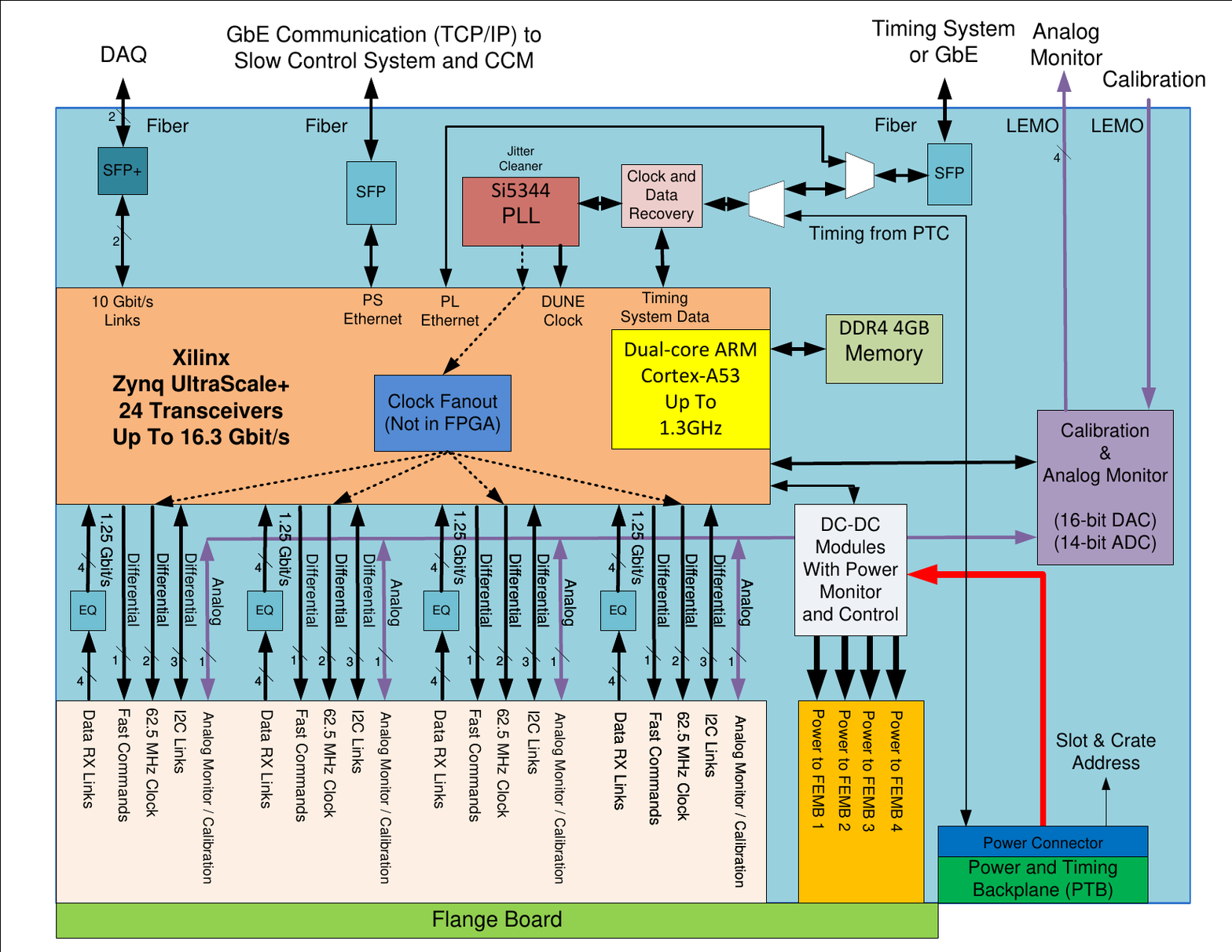}
    \caption{\label{fig:wib_block} Block diagram of the WIB components. Connections at the bottom go through the back of the WIEC, and connections at the top are externally accessible through the WIB's front panel. Input timing and power come through the PTB in the bottom right. The timing signal used by the FPGA is recovered from the input timing through a clock and data recovery chip and a Si5344 jitter cleaner. The FPGA interfaces to the four FEMBs through the flange board in the bottom left. With each FEMB, it manages four 1.25 Gbit/s data streams via on-board equalizers (EQ), as well as clock distribution, control, and configuration. Power for the FEMBs is regulated by FPGA-monitored DC–DC converters, detailed further in figure \ref{fig:wib_power}. The FPGA interfaces with four SFP modules at the top: two 10 Gbit/s modules for streaming data to the DAQ system, one module for timing that can be used when there is no timing signal through the PTB, and one module for a GbE interface to slow control systems and DAQ CCM. Calibration and analog monitoring paths to the FEMBs are shown in purple. These can be handled by the on-board 16-bit DAC and 14-bit ADC, or they can be routed to LEMO connectors shown in the top right for external pulse injection and monitoring.}
\end{figure}

The block diagram in figure \ref{fig:wib_block} shows the arrangement of these interfaces around the WIB's core in a Xilinx Zynq UltraScale+ MPSoC ZU6CG, supported by on-board DDR4 memory for running an embedded Linux OS. The operations of this core are described in section \ref{sec:wibfirmware}.

For electrical safety, a 5\,A fuse is installed at the 12\,V input used to power the WIB. This is followed by a unidirectional transient-voltage-suppression (TVS) diode for protection against electrical surges and electrostatic discharges. To meet Xilinx operating requirements for the Zynq FPGA and ensure stable, low-noise power delivery, the WIB uses an LTC2977 power system manager to sequence, trim, and supervise all voltage rails. This device implements a digital servo loop using an internal ADC, DAC, and processor operating in continuous trim mode. Its DAC output feeds the DC–DC converter’s feedback network, maintaining precise point-of-load regulation with minimal voltage fluctuation.

During normal detector operations, the WIB receives a 62.5\,MHz system clock from the PTC through the PTB and distributes it to the FEMBs for synchronized detector readout. For tests outside of normal detector operations, the WIB can receive an external timing signal through its front panel timing SFP. The WIB also contains on-board oscillators that can generate an internal clock to enable standalone operation when external timing is unavailable.

Given that the cold signal cables between the WIB and FEMBs can be up to 35\,m long, the WIB employs MAX3801UTG+ adaptive equalizers to compensate for up to 30\,dB of attenuation at 1.6\,GHz. The FPGA decodes 16 separate 1.25 Gbit/s data streams coming from the four FEMBs, aligns the data frames with a global time stamp from the DTS, forms data packets based on the UDP protocol, and sends them to the DAQ system over two 10 Gbit/s optical links using SFP+ modules.

The WIB allows for TCP/IP communication over GbE, which is used by both the DAQ CCM framework and the detector slow control system. DAQ CCM uses this interface to issue configuration commands to the WIB, which the WIB follows to configure its FEMBs and their ASICs. The detector slow control system uses this interface to control and access the status of various I2C-compatible devices and sensors on the WIB. The software interfaces for these systems are described in section \ref{sec:wibsoftware}.

The calibration and analog monitor module on the WIB is configurable through internal registers to perform either function, but not both simultaneously. When configured for calibration, it injects calibration pulses to the LArASICs of every FEMB that the WIB controls, using either the on-board 16-bit DAC or an external source through the WIB's LEMO connector. When configured for analog monitoring, the module connects to one of the various FEMB analog signals described in section \ref{sec:femb_calibmonitor}. These signals are then either digitized by the on-board 14-bit ADC or directly routed to dedicated LEMO connectors for measurement with external instruments.

\begin{figure}
    \centering
    \includegraphics[width=0.95\linewidth,trim=.2cm .2cm .2cm .2cm,clip]{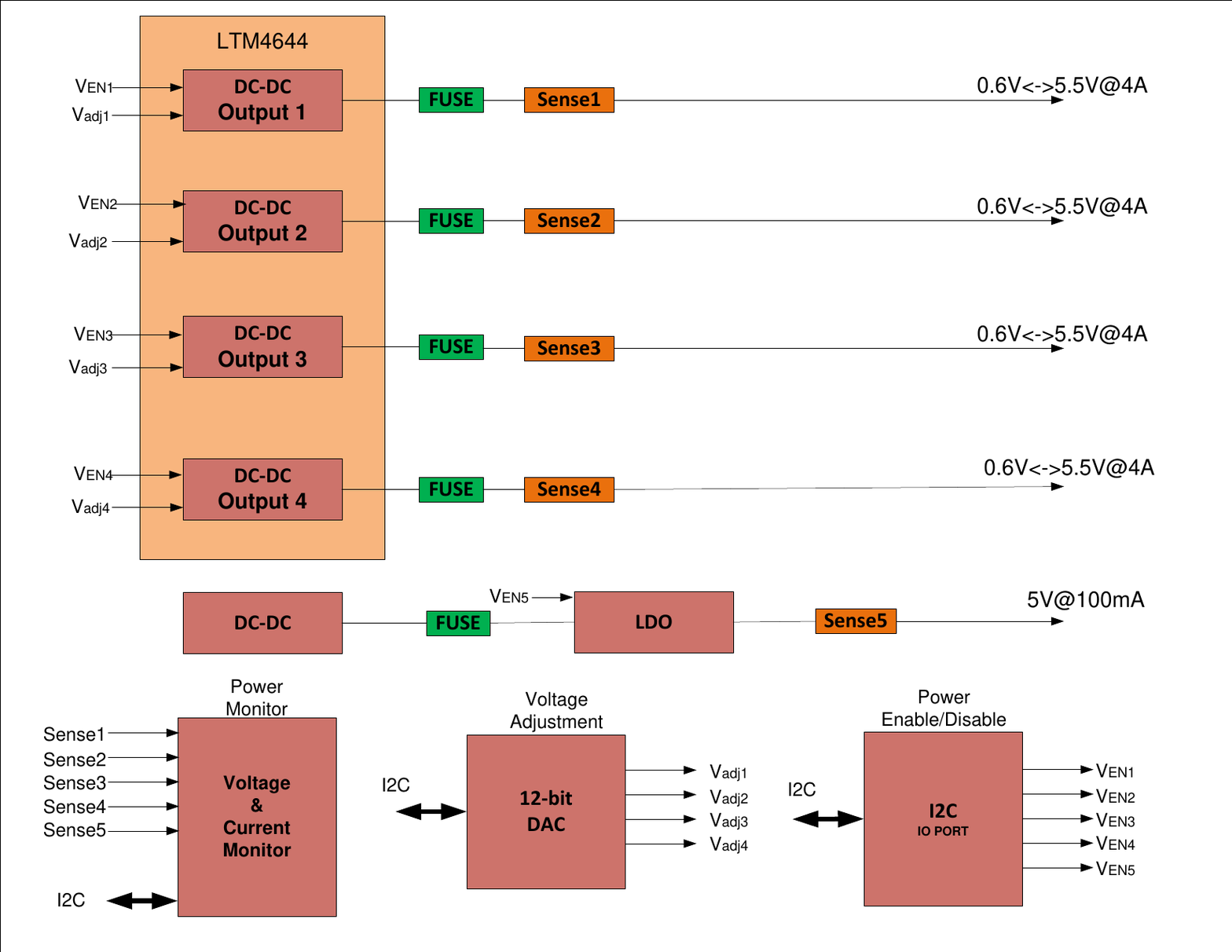}
    \caption{\label{fig:wib_power} Power distribution scheme from the WIB to FEMB. This block is replicated four times on each WIB, once for each FEMB it controls. The $V_{EN}$ outputs can be toggled via I2C to enable or disable their respective DC-DC or LDO units. The $V_{adj}$ outputs from the 12-bit DAC are similarly I2C-controlled and are used to adjust the LTM4644 DC-DC outputs. Three of the DC-DC outputs from the LTM4644 are used for the LArASIC, ColdADC, and COLDATA power rails shown in figure \ref{fig:femb_power}. The fourth is a spare and is normally deactivated.}
\end{figure}

Four LTM4644 DC–DC converter modules supply power to the FEMBs. Each module generates four rails capable of sustaining up to 4\,A. Figure \ref{fig:wib_power} shows the power distribution scheme from one of these modules to a FEMB, providing the inputs to the FEMB power scheme shown in figure \ref{fig:femb_power}. The LTM4644 voltage and current outputs can be monitored and adjusted through internal registers on the WIB. Each channel’s output can be fine-tuned through a voltage adjustment pin, driven by a resistor network combined with a dedicated 12-bit DAC under FPGA control. Under normal operating conditions, three of the configurable outputs are set to 4\,V, and the fourth is unused. Benchtop tests have shown that fluctuations in the FEMB 5\,V bias can affect the FEMB noise performance more than the other rails do; to combat this, a TPS73250DCQ LDO, powered by an LTM8029 converter, is used to provide an ultra-low-noise fixed 5\,V rail.

All DC–DC module inputs and outputs are isolated with common-mode chokes to suppress noise conduction toward the FEMBs. Downstream fuses protect connected circuits against short-circuit faults. Multiple LTC2991 chips measure voltages and currents across output sense resistors, using four differential channels per chip. Each chip’s I2C address is configured via pull-up or pull-down resistors, allowing up to eight unique addresses per WIB. Bidirectional I2C level shifters are used to enable communication between the FPGA and the 5\,V monitoring devices, adapting signal voltages to match the FPGA's I/O bank power settings.

\subsection{Firmware}
\label{sec:wibfirmware}
\begin{figure}
    \centering
    \includegraphics[trim=1cm 6cm 1cm 1cm,clip=true,width=0.75\linewidth]{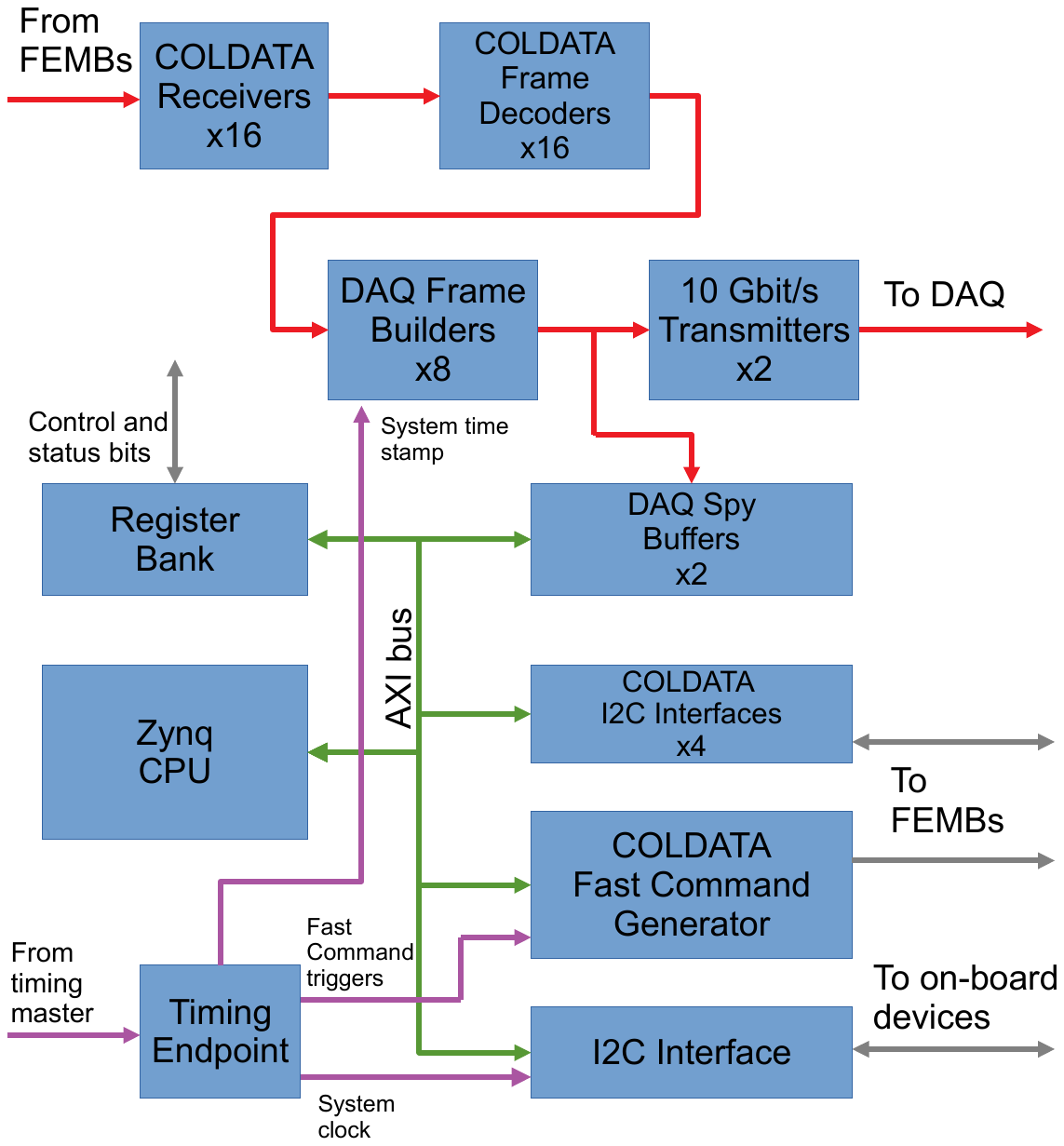}
    \caption{WIB firmware structure, showing notable blocks. Data from the FEMBs flows along the red arrows. Timing signals flow along the purple arrows. The AXI bus interfaces to the CPU are shown in green and are described further in table \ref{tab:AXI_modules}. Interfaces to control and check the status of devices on both the WIB and the FEMBs are shown in gray.}
    \label{fig:wib_firmware_structure}
\end{figure}

Each WIB functions under the control of a Xilinx Zynq UltraScale+ MPSoC ZU6CG. This device consists of a CPU and attached FPGA logic. The firmware for this FPGA logic must receive data from the FEMBs, decode those data frames, align them according to both the COLDATA-provided and DTS-provided timestamps, check data integrity within those frames, and form the data packets that are sent to the DAQ system. It also must receive and decode timing commands from the DTS, and it must provide slow control functions and monitoring for both the FEMBs and the WIB itself. A diagram of the structure of the WIB firmware is shown in figure \ref{fig:wib_firmware_structure}, and the rest of this section describes some of its notable blocks.

\paragraph{Data Flow}
Data from the FEMBs first enters the WIB through the COLDATA receivers, which are serial receivers that each operate at a bit rate of 1.25 Gbit/s. Each WIB receives data from four FEMBs, each FEMB has two COLDATA chips, and each COLDATA chip sends data via two serial links, for a total of 16 COLDATA receivers that each WIB must handle. These receivers are implemented using Zynq GTH IP gigabit transceiver cores. The output of each receiver is a 16-bit bus that carries the deserialized data.

The COLDATA receiver outputs are next sent to COLDATA frame decoders. The COLDATA frame decoders are capable of decoding any of the COLDATA data formats described in section \ref{sec:coldata}. The format type is automatically determined from the received data stream. The outputs of the COLDATA frame decoders are then gathered in DAQ frame builders. These are responsible for preparing data for transmission to the DAQ system, which expects the data to arrive in a specific format. The WIB has a total of eight DAQ frame builder modules, each of which prepares data arriving from one COLDATA. The output of each DAQ frame builder is a 64-bit data bus, a ``valid'' flag, and a ``last word'' flag. 
Lastly, the outputs of the DAQ frame builders are sent to one of two 10 Gbit/s transmitter modules. Each one of these transmits constructed DAQ frames to the DAQ system via one of the two 10 Gbit/s optical links on the WIB, following a protocol called Hermes developed by the DUNE DAQ \cite{Hermes}.

To check data integrity, each COLDATA frame decoder calculates an 8-bit checksum from the bytes in the received data and compares the calculated checksum against the checksum bytes received with the data. Errors are flagged in dedicated WIB registers and reported in headers of the final constructed DAQ frames. The headers of the DAQ frames are scanned to perform live checks for data corruption and are also carried forward to any offline saved data to allow for identification of corrupted data that slipped through. The errors recorded in the WIB registers can be periodically checked by the slow control system to look for rare but non-zero rates of data corruption occurring outside the windows of DAQ operation. During the ProtoDUNE-HD operational period, no checksum errors in the data were observed.

\paragraph{COLDATA timestamp synchronization}
As the data blocks are constructed and transmitted, each is required to contain a 64-bit timestamp that carries the precise time at which that data block was digitized. The timing information initially comes from two sources:
\begin{enumerate}
    \item The COLDATA chips have internal time counters. These counters are synchronized by dedicated Fast Commands issued periodically by the DTS. This procedure guarantees that each COLDATA's internal timestamp reflects the actual digitization time of the data blocks that it is transmitting. However, the COLDATA internal timestamps were only designed to carry information on the timescale of $O(1000)$ consecutive COLDATA data frames, which are spaced 512\,ns apart following the ColdADC digitization rate. These counters are thus only 15 bits long for space efficiency, causing them to wrap around every $\sim$524\,$\mu$s. Because of this, these counters cannot be used directly in the WIB-level data blocks.
    \item The WIB receives the full 64-bit timestamp from the DTS. However, data from the COLDATA has some non-zero latency arriving at the WIB, which results in an offset between the COLDATA 15-bit timestamp sent with the data and the DTS 64-bit timestamp. The magnitude of this offset is different for each COLDATA serial data link and may change from one power-up to another, so it cannot be corrected with a fixed value.
\end{enumerate}
To combine these two sources of timing information and correct for their individual deficiencies, the WIB implements a synchronization procedure that continuously analyzes the COLDATA 15-bit time counters and compares them with the lower 15 bits of the DTS timestamps. Based on that comparison, each data frame is delayed by the appropriate number of clock ticks so that its 15-bit counter aligns precisely with the DTS timestamp. All delays are calculated automatically and continuously in the firmware. These delays are available for monitoring via registers, and the values of these delays can be used to detect COLDATA link errors.

\paragraph{Zynq CPU module and AXI modules}
The Zynq CPU module is a hard CPU IP core provided in the Zynq device. Its main purpose is to control and monitor the firmware modules. Access to all modules in the WIB firmware is provided via the standard Zynq AXI bus interface, which the CPU uses to perform some of its tasks. Section \ref{sec:wibsoftware} describes the software that runs on the Zynq CPU. The full list of AXI modules in the FPGA logic is shown in table \ref{tab:AXI_modules}.

\begin{table}
    \centering
    \caption{AXI modules used in the WIB firmware, also depicted in figure \ref{fig:wib_firmware_structure}. These are used by software in the CPU module for various control and monitoring tasks.}
    \label{tab:AXI_modules}
    \begin{tabular}{|c|c|p{8cm}|}\hline
         Module Name&  Count& Function\\\hline
         Fast Command Generator&  1& Generates Fast Commands for all COLDATA chips\\\hline
         COLDATA I2C Interface&  4& I2C interfaces for the primary COLDATA on each FEMB\\\hline
         I2C Interface&  1& I2C interface for devices on the WIB\\\hline
         Register Bank&  1& Access to the WIB's control and status registers\\\hline
         DAQ Spy Buffer&  2& Spy memory for formatted data being prepared for transmission to the DAQ system, which can be used for debugging\\ \hline
    \end{tabular}

\end{table}

\subsection{Software Interfaces}
\label{sec:wibsoftware}
While the WIB firmware handles all the WIB's time-sensitive functions, any other functions without strict latency or timing requirements are relegated to the WIB software. The software is divided into two main parts: a DUNE DAQ module called ``\texttt{wibmod}'' which configures the WIB via the DUNE DAQ's CCM framework, and a WIB server (\texttt{wib\_server}) that runs within the Zynq CPU core residing on each WIB and is responsible for local (on-board) control and monitoring.  The details of \texttt{wibmod} and its integration within the full DUNE DAQ software framework will be described in a future publication detailing the entire DUNE DAQ suite; this section describes only the \texttt{wib\_server} that resides on the WIB. The software comprising the \texttt{wib\_server} is publicly available \cite{WIB_github}.

The Zynq CPU core runs PetaLinux 2020.1, customized to contain all software dependencies for building and running the \texttt{wib\_server}. The \texttt{wib\_server} is a C++ application which communicates with the WIB firmware and the FEMBs via memory-mapped registers. The server also interfaces with the various on-board hardware sensors and power regulators via I2C buses using standard kernel drivers. The server exposes these devices to the external detector slow control system and mediates communications between them.  

The \texttt{wib\_server} includes all the logic for configuring and monitoring the WIB and FEMBs, and it accepts high level commands on a ZeroMQ REP socket from \texttt{wibmod} when interfacing with the DUNE DAQ system. The use of a ZeroMQ REP socket provides increased flexibility, allowing a variety of languages and systems to interface with the WIB via a remote procedure call (RPC) scheme. 

\section{Power and Timing Card}
\label{sec:ptc}

Each WIEC contains one Power and Timing Card (PTC) that is primarily responsible for distributing power and timing to the WIBs in the crate. It must be able to supply 12\,V power at up to 6\,A to each of the six WIBs in a WIEC over the PTB. On the timing side, it distributes timing information received through its own SFP to all WIBs in the WIEC over the PTB. The PTC also must priority encode timing data transmissions from one WIB at a time back through itself, as only one WIB at a time can transmit through the single timing SFP on a PTC.

\begin{figure}
    \centering
    \includegraphics[width=0.9\linewidth]{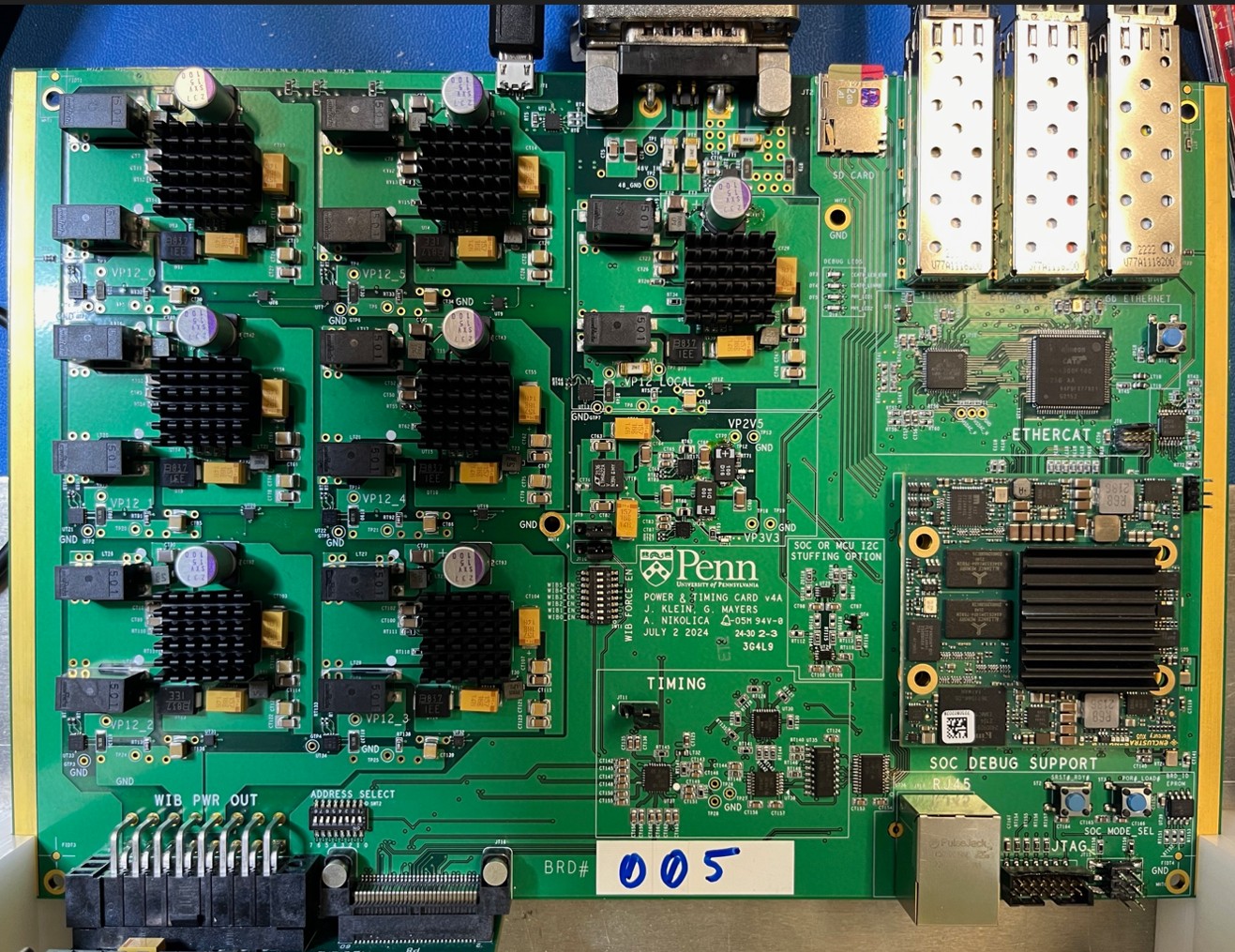}
    \caption{Revised PTC containing new functions relative to the ProtoDUNE-HD PTCs. The connections to the PTB are through the connectors shown in the bottom left, and front panel connections to external systems are at the top.}
    \label{fig:PTC_PCB}
\end{figure}

ProtoDUNE-HD used a version of the PTC that operated as an entirely passive device with only the aforementioned functions, identical to what was used in ProtoDUNE-SP \cite{PDSP_design}. Operational experience from ProtoDUNE-SP showed that it would be useful for the PTC to serve a number of active functions as well. While these upgrades were not completed in time for \mbox{ProtoDUNE-HD}, a new version of the PTC has since been developed for use in the DUNE far detectors. This updated PTC is pictured in figure \ref{fig:PTC_PCB} and will be the subject of the rest of this section. The redesigned PTC implements the ability to power individual WIBs on and off, as well as to monitor on-board voltages and currents, including the main power, individual WIB power, and local power. It can also monitor its PCB surface temperatures and read an assortment of sensors from the WIBs over a dedicated I2C bus on the PTB. All of this sensor information, as well as the status of the timing signal, can be transmitted from the PTC to the detector slow control system and the detector safety system.

\subsection{Hardware}

The heart of the PTC consists of six LTM8064 DC-DC converters that generate 12\,V for each WIB from the main 48\,V input to the PTC. Each LTM8064 has input and output power conditioning filters, and the switching frequency was chosen to be 970\,kHz, outside the frequency bands of concern for noise pickup on the FEMBs.

The PTC contains a timing SFP that receives an external timing signal, connecting it to a clock fanout that distributes copies of the timing signal to all WIBs. A priority encoder is implemented in hardware, which takes as input a single line per WIB that is asserted when a WIB wants to transmit, and then routes the highest numbered WIB's transmission to the SFP.

\begin{figure}
    \centering
    \includegraphics[width=\linewidth,trim=1.5cm 1.2cm 1.5cm 1.5cm,clip]{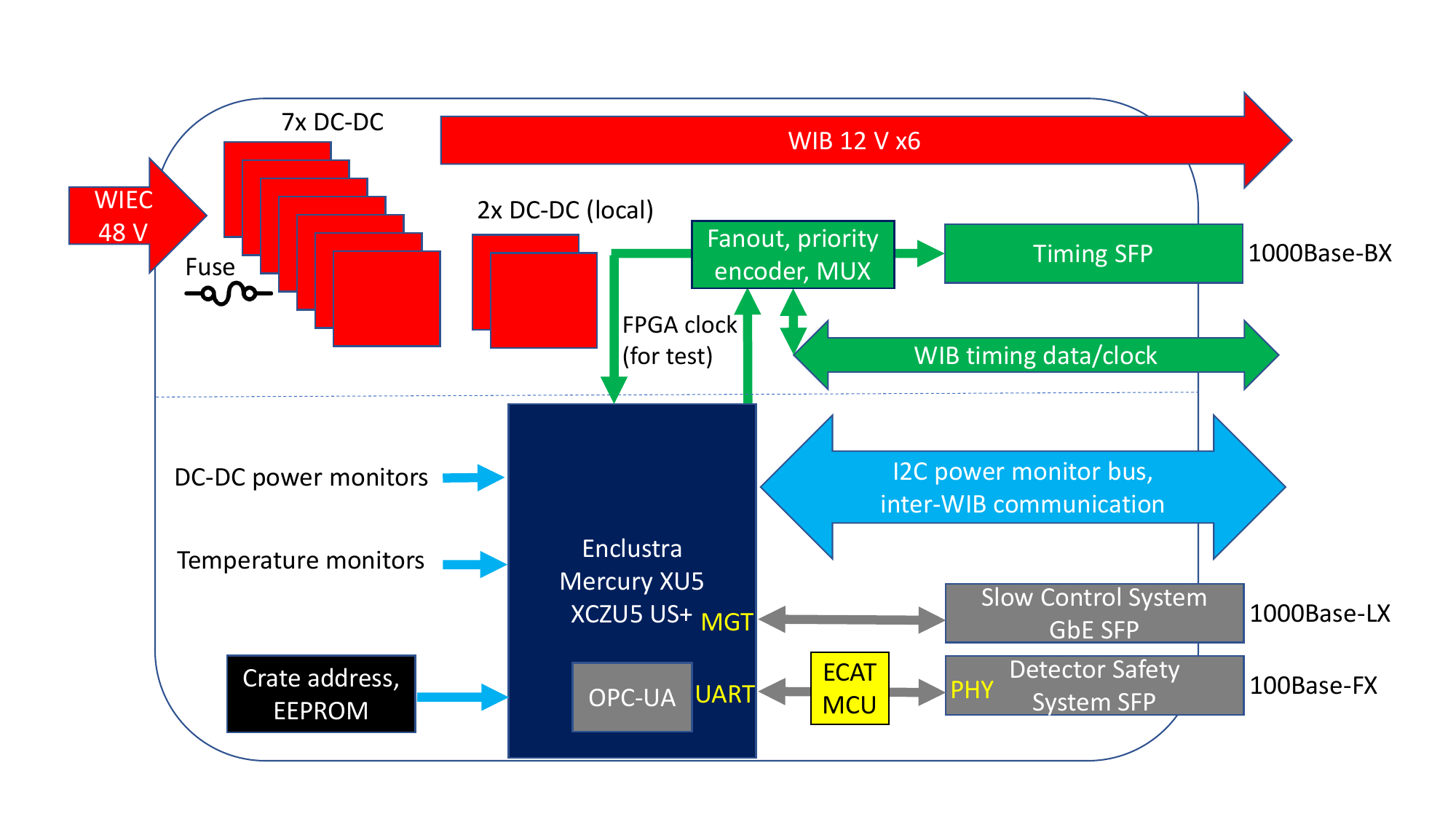}
    \caption{PTC hardware block diagram. The ProtoDUNE-HD PTCs only included the voltage (red) and timing (green) paths shown in the upper half. Functions in the bottom half, including the FPGA and associated monitoring lines (blue) and communication interfaces (gray), were added in a recent PTC redesign and will be used in DUNE.}
    \label{fig:PTC_HW}
\end{figure}

A block diagram of the PTC hardware is shown in figure \ref{fig:PTC_HW}, which now contains the following interfaces:
\begin{itemize}
    \item A main 48\,V power connector that powers the PTC and the rest of the WIEC, which includes voltage sense wires that can be connected to a remote power supply.
    \item A 1000Base-BX SFP for receiving timing signals from the DTS and transmitting back to it.
    \item A 1000Base-LX SFP for a GbE connection to slow control systems.
    \item A 100Base-FX SFP for communication with the detector safety system using EtherCAT \cite{ethercat}.
    \item A mini-USB connector for serial UART debugging
    \item A removable, bootable micro-SD card for the on-board system-on-chip (SoC), which controls most of the PTC's functions.
    \item Front panel status LEDs to display power status (WIBs and SoC), timing status (lock, transmit), SoC configuration status, SFP fiber link statuses, over-temperature indicator, GbE link status, and EtherCAT link status. 
    \item An 8-position DIP switch for setting a unique WIEC crate address.
    \item The PTB interfaces which include: power to WIBs, timing information to and from WIBs, timing transmit flags for priority encoding, I2C bus, and addressing.
\end{itemize}

The SoC is a commercial Enclustra ME-XU5-5EV-2I-D12E mezzanine card, which is replaceable during the lifetime of a DUNE detector. It is powered by a seventh DC-DC module, which also feeds smaller DC-DC converters that power other integrated circuits on the PTC. The GbE interface to the slow control system is implemented with hardware IP in the SoC and connected through the multi-gigabit transceivers (MGT) to the SFP. The EtherCAT interface is implemented with a hardware IP block in an Infineon XMC-4300 microcontroller unit (MCU) interfaced to a Marvell 88E3015 fast Ethernet transceiver (PHY) connected to the SFP. The MCU is connected to the SoC with UART links which transfer data to be packetized and sent to the detector safety system. The packet format can be configured using C code on the MCU. This C code is a combination of Beckhoff-provided code for the EtherCAT IP block and user-generated code. A JTAG interface between the SoC and MCU is provided for field upgrades, which can be uploaded through the GbE interface on the front panel. 

In addition to several I2C buses on the PTC that the SoC can use to read local sensors, there is a dedicated I2C bus to the WIB through a Texas Instruments (TI) TXS0102 level translator which connects to the PTB. On each WIB, a TI TCA9406 level translator connects to the shared I2C bus. This part reverts to a high-impedance state when powered off so that the I2C bus will not be affected if a misbehaving WIB is disabled by the PTC. Each WIB also has an MAX7357 bus multiplexer which can be addressed to switch between several IC buses on the WIB. 

Each PTC has an EEPROM with a globally unique serial number. MAC addresses can be derived either from this serial number or from geographical addressing (crate number) set by a DIP switch on the PTC. The SoC runs Xilinx Petalinux from a bootable SD card, which can be configured for NFS mounting of a filesystem.

The PTC uses a 10-layer PCB, with thick 2\,oz~copper for power planes and fusing for the main power and SoC mezzanine power connectors. The power distribution grounds and power planes are isolated from the digital electronics (SoC, MCU, timing components). TI TMP117 temperature sensors are placed near each DC-DC converter and report PCB temperature at those locations. Threshold registers in the temperature sensors and the LTC2945 voltage/current monitors can be set to trigger an immediate alert line which goes to the FPGA. Alert lines can be OR'ed in FPGA logic to turn on a red overtemperature LED on the front panel, set a register bit, or transmit error flags to the slow control or detector safety systems. Heatsinks are applied to each DC-DC converter ball grid array (BGA) package as well as the SoC package on the mezzanine.

\subsection{Firmware and Software}

\begin{figure}
    \centering
    \includegraphics[width=\linewidth,trim=3.5cm .2cm 3.5cm 2cm,clip]{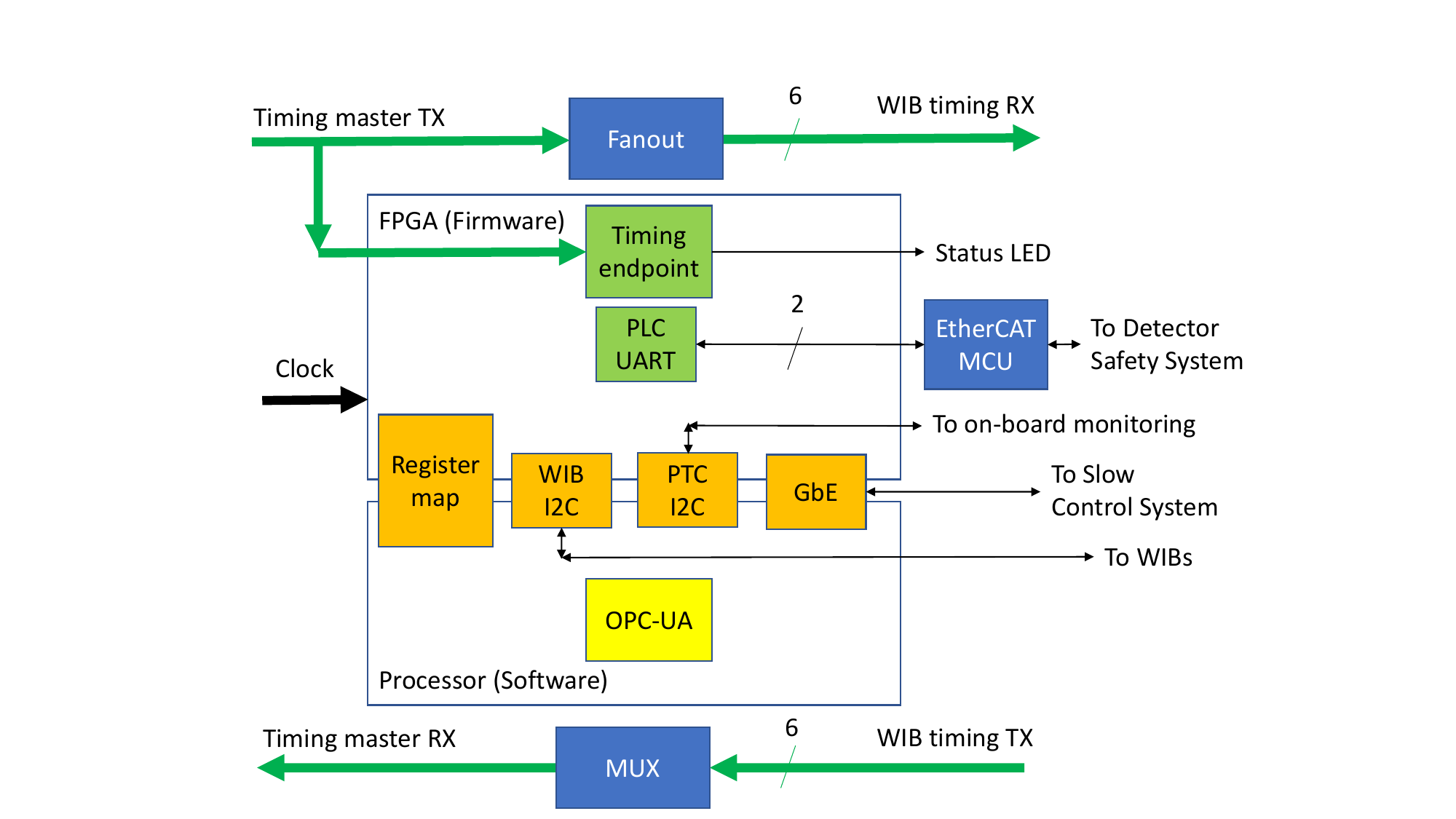}
    \caption{PTC firmware block diagram, showing major interfaces and flow of timing signals. Timing master RX/TX both flow through the PTC's front panel timing SFP, and WIB timing RX/TX both flow through the PTB.}
    \label{fig:PTC_FW}
\end{figure}

A block diagram of the PTC firmware is shown in figure \ref{fig:PTC_FW}. The register map is the main interface to all systems external to the firmware. It provides read-write bits for enabling and disabling \mbox{DC-DC} converters, as well as resetting or enabling firmware blocks and external components. It also contains read-only bits for crate addressing, alert indicators, SFP status, and timing lock status. A copy of the firmware timing endpoint receives timing information for the purpose of indicating the timing lock status. 

The firmware implements dedicated I2C controllers that allow software on the PTC to interface with and issue read commands to the PTC and WIB buses. Firmware-only control of these buses can also be implemented with reprogramming of the SoC. A GbE interface is implemented with a dedicated hardware IP block on the SoC. A firmware UART controller forwards information relevant to the detector safety system to the EtherCAT MCU. 

The processor is able to run custom software, such as an OPC-UA server that can be used for communication with the slow control system. In addition to all the monitoring functions that the SoC performs, it is able to override the timing signal and provide its own timing stream to WIBs for testing purposes. It can also override the priority encoder in cases where a misbehaving WIB erroneously keeps its transmit line asserted. Current versions of the firmware use much less than 50\% of available resources, allowing accommodation of future feature requests.

\subsection{Performance}

The primary function of the PTC is to provide clean, filtered power to the WIBs and, by extension, the FEMBs, which have the most stringent noise requirements. Tests with the upgraded PTC in standalone test benches and with a single WIEC in ProtoDUNE-HD have shown very similar performance to previous versions of the PTC, demonstrating that the addition of the SoC has not resulted in additional noise being coupled to the FEMBs. 

The PTC's EtherCAT interface has been tested with Beckhoff TwinCAT software in a standalone test crate, showing successful transmission of voltage and current readings from the PTC's main 48 V line to a Beckhoff PLC system. Future work includes further development of MCU software and FPGA firmware to expand the number of sensors being monitored, as well as potential inclusion of error detection. This interface will eventually be integrated with the DUNE detector safety system to implement various hardware interlocks.

The PTC's temperature performance was tested with a fully populated WIEC. The higher density of components on the upgraded PTC relative to previous versions results in generally increased temperatures when operated inside a WIEC, particularly around the DC-DC converters. This was addressed with a minor layout change on the board, changing the orientation of tall common-mode chokes to improve airflow. This was complemented by improvements to the overall airflow in the WIEC with modifications to have larger air holes in the side of the crate and more fans. With these changes, the PTC meets temperature specifications for both its DC-DC converters and the SoC die temperature, as measured via Xilinx JTAG.

\section{Electronics Performance in ProtoDUNE-HD}
\label{sec:performance}

\begin{figure}
    \centering
    \includegraphics[width=0.95\linewidth]{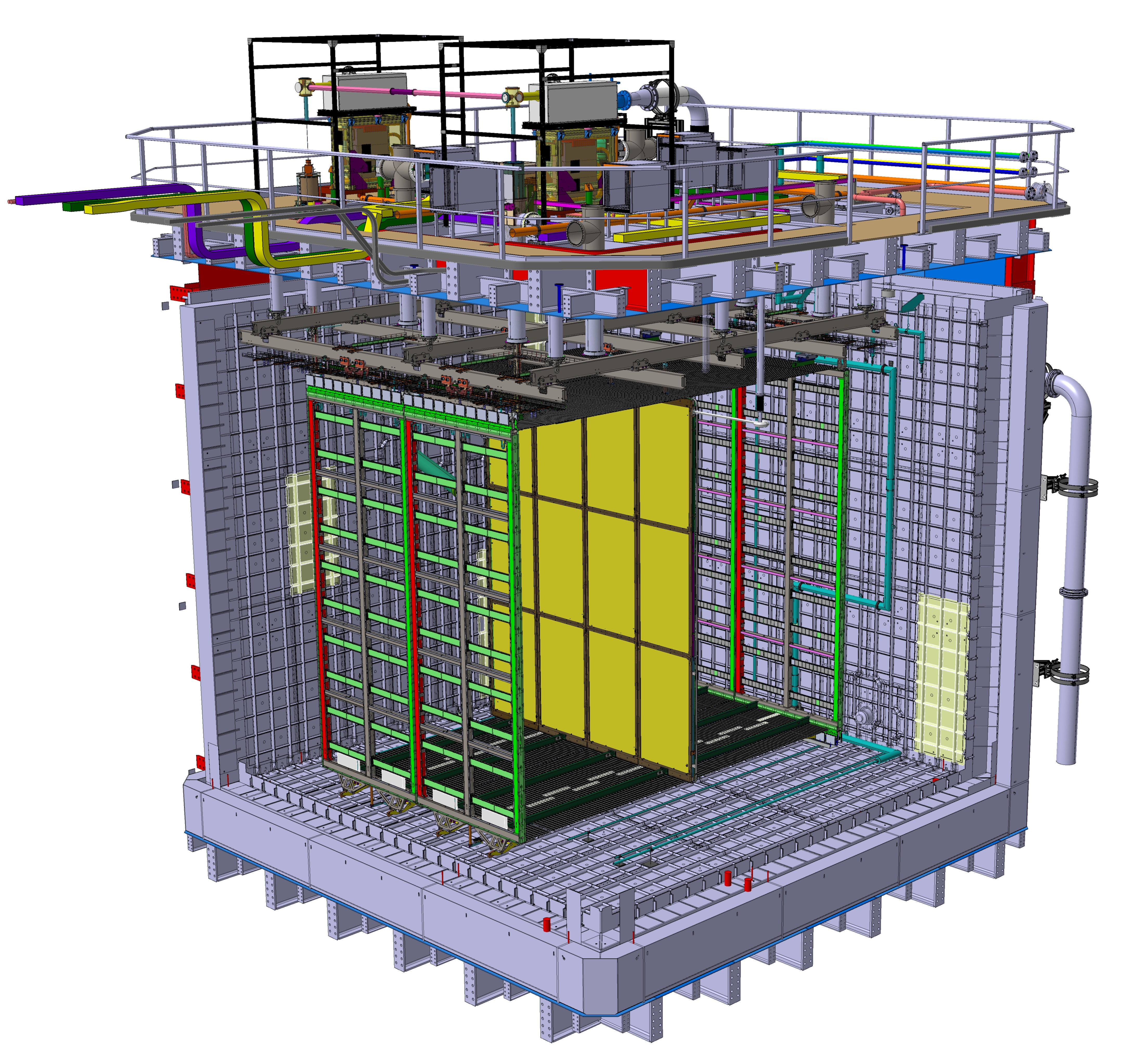}
    \caption{\label{fig:np04} Schematic of the ProtoDUNE-HD cryostat and interior TPC components. The cathode separating the two TPC volumes is shown as a solid yellow vertical plane in the middle. The two upper APAs can be seen in front of the cathode, and the two lower APAs can be seen behind the cathode.}
\end{figure}

The ProtoDUNE-HD detector contained a total of four APAs, whose layout within the cryostat is shown in figure \ref{fig:np04}. Two of the APAs were ``upper'' APAs, with the FEMBs physically installed on the top of the APA. The other two were ``lower'' APAs, which are effectively flipped upside down, resulting in the FEMBs being placed at the bottom of the detector. The dimensions of the ProtoDUNE-HD cryostat did not permit the upper and lower APAs to be stacked on top of each other like they would be in the DUNE HD far detector, but they were instrumented using the same lengths of cables that would be used in DUNE. This corresponded to 9 m data and power cables for the upper APA FEMBs and 22 m cables for the lower APA FEMBs.

ProtoDUNE-HD collected 10 weeks of beam data, as well as data from other subsystems, which will be the subject of future analyses. Here, we focus only on data used to calibrate and evaluate the TPC electronics. This includes pulser data from the on-board LArASIC pulsers, as well as a subset of cosmic-ray muon data that was collected using random triggers.

In the following results, 33 of the 10,240 electronics channels ($\sim$0.3\%) are excluded from analysis. Out of these, 17 channels were known to simply have no wire input because the wire was broken during APA assembly or handling. Another 6 channels lost connection to their wire under cryogenic conditions, but were otherwise normal in warm conditions. These are believed to have been caused by imperfections in the mechanical interface between the FEMBs and the APA that worsen under thermal stress; this interface has been improved in more recent APAs constructed for the DUNE far detector. These 23 channels all remained functional from the perspective of the electronics. The last 10 excluded channels were damaged by a high voltage discharge incident during the course of detector operations, which was large enough to overwhelm the protection diodes on the electronics and render their front-end amplifiers unusable. Otherwise, detector conditions were stable, and the results presented here are from representative periods of detector operation.

\subsection{Pulser-Based Calibration}
\label{sec:pulsers}

To characterize the analog response of the TPC front-end electronics and correct for electronics-induced distortions, we analyze internal pulser waveforms channel by channel. We start by injecting calibration pulses to each channel through the LArASIC pulser. We then fit the resulting raw waveforms in the time domain with an analytical electronics-response model adapted from studies of the MicroBooNE electronics~\cite{Adams_2018}, which used an earlier version of the \mbox{LArASIC}. For clarity, we report the model in the frequency domain. Here, the MicroBooNE electronic transfer function \(R(s)\) is generalized to the form of a realistic CR–(RC)\(^{n}\) shaper with imperfect pole–zero cancellation and finite decay:

\begin{equation}
    T(s)=A\,
\frac{(s+k_{3})(s+k_{5})}{(s+k_{4})(s+k_{6})\,(s+p_{0})\big[(s+p_{1r})^{2}+p_{1i}^{2}\big]\big[(s+p_{2r})^{2}+p_{2i}^{2}\big]}
\end{equation}

Some of the parameters in this transfer function are defined as functions of the peaking time $t_{p}$ and the response amplitude $A_{0}$: 
\begin{equation}
\begin{aligned}
    p_{0}&=1.477/(t_{p}C_{T}) \\
    p_{1r}&=1.417/(t_{p}C_{T}) \\
    p_{1i}&=0.598/(t_{p}C_{T}) \\
    p_{2r}&=1.204/(t_{p}C_{T}) \\
    p_{2i}&=1.299/(t_{p}C_{T}) \\
    C_{T}&=1/1.996 \\
    A&=A_{0}*2.7433/(t_{p}C_{T})^{4} 
\end{aligned}
\end{equation}

The terms \(p_{1r}+i\,p_{1i}\) and \(p_{2r}+i\,p_{2i}\) are complex-conjugate pole pairs; \(k_{3},k_{5}\) are real zeros; \(k_{4},k_{6}\) are real poles. These added pole–zero pairs \((k_{3},k_{4})\) and \((k_{5},k_{6})\) control the small undershoot/overshoot tails observed in data: exact cancellation (\(k_{3}=k_{4}\), \(k_{5}=k_{6}\)) reduces \(T(s)\) to the ideal form, while controlled mismatches between the parameters set the tail's sign and timescale. Mismatches in these parameters correspond to imperfect pole-zero cancellation in the amplifier as a result of natural variations in the fabrication process. This parametrization reproduces the leading edge, peaking time, and the characteristic overshoot or undershoot of the pedestal after a pulse. We fit this function to each channel's average response independently, allowing us to extract the per-channel gain, peaking time, and a few terms that control the undershoot or overshoot amplitude and recovery.

\begin{figure}[!htbp]
  \centering
  \includegraphics[width=0.8\linewidth]{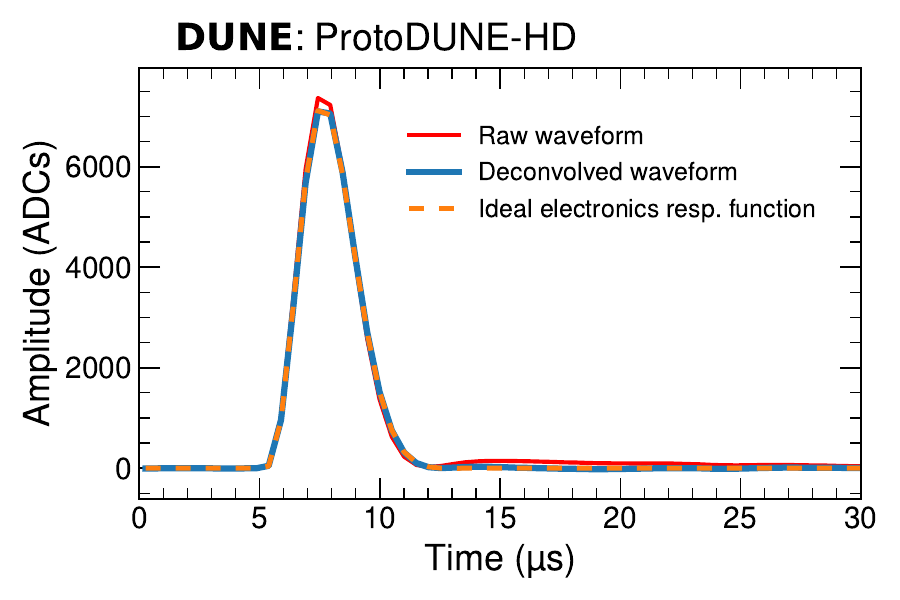}
  \caption{\label{fig:deconvolved_wf}
  Example pulser waveform from a typical channel, showing the raw data (red), deconvolved waveform using the fitted channel response (blue), and the chosen ideal electronics response (orange, dashed). One time tick is 512\,ns, following the digitization rate of the ColdADC. Deconvolution corrects for the over/undershoot and recovers a standardized shape suitable for precise gain and peaking-time extraction.}
\end{figure}

After fitting, we apply a measured-response deconvolution to each channel to correct for non-ideal variation in the channel response shapes, allowing us to standardize them to a common nominal response. We use the fitted per-channel response to remove channel-dependent shaping and then re-express the waveform with a chosen nominal response, while lightly tapering high frequencies to avoid noise amplification. This suppresses over/undershoot without altering the physical timing, which enables consistent gain and peaking-time determination across channels. Figure~\ref{fig:deconvolved_wf} shows the result of this fit to a typical channel. The raw waveform exhibits a fast rise, decay, and visible overshoot of the pedestal after the pulse, which is consistent with imperfect pole–zero cancellation. The deconvolved waveform is nearly symmetric about the peak and closely follows the target ideal response, indicating that the observed distortions are dominated by the electronics and are well corrected by this fit procedure.

\begin{figure}[!htbp]
  \centering
  \includegraphics[width=0.8\linewidth]{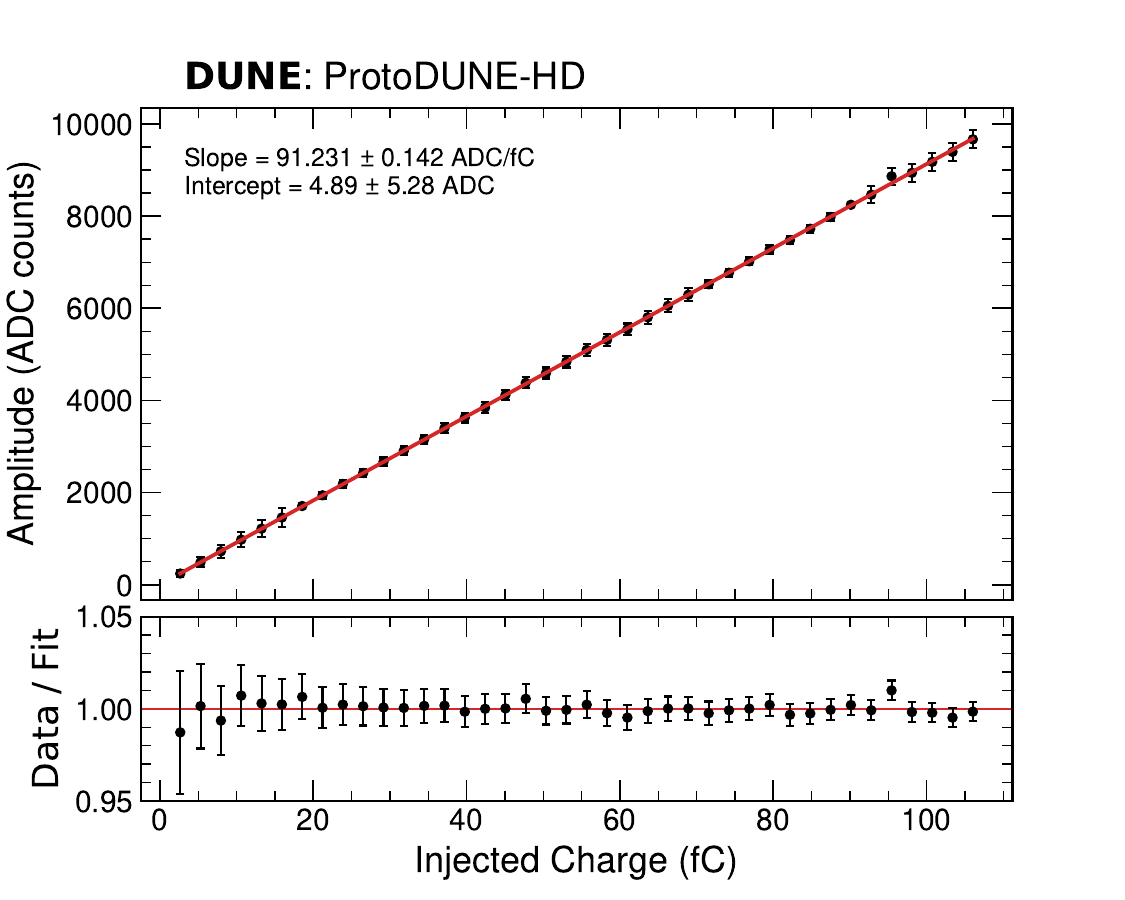}
  \caption{\label{fig:fit_example_7p8}
  Linear fit of peak amplitude $A_0$ versus injected charge $Q$ using pulser data collected for a typical channel. The depicted channel was configured with settings of $7.8\,\mathrm{mV/fC}$ gain and 2\,$\mu$s peaking time. The slope $dA_0/dQ$ (ADC/fC) gives the calibrated gain for this channel; the intercept is consistent with zero as expected. Error bars show per-point uncertainties on the fitted peak amplitude.}
\end{figure}

We apply this pulse shape fit and correction procedure to data taken with a peaking time of 2\,$\mu$s and at gains of $7.8$ and 14\,mV/fC. For each channel, we collect data with a range of different pulser DAC settings. The DAC setting determines the amplitude of internal test pulses, with each DAC value corresponding to a known injected charge $Q$ through the on-chip injection capacitor and the DAC transfer function. At every setting, we record the fitted peak amplitude $A_0$ and the peaking time $t_p$.  For each channel, the slope $dA_0/dQ$ obtained from a linear fit of the fitted pulse amplitude versus injected charge $A_0(Q)$ over the full charge scan determines its calibrated gain, allowing us to convert between ADC units and charge. Figure~\ref{fig:fit_example_7p8} shows a representative $A_0$–vs–$Q$ fit at a nominal gain of 7.8\,mV/fC. We extract the calibrated peaking time of each channel from the same pulse shape fits, calculating it as the uncertainty-weighted average of the fitted peaking time values over the full charge range.

\begin{figure}[htbp]
  \centering
  \begin{subfigure}[b]{0.48\linewidth}
      \includegraphics[width=\linewidth]{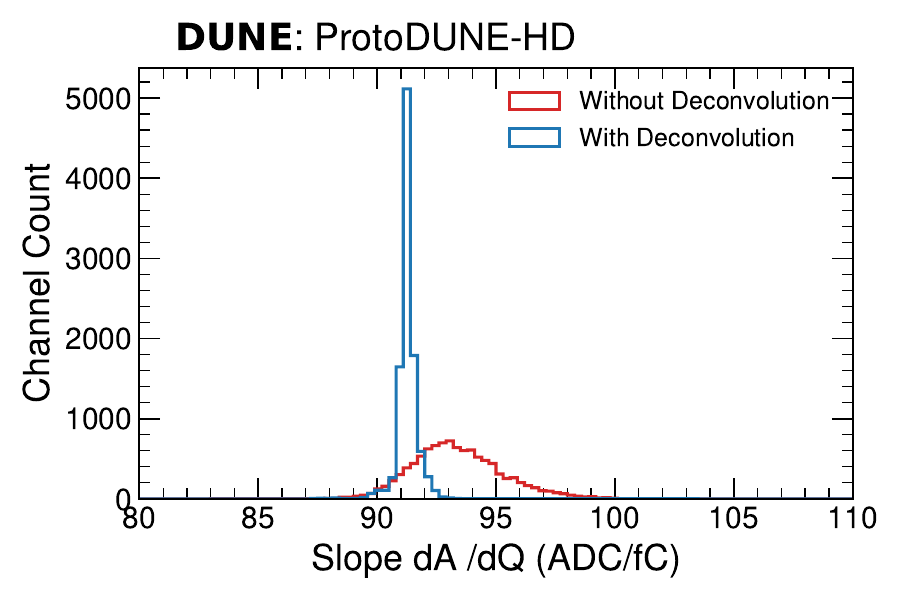}
      \caption{$7.8\,\mathrm{mV/fC}$ gain, 2\,$\mu$s peaking time}
      \label{fig:gain_hist_7p8}
  \end{subfigure}
  \hfill
  \begin{subfigure}[b]{0.48\linewidth}
      \includegraphics[width=\linewidth]{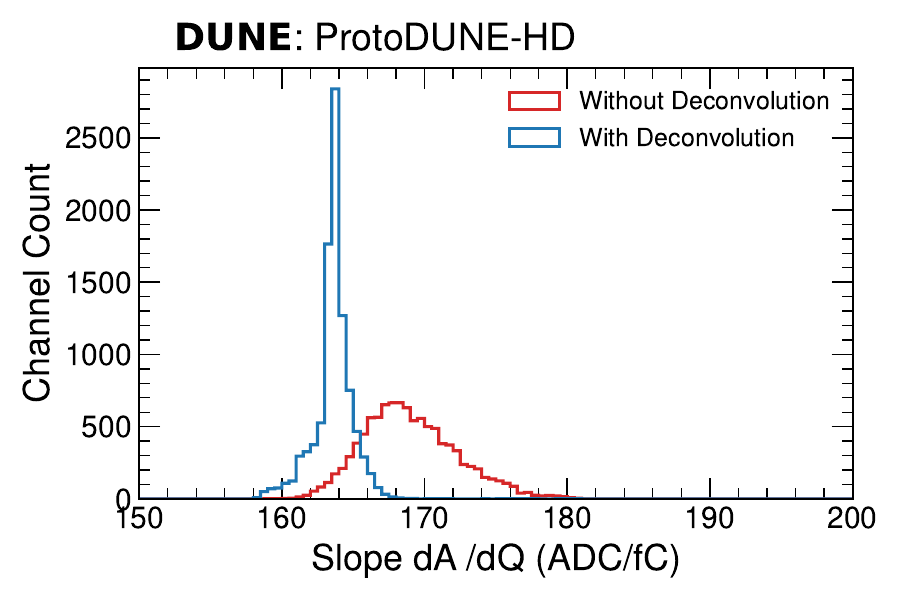}
      \caption{$14\,\mathrm{mV/fC}$ gain, 2\,$\mu$s peaking time }
      \label{fig:gain_hist_14}
  \end{subfigure}

  \vspace{0.8em}

  \begin{subfigure}[b]{0.48\linewidth}
      \includegraphics[width=\linewidth]{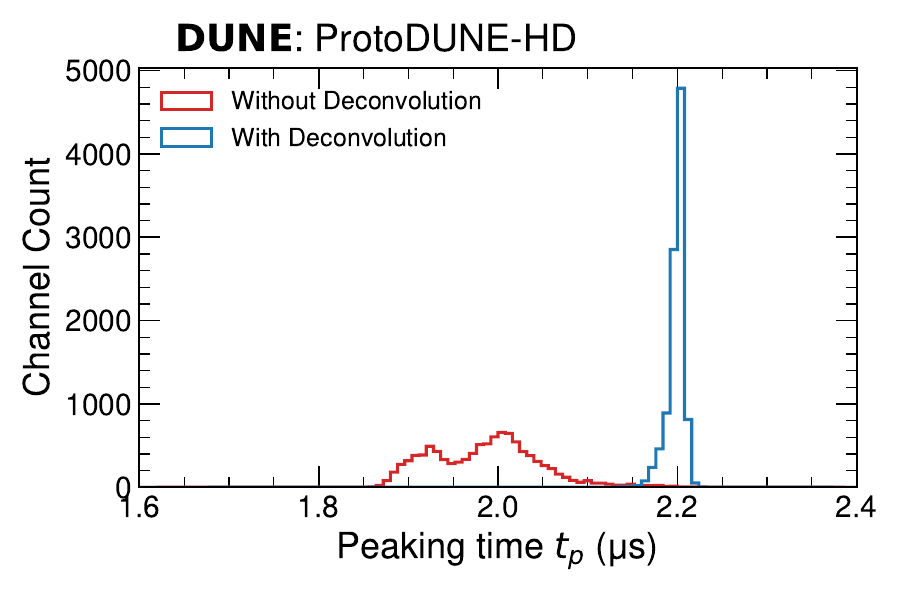}
      \caption{$7.8\,\mathrm{mV/fC}$ gain, 2\,$\mu$s peaking time }
      \label{fig:tp_hist_7p8}
  \end{subfigure}
  \hfill
  \begin{subfigure}[b]{0.48\linewidth}
      \includegraphics[width=\linewidth]{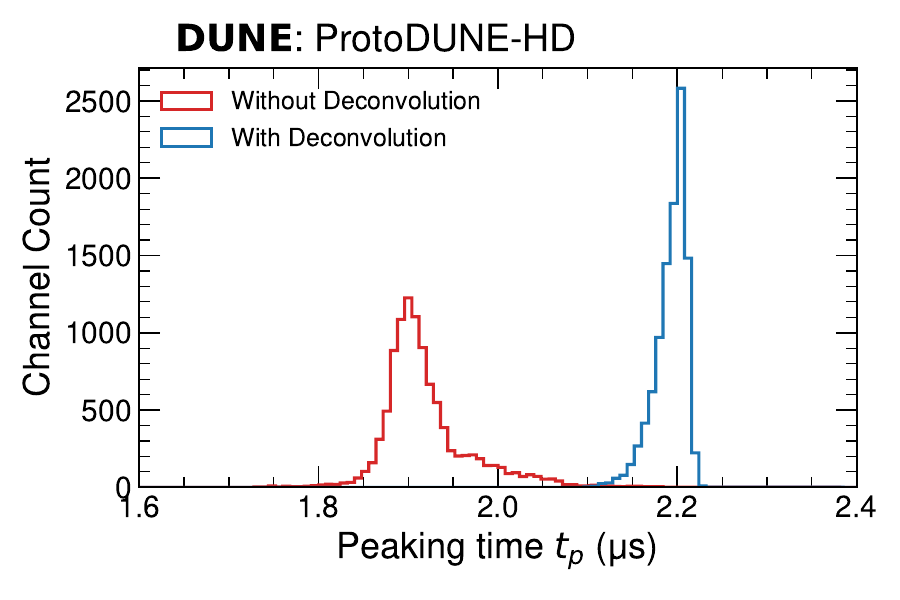}
      \caption{$14\,\mathrm{mV/fC}$ gain, 2\,$\mu$s peaking time }
      \label{fig:tp_hist_14}
  \end{subfigure}

  \caption{Calibrated results for both gain settings. After deconvolution, the $dA/dQ$ histograms peak at calibrated gains of $7.814 \pm 0.095\,\mathrm{mV/fC}$ for the nominal $7.8\,\mathrm{mV/fC}$ configuration and $14.063 \pm 0.146\,\mathrm{mV/fC}$ for the nominal $14\,\mathrm{mV/fC}$ configuration. Conversions from $dA/dQ$ to units of mV/fC are obtained with a factor of $0.08545\,\mathrm{mV/ADC}$, coming from equal division of the ColdADC's 1.4 V operating range into $2^{14}$ bins for its 14-bit outputs. The peaking time histograms peak at $2.198 \pm 0.009\,\mu\mathrm{s}$ for the nominal $7.8\,\mathrm{mV/fC}$ configuration and $2.192 \pm 0.017\,\mu\mathrm{s}$ for the nominal $14\,\mathrm{mV/fC}$ configuration.  }
  \label{fig:decon_params}
\end{figure}

Figure~\ref{fig:decon_params} shows the per-channel distributions of the fitted slope $dA_0/dQ$ and peaking time for the two nominal gain settings. Before deconvolution, the distributions are broad and biased away from the nominal values, consistent with channel-dependent over/undershoot. After deconvolution, they are narrowed and centered at the expected values, indicating removal of electronics-driven distortions. The peaks of the gain distributions are $7.814 \pm 0.095\,\mathrm{mV/fC}$ for the $7.8\,\mathrm{mV/fC}$ configuration and $14.063 \pm 0.146\,\mathrm{mV/fC}$ for the $14\,\mathrm{mV/fC}$ configuration. The conversion from the slope in $\mathrm{ADC/fC}$ to gain in $\mathrm{mV/fC}$ is obtained by multiplying by the ADC least-significant-bit size (volts per count), which is determined from the ADC full-scale input range and resolution. With the ColdADC's 1.4~V operating range and 14-bit resolution, this corresponds to a conversion factor of $0.08545\,\mathrm{mV/ADC}$. For peaking time, the distributions after deconvolution are peaked at $2.198 \pm 0.009\,\mu\mathrm{s}$ for the $7.8\,\mathrm{mV/fC}$ configuration and $2.192 \pm 0.017\,\mu\mathrm{s}$ for the $14\,\mathrm{mV/fC}$ configuration. The cited uncertainties are the standard deviation of results across all calibrated channels.

\begin{figure}[!htbp]
  \centering
  \includegraphics[width=0.75\linewidth]{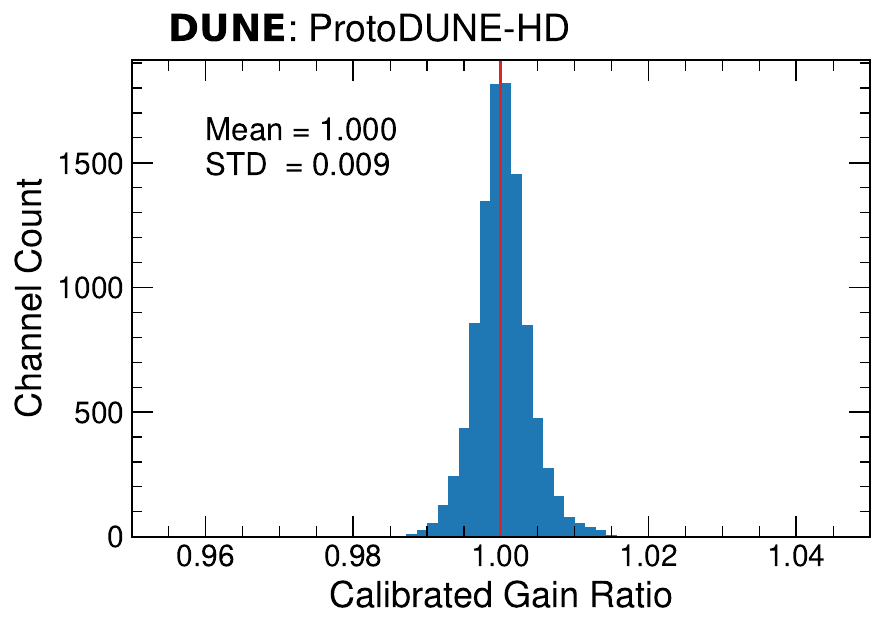}
  \caption{\label{fig:gain_ratio} Calibrated gain ratio between the no-gain-matching and gain-matching modes. For each channel $i$, we compute a ratio $R_i = (\mathrm{d}A_0/\mathrm{d}Q)^{\text{no-gain}}_i \big/ (\mathrm{d}A_0/\mathrm{d}Q)^{\text{gain}}_i$, with the uncertainty coming from standard error propagation. The distribution is centered at unity with a width of $0.9\%$, consistent with the per-channel fit uncertainties.}
\end{figure}

To validate the consistency of the gain calibration, we perform a cross-check by repeating the analysis with and without the internal LArASIC gain-matching correction applied. This setting changes the step sizes in the LArASIC's internal DAC output, resulting in different amounts of injected charge. The resulting per-channel gain ratios, shown in figure~\ref{fig:gain_ratio}, quantify the variation in per-channel calibrations obtained with the two pulser configurations. The distribution is centered at unity with a standard deviation of~$0.9\%$, indicating that these two pulser configurations yield calibration results consistent with each other. The narrow width of the ratio distribution confirms that the observed spread in calibrated gains is fully consistent with the expected per-channel fit uncertainties, demonstrating that the electronics response model we used provides an accurate description of the measured waveforms.

After gain calibration, another important quantity is the linearity of the channel response, which will determine the error we incur by assuming a constant gain over the entire available dynamic range. We quantify the channel response linearity using the same pulser scans, comparing the measured pulse amplitude for each amount of injected charge to the prediction from a linear model that assumes constant gain. We model a perfect linear response for each channel $c$ as

\begin{equation}
\hat{A}_{c}(Q) \;=\; a_c \;+\; b_c\, Q,
\label{eq:linfit}
\end{equation}

where $Q$ is the injected charge (fC), $\hat{A}_c$ is the expected reconstructed amplitude in ADC counts for channel $c$, and $(a_c,b_c)$ are per-channel fit parameters. We fit eq.~(\ref{eq:linfit}) over the entire available $Q$ range using weights $w_{i,c}=1/\sigma_{A_{i,c}}^2$, where $\sigma_{A_{i,c}}$ is the standard deviation of each measured amplitude $A_{i,c}$ over repeated pulses with the same nominal $Q_i$. For every measurement $(Q_i,A_{i,c})$ on channel $c$, we then compute the (percentage) nonlinearity $\mathrm{NL}$ by the measured deviation of $A_{i,c}$ from the predicted linear response $\hat{A}_{c}(Q_i)$:

\begin{equation}
\mathrm{NL}_{c}(Q_i)\;[\%] \;=\; 
100 \times \frac{A_{i,c} - \hat{A}_{c}(Q_i)}{\hat{A}_{c}(Q_i)}
\label{eq:nl}
\end{equation}

\begin{figure}[!htbp]
  \centering
  \includegraphics[width=0.85\linewidth]{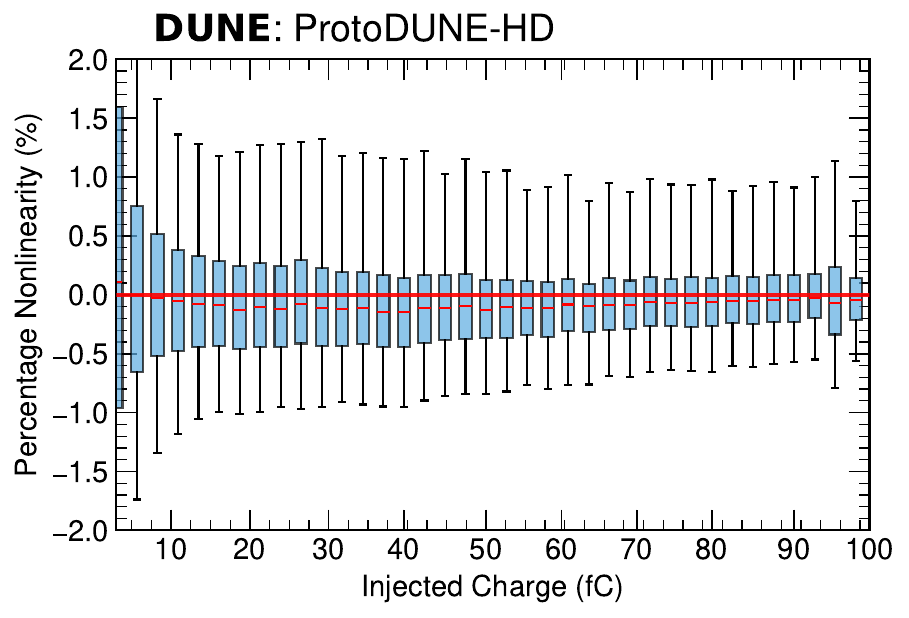}
  \caption{\label{fig:non-linearity} Percentage nonlinearity, as defined in eq.~(\ref{eq:nl}), versus injected charge. The interquartile range (boxes), median (red lines), and 5th--95th percentile results (whiskers) across all channels are indicated. Outliers are suppressed for visual clarity.}
\end{figure}

Since $\hat{A}_{c}(Q_i)$ is obtained with a linear fit of the whole charge range in the same way that is done for gain calibration, this gives us a measure of the error we incur by assuming a perfectly linear response during calibration. Figure~\ref{fig:non-linearity} displays the resulting percentage nonlinearity as a function of injected charge for the 7.8\,mV/fC gain setting. To summarize behavior across channels, we group the $\mathrm{NL}$ values by injected charge $Q$ (one box per $Q$ setting). Each box indicates the distribution across channels, showing the median (red line), interquartile range (box), and the 5th--95th percentile range (whiskers); points outside that interval are treated as outliers and visually suppressed for clarity in the final figure.

Across most of the charge range, the median nonlinearity per charge point deviates from zero by less than $0.1\%$, indicating good average linearity of the collection response with respect to injected charge. The spread across channels, as shown by the whiskers and interquartile range boxes, increases at small $Q$ because electronic noise is no longer negligible compared to the signal size, affecting the ratio in eq.~(\ref{eq:nl}). At higher $Q$, the impact of electronic noise and digitization is negligible, and the spread among channels reflects the actual spread in nonlinearity.

\begin{figure}[htbp]
    \centering
    \includegraphics[width=0.95\linewidth]{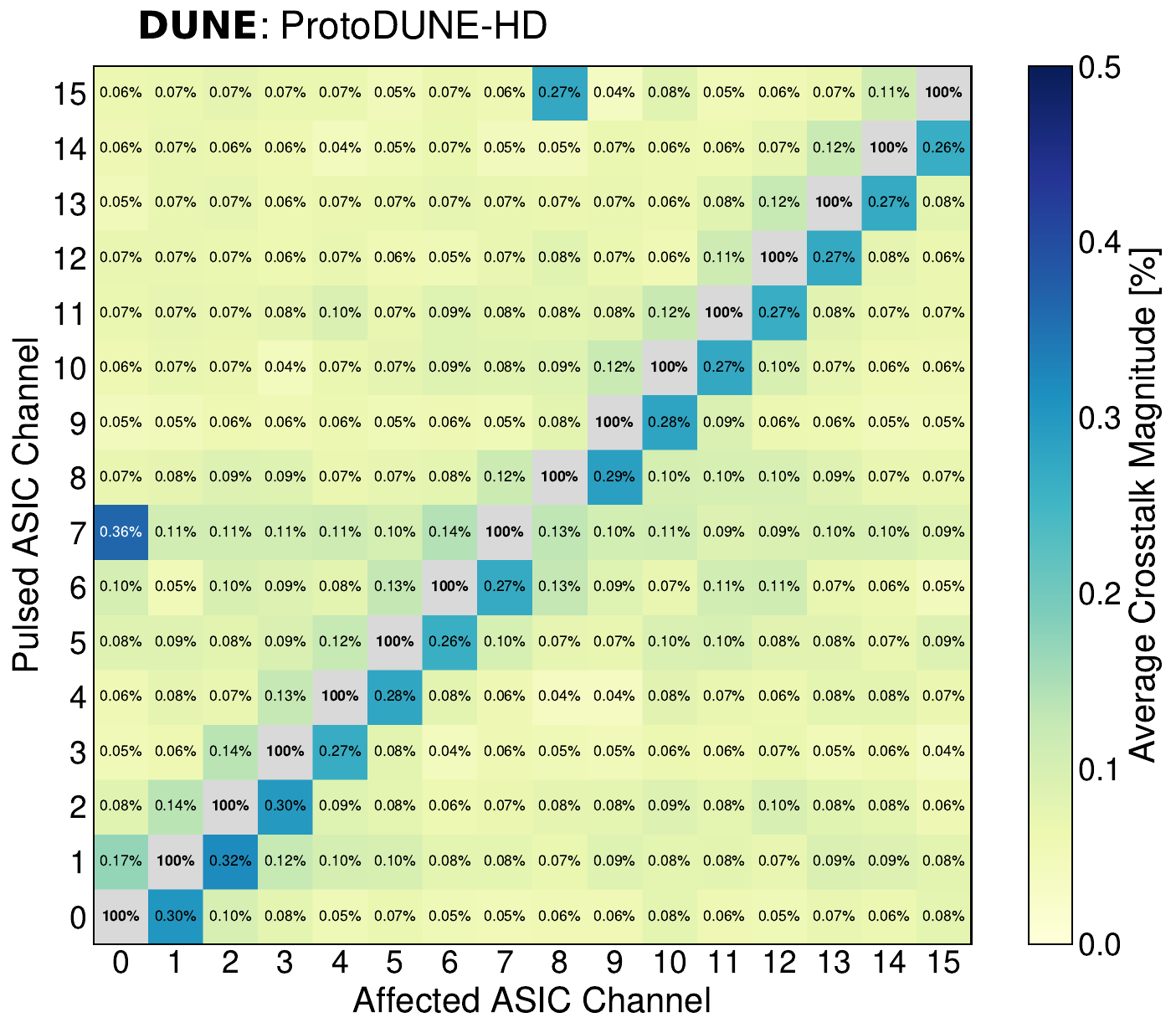}
    \caption{Matrix of average cross-talk values for a single ASIC, obtained by pulsing individual channels (y-axis) and observing the induced amplitude in the remaining channels (x-axis). For each pulsed–affected channel pair, the cross-talk magnitude is defined as the slope of a linear fit of the affected channel's response amplitude versus pulsed amplitude across multiple DAC settings. Values are then averaged across all suitable ASICs.}
    \label{fig:xtalk_matrix}
\end{figure}

To evaluate channel-to-channel cross-talk within the LArASIC, we analyze dedicated pulser runs in which a single channel of each ASIC (out of channels 0 through 15) was pulsed individually, while the remaining 15 channels on the same ASIC remained unpulsed and are referred to as the affected channels. This allows us to quantify how much of the injected signal leaks into other channels of the same ASIC. We note that this scheme primarily measures cross-talk within the electronics, and cross-talk between the physical APA wires is expected to be negligible. For each pulsed event, we average the waveform of every channel across all saved samples to suppress the effects of noise. We then identify the peak of the pulsed (source) channel's response and measure the corresponding amplitude of each affected channel's response in the same time bin. This measurement is repeated for multiple different amplitudes of injected pulses, and the cross-talk magnitude is extracted from a linear fit of the affected channel's response amplitude versus pulsed amplitude for each pair of channels. The slope of this fit defines the cross-talk ratio for each pulsed–affected channel pair. The intercept of the fit is allowed to float to account for crosstalk from the digital pulse used to activate the LArASIC pulsers, which is separate from the analog crosstalk we are trying to measure. ASICs with saturated pulsed channels or with insufficient statistics are excluded from the analysis. The cross-talk matrix shown in figure~\ref{fig:xtalk_matrix} contains the results averaged across all analyzed ASICs. Variation between ASICs was small, with the standard deviation of extracted cross-talk values being less than $0.01\%$.

The observed cross-talk levels in non-adjacent channels are typically less than $0.1\%$. For non-saturating pulses, an effect of this magnitude is mostly indistinguishable from noise in affected channels. There is slightly higher cross-talk from each channel $N$ to its adjacent channel $N-1$, which is attributed to the small capacitive coupling on the FEMB between neighboring inputs to the LArASIC. The largest cross-talk values, up to $\sim$0.3\%, come from the effect of channel $N$ on $N+1$ within 8-channel cycles, with channel 7 affecting channel 0 and channel 15 affecting channel 8. These pairs stem not from typical charge sharing, but from the design of the ColdADC, which is internally divided into two 8-channel ADCs. Each one of these ADCs sequentially samples its 8 input SHAs in a cycle, and bandwidth limitations from the internal multiplexer allow kickback from one sample to affect the next sample to a small degree \cite{ColdADC}, producing the cross-talk structure that we see here from channels $N$ to $N+1$ within the ColdADC 8-channel groupings. However, even the maximum level of cross talk seen is well below 1\%, satisfying DUNE requirements to have minimal effect on calorimetric analysis \cite{VD_TDR}. 


\subsection{Noise Performance}
\label{sec:performance_noise}

The noise performance of the electronics is defined as a measure of the fluctuations in their response independent of any signal, contributing some uncertainty to the amplitude of each signal we observe. We quantify it as the RMS of the pedestal-subtracted digitized electronics output in the absence of any charge signals from ionization events. As described in section \ref{sec:femb_performance}, it is expressed as an ENC in number of electrons.

The two front-end settings with the most impact on the electronics response are the gain and peaking time, with the gain controlling the signal amplitude and the peaking time controlling the signal width. For gain, we seek to maximize the amplifier's available dynamic range without degradation in noise performance. We discard the option of 25\,mV/fC because the resulting dynamic range is too limited, and we discard the option of 4.7\,mV/fC gain because it is low enough that the ADC and quantization noise are no longer negligible. This leaves us with 7.8\,mV/fC gain or 14\,mV/fC gain as the primary options. For the peaking time, 0.5\,$\mu$s is too short given our chosen digitization rate of ${\sim}1.95$\,MHz. 3\,$\mu$s is long enough that pileup of charge along the drift direction in a single channel can become a concern. This leaves us with 1\,$\mu$s and 2\,$\mu$s as the primary options. For reference, the MicroBooNE and ProtoDUNE-SP experiments, which used earlier versions of the LArASIC, both operated with 14\,mV/fC gain and 2\,$\mu$s peaking time \cite{Adams_2018,PDSP_results}.

\begin{figure}
    \centering
    \includegraphics[width=0.48\linewidth]{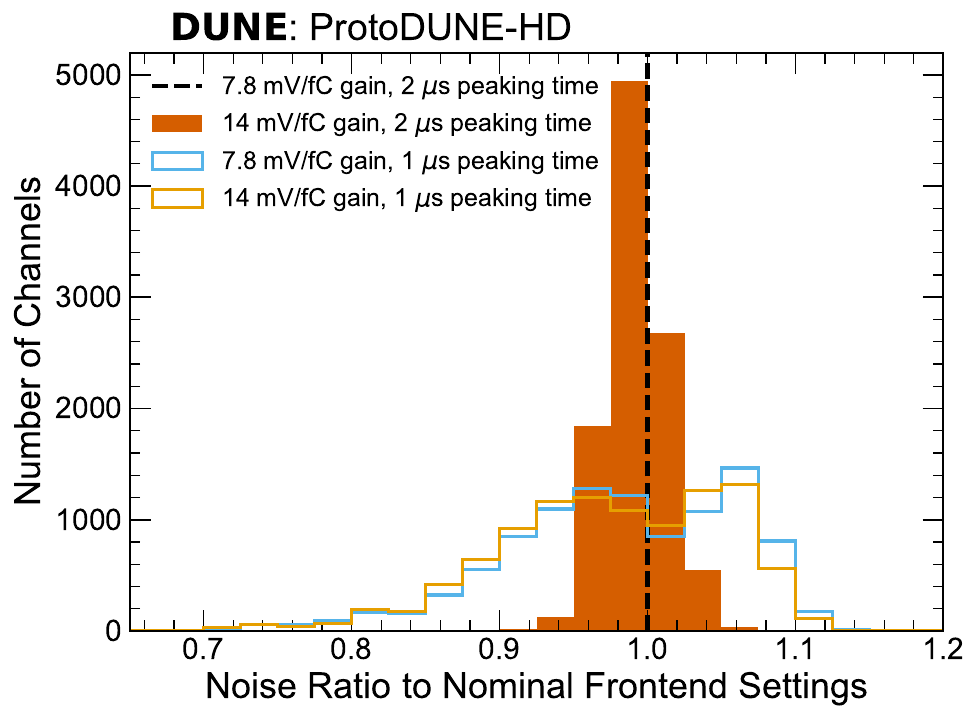}
    \includegraphics[width=0.48\linewidth]{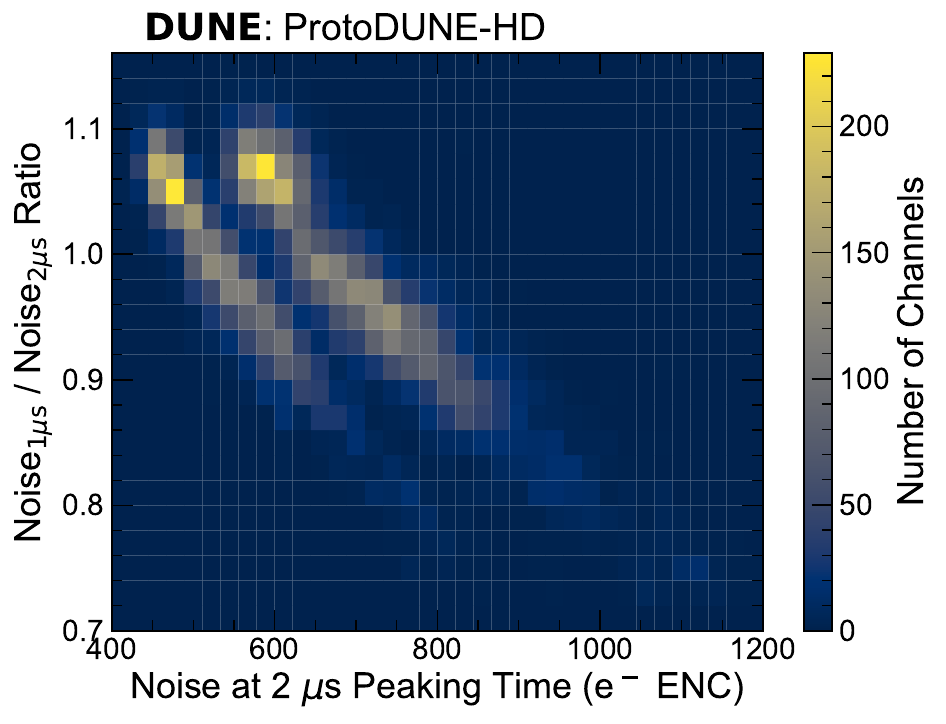}    \caption{\label{fig:ce_settings_ratios} \textbf{Left:} ratios of noise levels per channel for settings of 7.8 or 14\,mV/fC gain and 1 or 2\,$\mu$s peaking time in the LArASIC. Comparisons are against the nominal settings, which were 7.8\,mV/fC gain and 2\,$\mu$s peaking time. \textbf{Right:} ratio of noise at 7.8\,mV/fC gain with 1\,$\mu$s peaking time to noise with 2\,$\mu$s peaking time, plotted against noise with the nominal settings. Typical channels have slightly worse noise with 1\,$\mu$s peaking time, but channels in the higher noise tail are improved with 1\,$\mu$s peaking time.}
\end{figure}

ProtoDUNE-HD ran with nominal front-end settings of 7.8\,mV/fC gain and 2\,$\mu$s peaking time for all of the LArASICs. Following the calibration procedure described in section \ref{sec:pulsers}, comparisons of the noise levels from alternative settings against the final nominal settings are shown in figure~\ref{fig:ce_settings_ratios}. There is very little difference in noise between the two gain settings, with the 7.8\,mV/fC gain showing on average $<1\%$ higher noise. This is attributable to the performance of the ColdADC, which has generally low noise levels of around 130\,$\mu$V RMS, or 1.5 14-bit ADC units \cite{ColdADC}. This negligibly small difference in noise leads to the choice of 7.8\,mV/fC as the gain setting, to maximize the amplifier's available dynamic range. We observe that using a 2\,$\mu$s peaking time reduces noise for the bulk of channels but produces a longer high-noise tail. We choose a 2\,$\mu$s peaking time to optimize the performance of the bulk of channels.

\begin{figure}
    \centering
    \includegraphics[width=0.48\linewidth]{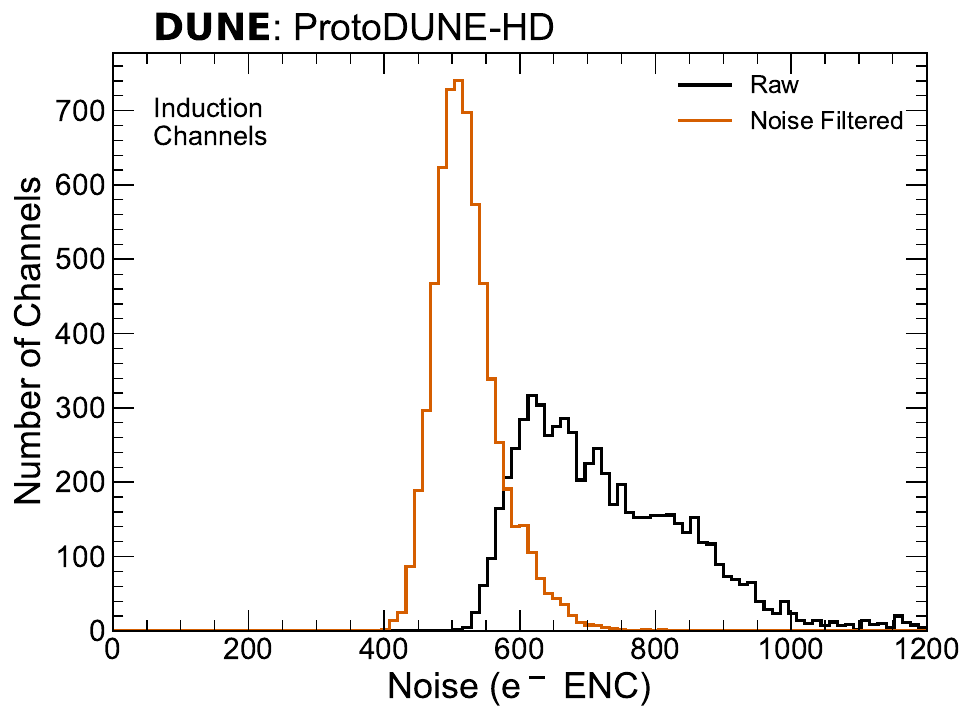}
    \includegraphics[width=0.48\linewidth]{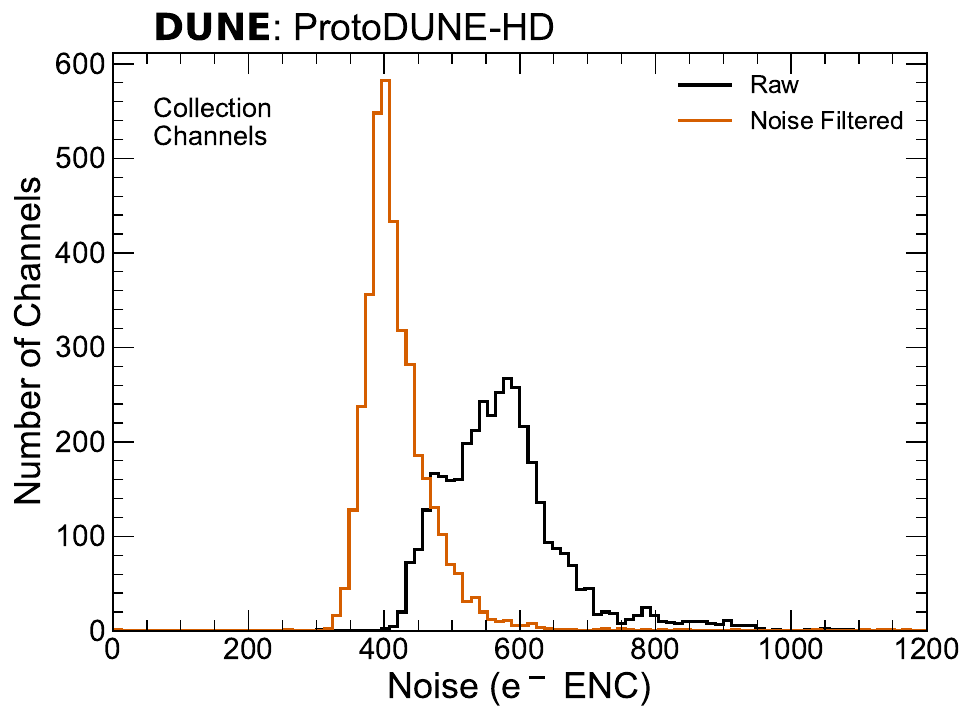}
    \caption{\label{fig:noise_rms} Observed noise levels across all channels in beam data, plotted separately for channels connected to induction wires (\textbf{left}) and channels connected to collection wires (\textbf{right}). Wires on the induction layers have different lengths from wires on the collection layer, and so in general their expected noise levels are different. Noise levels are shown before and after offline filtering.}
\end{figure}

The noise levels achieved across all channels with the nominal front-end settings are shown in figure \ref{fig:noise_rms}. These noise values were calculated using a period of active test beam data collection, so that they reflect the charge readout electronics performance under conditions consistent with later beam data analysis. Wires on both induction planes are longer than those on the collection planes, corresponding to higher capacitive loads and thus higher noise on their electronics channels. Induction wires also have more variance in their capacitance due to the manner in which they wrap around the APAs, leading to more variation in their noise levels relative to channels connected to collection wires. Noise results are shown both before and after offline noise filtering, which removes correlated noise from channels on the same FEMB \cite{Acciarri_2017,PDSP_results}. The filtered noise is most relevant for projected performance in event reconstruction and any offline analysis, where the filter will have been applied. The unfiltered noise is relevant for determining online trigger thresholds, as latency requirements prohibit the application of the full filtering algorithm there. In \mbox{ProtoDUNE-HD}, these unfiltered noise levels allowed for the use of online trigger thresholds of 60 ADC counts per collection channel, corresponding to signal sizes of 4000-4200 electrons.

\begin{figure}
    \centering
    \includegraphics[width=0.48\linewidth]{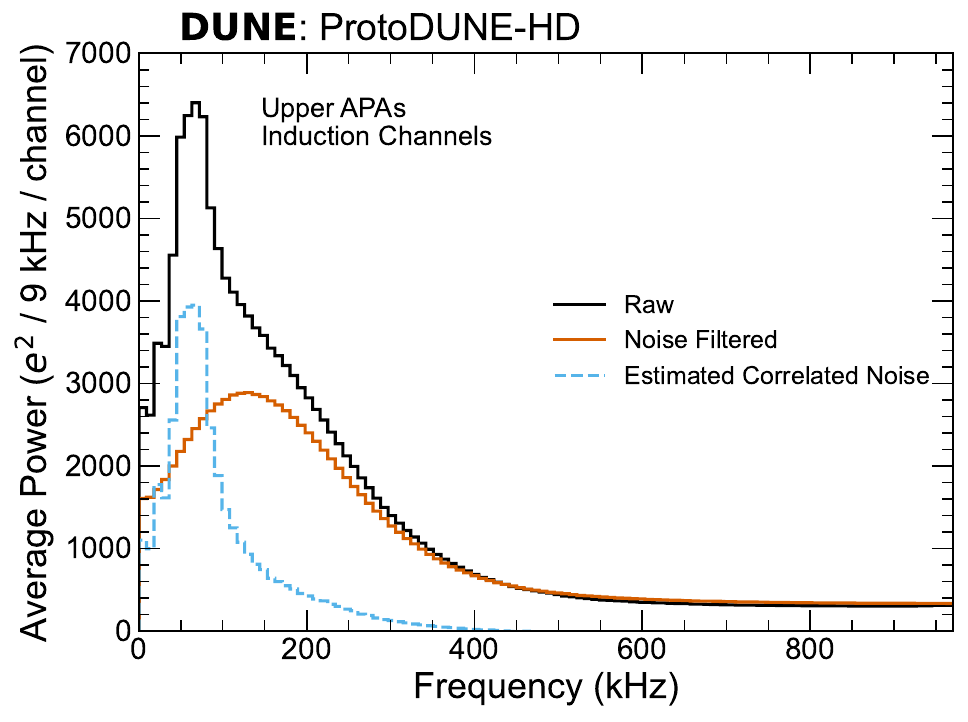}
    \includegraphics[width=0.48\linewidth]{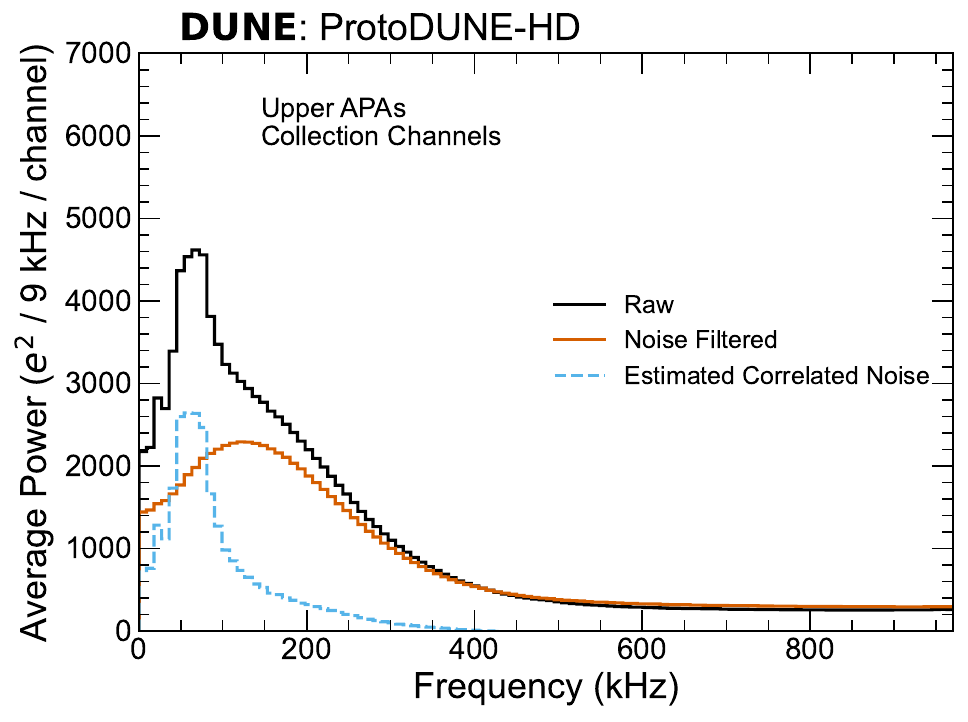} \\
    \includegraphics[width=0.48\linewidth]{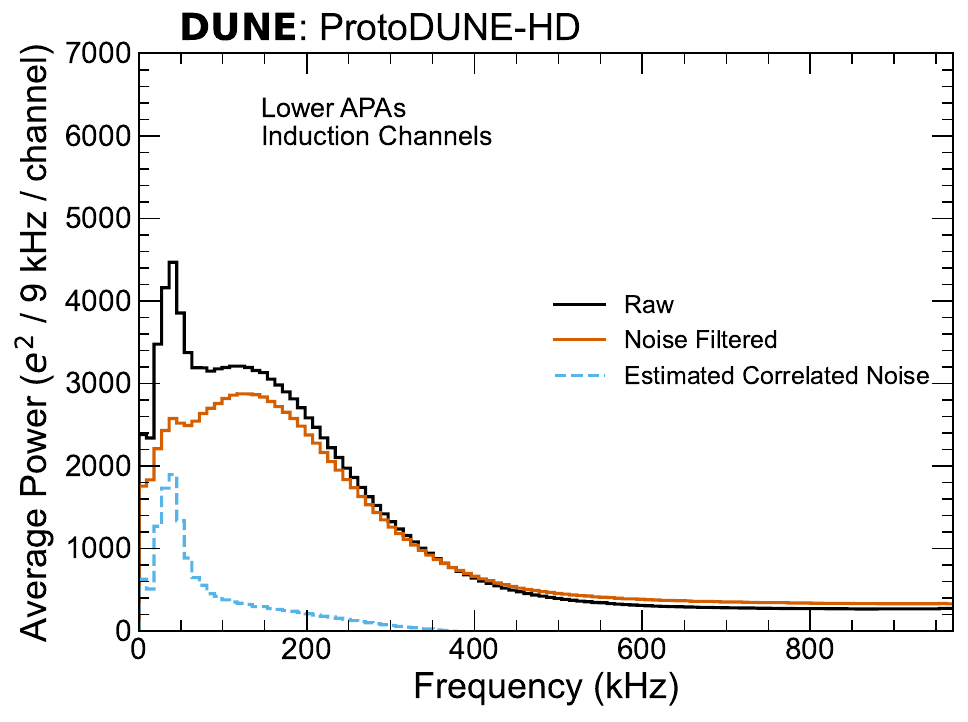}
    \includegraphics[width=0.48\linewidth]{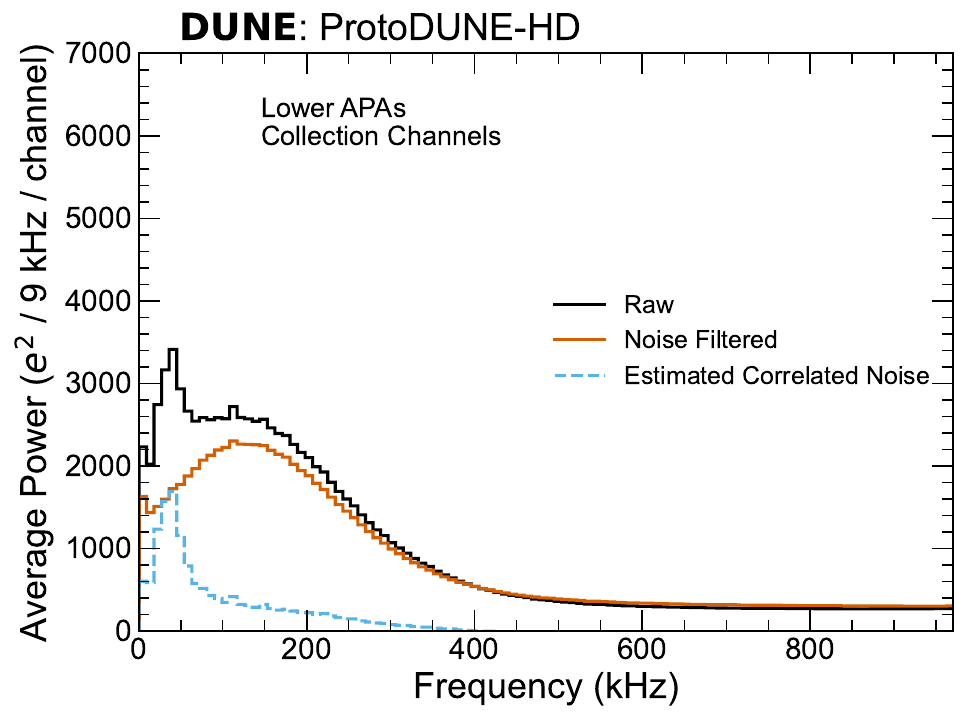}
    \caption{\label{fig:cnr_ffts} Average noise power spectra for charge readout electronics channels, plotted separately for upper APAs (\textbf{top}) and lower APAs (\textbf{bottom}), and for induction channels (\textbf{left}) and collection channels (\textbf{right}). The estimated correlated noise is obtained by subtracting the filtered noise power spectrum from the raw noise power spectrum. Correlated noise pickup is seen more strongly on upper APAs, which are closer to warm noise sources, and on induction channels, which are less shielded than collection channels.}
\end{figure}

The effect of the correlated noise removal in the frequency domain is shown in figure \ref{fig:cnr_ffts}. It primarily eliminates pickup in the lower frequency region, with the strongest peaks in the \mbox{25 to 100\,kHz} range. We estimate the power spectrum of the correlated noise by subtracting the filtered noise power spectrum from the raw noise power spectrum. We observe that the correlated pickup takes different shapes and magnitudes between the lower and upper APAs, which we attribute to the difference in cable lengths connecting their FEMBs to the warm electronics. In particular, at least some of the correlated pickup is conducted down to the FEMBs from the warm electronics, and this noise is more attenuated in the longer cables of the lower APAs. In both cases, the offline noise filter is able to eliminate almost all of the extra pickup, leaving us with just the expected intrinsic noise from the front-end electronics.

Detector performance is generally quantified with the signal-to-noise ratio (SNR), since per-channel noise levels depend on aspects of the detector design that also impact signal sizes. For LArTPCs, this is often evaluated using the signals seen from minimum ionizing particles. One of the APAs in ProtoDUNE-HD suffered from an installation error that resulted in the high-voltage bias for its collection layer being disconnected. While its TPC electronics remained unaffected and were thus included in the pulser and noise analyses up to this point, this missing bias affected signal formation on that APA's wires. For simplicity, we omit all of the channels on that APA from the following SNR analysis; they will require special treatment in future work.

We evaluate the SNR of the TPC electronics by manually scanning data taken with a random trigger for cosmic muon tracks. We select five tracks that are traveling both perpendicular to the drift direction of the TPC and parallel to the floor of the TPC, and which traverse the entire ${\sim}$4.6\,m length of the TPC. Purity monitors in ProtoDUNE-HD measured the electron lifetime to be well above 30 ms, similar to the argon purity achieved in ProtoDUNE-SP \cite{PDSP_Purity}. Since this is much larger than the maximum possible electron drift time of ${\sim}3$\,ms in the TPC, electron lifetime effects are ignored in this analysis. We evaluate the signal as the amplitude of the pulse that appears on each channel from the selected track, angle-corrected to the expected amplitude for a track traveling perpendicular to the channel's wire. This standardizes the comparison across channels of different wire layers, which are oriented at different angles. We evaluate noise by calculating the RMS response of the waveform in time ranges outside the signal region in the selected events, so that the signal and noise are evaluated under the same detector conditions.

\begin{figure}
    \centering
    \includegraphics[width=0.9\linewidth]{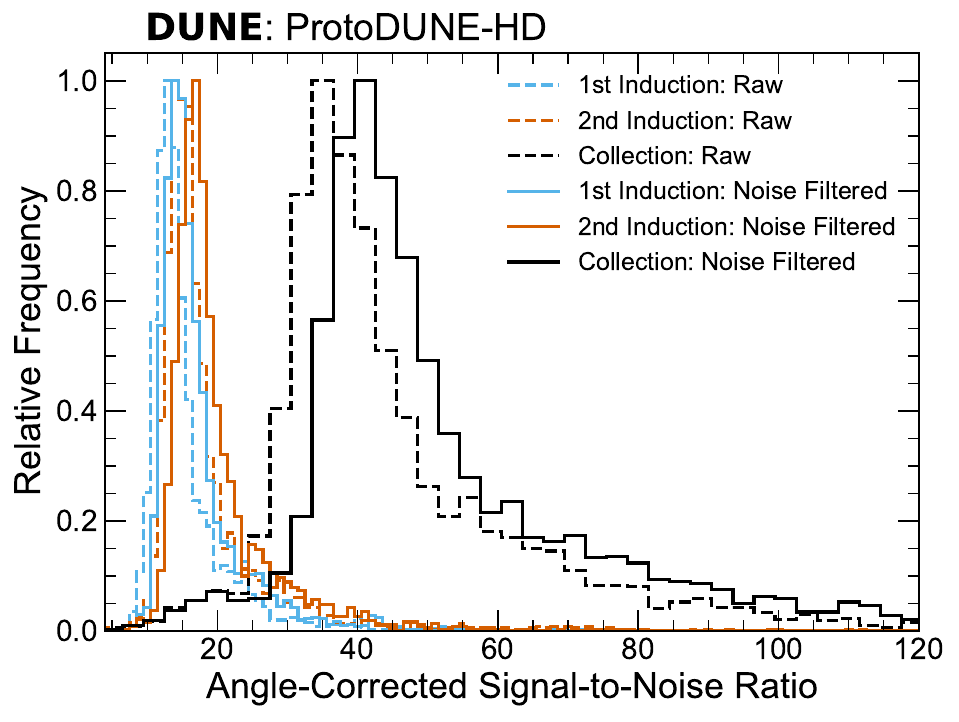}
    \caption{\label{fig:snr} Signal-to-noise ratio for an assortment of channels, evaluated by selecting muons traveling perpendicular to both the drift direction of the TPC and the orientation of the collection wires. Results are plotted separately for the 1st induction, 2nd induction, and collection layers, each of which expects different signal shapes and thus different SNRs. In order to standardize the SNR comparison, signal sizes for each wire are angle-corrected to the expected value for a track traveling perpendicular to the wire. Results are shown both before and after offline noise filtering.}
\end{figure}

The resulting SNR values before and after offline noise filtering are shown in figure \ref{fig:snr}. The most probable values for the SNR are 13, 14, and 34 for the 1st induction, 2nd induction, and collection planes respectively before noise filtering, and they are 14, 17, and 40 after filtering. Due to differences in the local drift velocities, the highest signal sizes are expected on the collection plane, and the smallest signal sizes are expected on the 1st induction plane.

\begin{figure}
    \centering
    \includegraphics[width=0.47\linewidth]{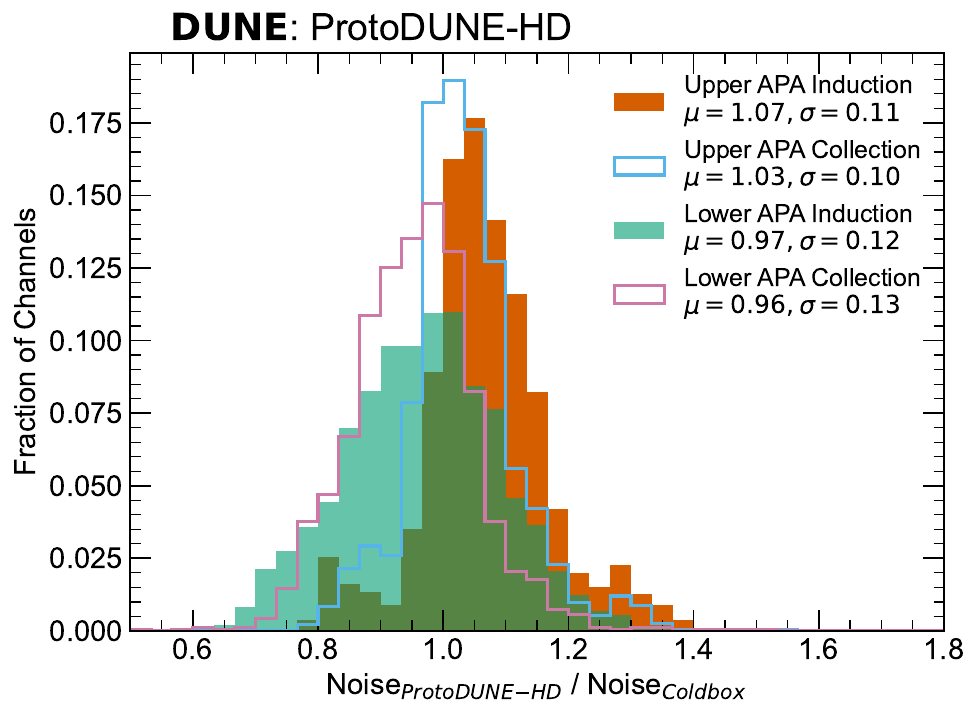}
    \includegraphics[width=0.47\linewidth]{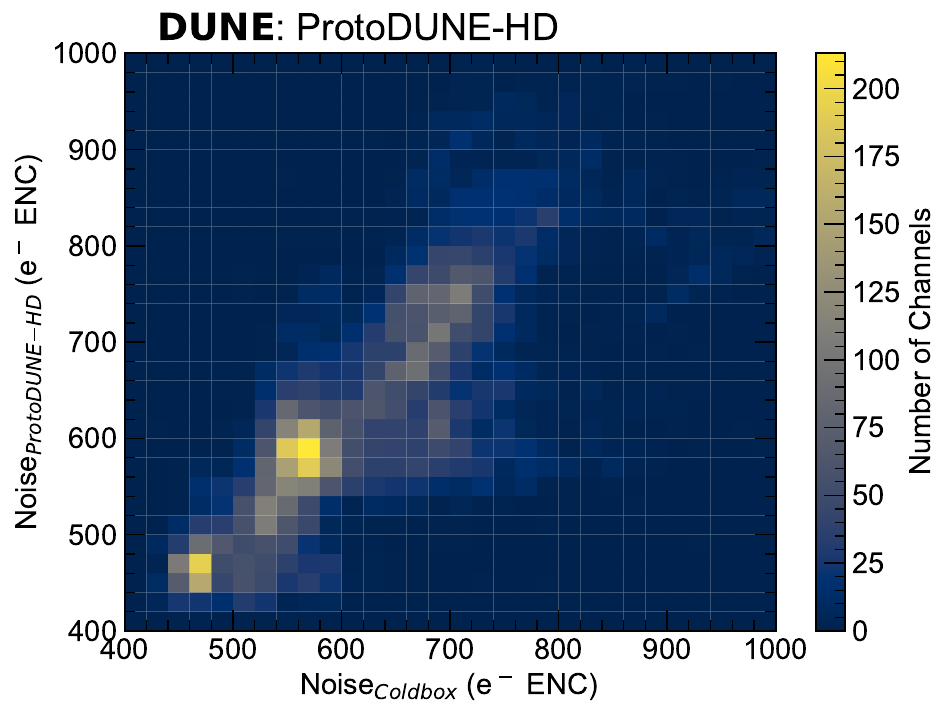}
    \caption{\label{fig:ColdboxRatio} \textbf{Left:} ratio of raw noise levels observed in ProtoDUNE-HD compared to raw noise levels observed for the same electronics channels in their respective coldbox tests. The multi-peak structure for the upper APAs is due to their higher sensitivity to noise pickup, as shown in figure~\ref{fig:cnr_ffts}. \textbf{Right:} raw noise of each channel as measured in ProtoDUNE-HD plotted against the noise measured in the corresponding coldbox test. Most of the observed variation in noise comes from differences in the coldbox and ProtoDUNE-HD environmental conditions.}
\end{figure}

Another metric of interest comes from the coldbox tests that are conducted for each APA module before installation in the ProtoDUNE-HD cryostat, similar to what was done for ProtoDUNE-SP \cite{Adams_2020,PDSP_design}. These coldbox tests are performed with all of the electronics and other detector components installed on the APA, using nitrogen gas to cool the APA down to temperatures around 140--170\,K. This allows us to screen for anomalous behavior in any of the detector components ahead of the full detector operations, and this same procedure will be used for DUNE HD far detector installation as well. While performance under cold nitrogen gas conditions should not be identical to performance under liquid argon conditions, they should still be correlated in a useful manner. A comparison of measured raw noise performances in ProtoDUNE-HD for each electronics channel compared against what was seen in their respective coldbox tests is shown in figure \ref{fig:ColdboxRatio}. The average ratio being close to unity is a coincidence arising from two effects that almost cancel out: higher noise in the coldbox from higher temperatures and lower noise in the coldbox from lower wire capacitances due to the use of gas instead of liquid. We see that the spread in the distributions, which stems from generally different noise conditions in the coldbox, is reasonably contained. In particular, no channels differ drastically in noise performance between the coldbox environment and the ProtoDUNE-HD environment. This provides confidence that coldbox tests for the DUNE far detector modules can usefully inform our expectations for their performance in the full TPC.

\subsection{Saturation Behavior}

\begin{figure}
    \includegraphics[width=0.49\linewidth]{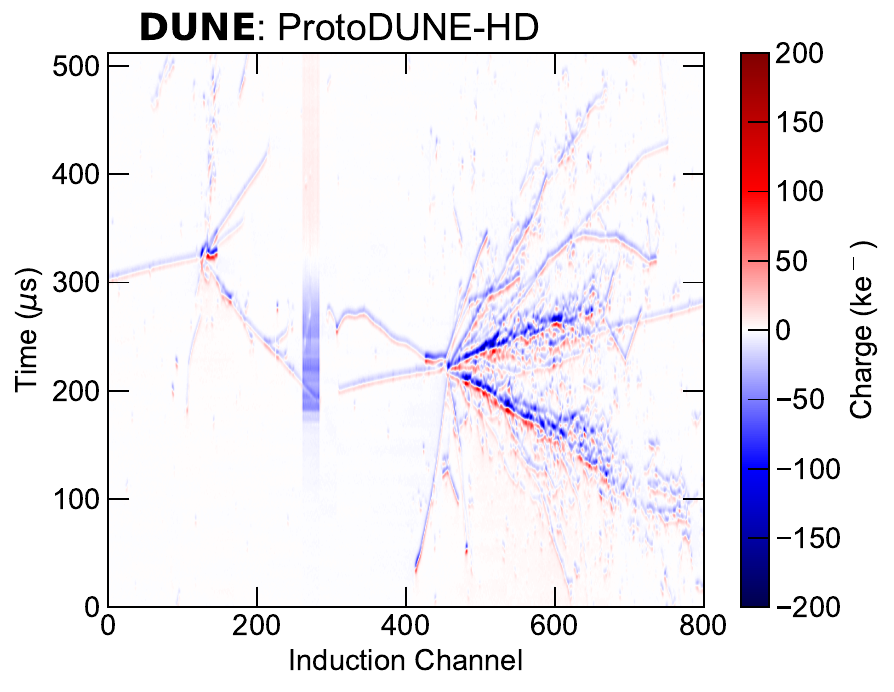}
    \includegraphics[width=0.49\linewidth]{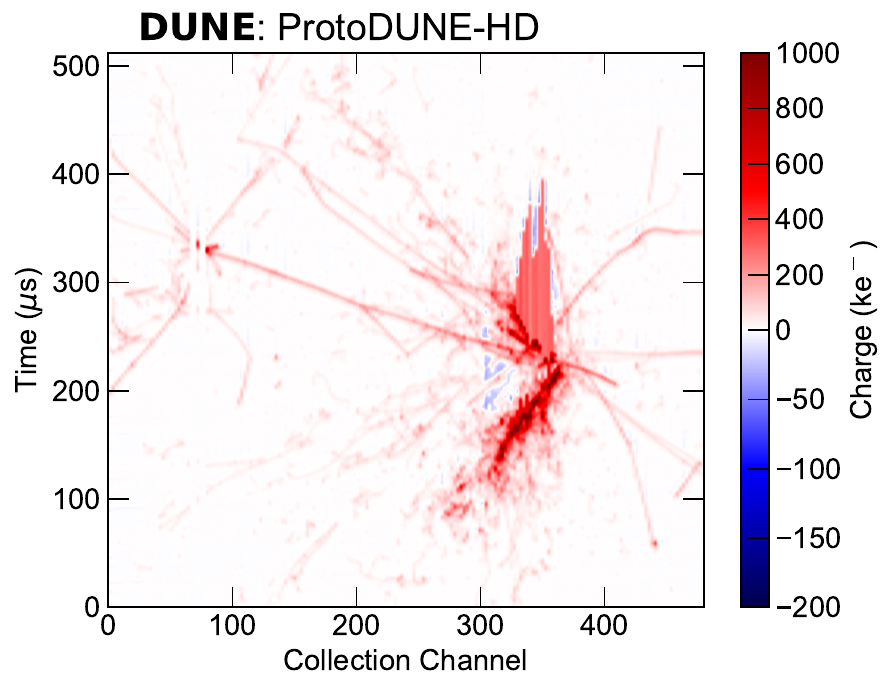}
    \caption{\label{fig:SaturatingEvent} Example of an event showing anomalous behavior coinciding with highly saturated channels, for one of the induction views (\textbf{left}) and the collection view (\textbf{right}). In the induction view, several channels disconnected from the main interaction point suddenly experience simultaneous negative polarity pulses, coinciding in time with highly saturating signals appearing on collection channels that share the same power rail. Note that the two views use different color scales, and note that collection channels at the maximum of the color scale have been saturated.}
\end{figure}

While the TPC electronics were designed with sufficient dynamic range to avoid saturation issues in the large majority of events that DUNE expects to see, there will invariably be some energy deposits that are dense enough to saturate the front-end amplifiers. Saturated channels can be easily identified in offline analysis and so in most cases the only consequence is the loss of calorimetric information for a particular event or particle. However, we observed in ProtoDUNE-HD data that highly saturated channels could induce anomalous responses in other channels sharing the same power rails. Figure \ref{fig:SaturatingEvent} provides an example of an event where this effect was particularly visible, with saturating signals on the collection plane inducing simultaneous opposite polarity responses on induction channels.

Benchtop tests showed that this was an effect of a transient surge in current draw when the amplifiers enter the highly saturated regime. For sufficiently large input signals, the installed bulk capacitance on the LDO-supplied LArASIC power rails cannot handle this current surge, causing these power rails to sag. As a result, when a channel sees an over-saturating positive polarity signal (corresponding to collection of electrons in the channel), other channels sharing the power rail will see an induced negative polarity signal in the immediate aftermath of the instigating saturating signal. As shown in figure \ref{fig:femb_power}, the LArASIC power rails are each shared by blocks of 64 channels (four LArASICs). Since these channels can in general be reading different regions of the detector, this introduces a possible risk of saturated channels impacting reconstruction of unrelated events.

\begin{figure}
    \includegraphics[width=0.49\linewidth]{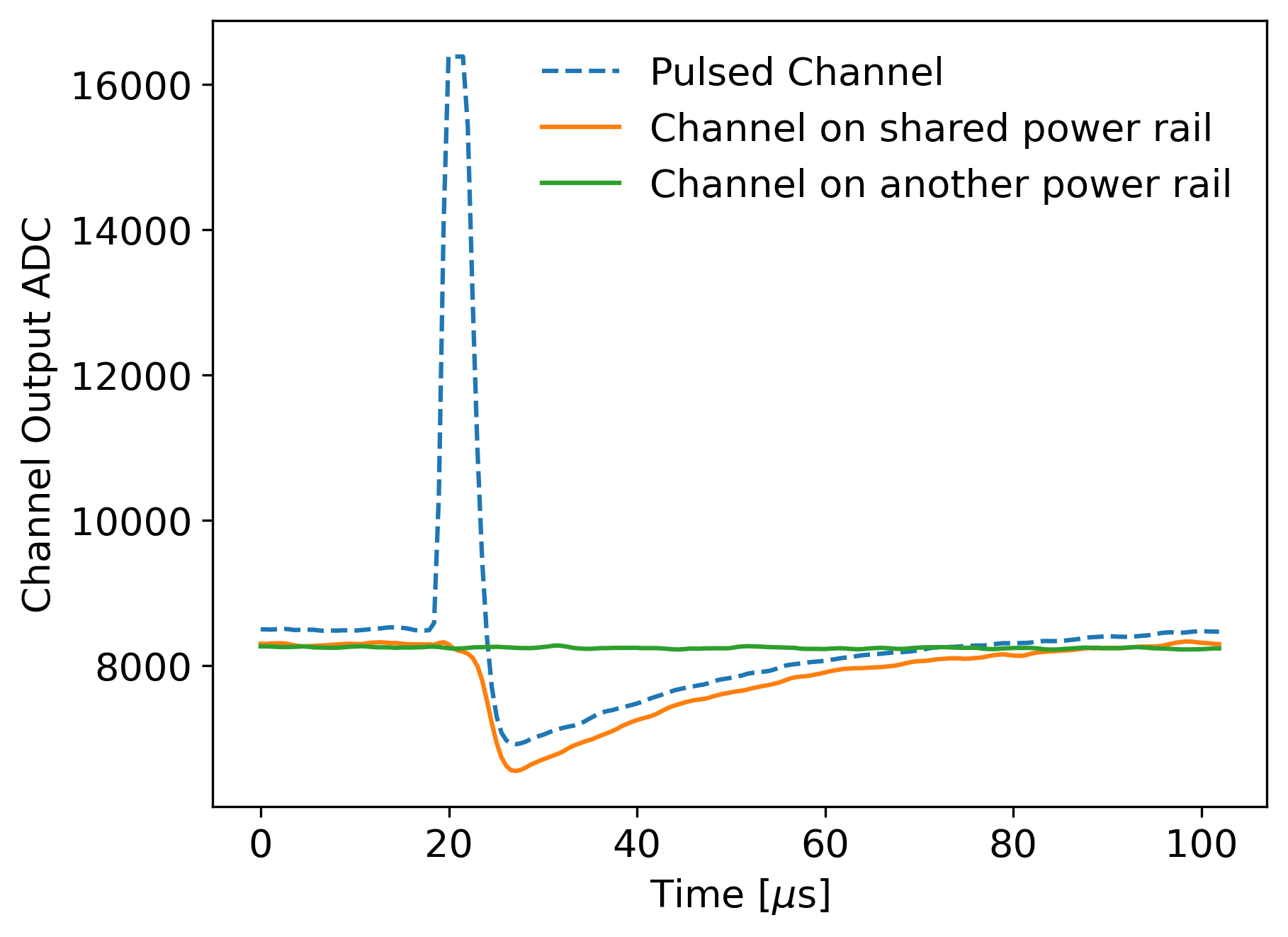}
    \includegraphics[width=0.49\linewidth]{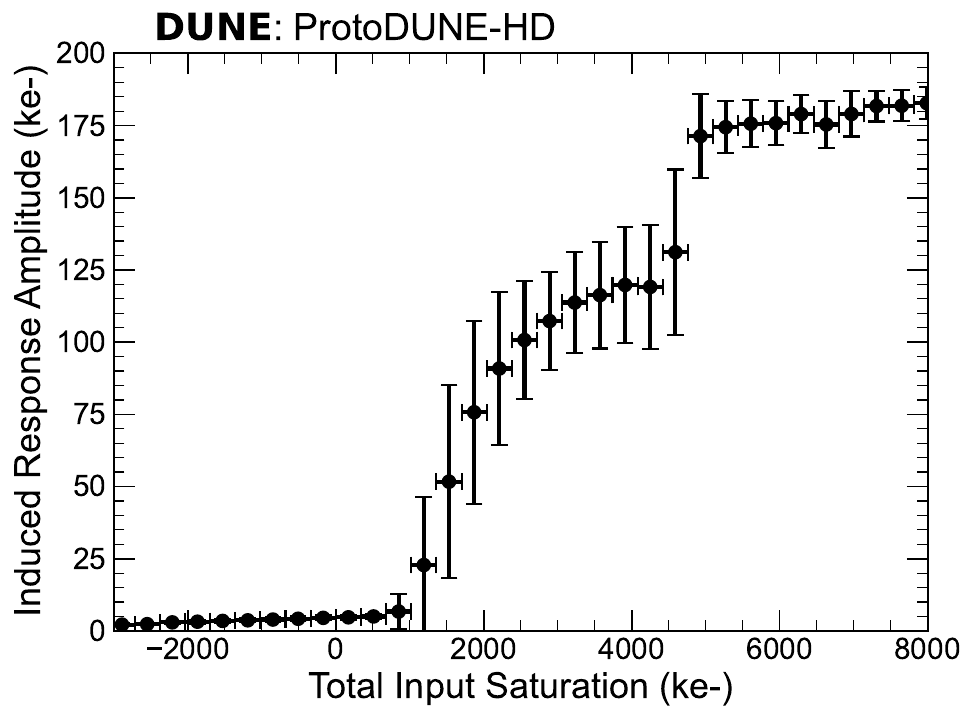}
    \caption{\label{fig:SaturationScan} \textbf{Left:} typical response of channels on a FEMB in response to an overly saturating pulse on a subset of channels. \textbf{Right:} magnitude of the induced response in channels sharing a power rail with channels that receive highly saturating signals, for different degrees of over-saturation in the pulsed channels. Error bars are the standard deviation of response magnitude across different ASICs.}
\end{figure}

To characterize this effect, we ran pulser scans where only six channels per LArASIC received an input pulse, while the other channels were observed for any induced responses. The results are shown in figure \ref{fig:SaturationScan}, where we can see the effect only appears once the input pulses have sufficient amplitude to over-saturate their channels. The magnitude of the induced responses increases as the initial input signal reaches further into the saturated regime, although it appears to stop increasing after the degree of over-saturation reaches several million electrons.

It is likely that the LArASICs used in ProtoDUNE-SP also exhibited this effect. However, that earlier version of the LArASIC suffered from a different issue with its response to saturating pulses known as the ``ledge effect'' \cite{HD_TDR}, which may have masked this effect. The ledge effect issue was fixed in the current versions of the LArASIC, allowing us to now see this more subtle effect.

Operating the LArASICs at lower gain reduces the probability that any particular event saturates a channel, and so our switch to 7.8\,mV/fC gain mentioned in section \ref{sec:performance_noise} reduces the fraction of events that are impacted by this effect. This effect can be mitigated by substantially increasing the capacitance on the power rails that supply the LArASIC on the FEMB, but this approach is not feasible given the limited available space on the FEMBs for additional capacitors. Instead, we note that events with highly saturated channels are generally excluded from analysis anyway due to the ensuing loss of calorimetric information. As a result, this effect only impacts scenarios where there is a non-saturating event in close proximity to a saturating event in both space and time, and where we wish to analyze the non-saturating event on its own and must correct for the distortions imposed by the saturated channels. We expect much fewer than $10\%$ of far detector beam neutrino events in DUNE to have saturated channels \cite{HD_TDR,VD_TDR}, and the rate of coincidences requiring this special treatment will consequently be low. Nonetheless, we expect this correction to be possible in offline analysis; saturated channels and unipolar negative pulses are both simple to flag, and the magnitude of the induced effect can be estimated by inspecting affected channels that did not see a true charge signal. 

\section{Conclusion}
\label{sec:conclusion}

This paper has presented a full description of the final custom-built TPC electronics system that will be used in the DUNE HD far detector, covering details of the individual cryogenic ASICs, the cryogenic front-end motherboards, and the warm interface electronics. The design of these electronics represents the culmination of nearly two decades of research into the use of cryogenic front-end electronics for LArTPC-based neutrino detectors. Over the course of this process, each of the components comprising this TPC electronics system has been thoroughly tested individually. However, ultimate validation required testing them all together in a large-scale prototype, integrated with other subsystems under realistic detector conditions.

The successful operation of ProtoDUNE-HD has provided this validation and represents a major milestone towards the realization of the DUNE HD far detector. We have demonstrated that the TPC electronics are able to continuously deliver data at the required rates over several months of detector operations without any digitization or transmission errors. The electronics have shown exceptional reliability and consistency in performance; while there were 33 channels that were unusable due to mechanical and high-voltage issues unrelated to the electronics, none of the remaining 10,207 channels required special treatment or exclusion from analysis. We have quantitatively studied the noise, linearity, cross-talk, and SNR in all of these channels, showing that they exceed the requirements for DUNE's physics program \cite{HD_TDR}. Pre-filter raw noise levels were low enough to comfortably permit online trigger thresholds of 4000--4200 electrons per channel. After offline filtering, we measured typical noise levels of 450--600\,e$^-$ ENC on induction channels and 350--500\,e$^-$ ENC on collection channels. This corresponded to most probable SNRs of 14 on 1st induction channels, 17 on 2nd induction channels, and 40 on collection channels. We found that non-linearity and cross-talk within the electronics is small enough to be completely negligible for most channels. The worst-case channels see effects of up to $0.3\%-0.4\%$ on their response, but this is still generally negligible compared to other detector effects, and they have been characterized so that they can be corrected for in offline analysis. With the results from ProtoDUNE-HD described in this paper, the TPC electronics design has been finalized, and we have begun production of components for the DUNE far detectors.

After the conclusion of ProtoDUNE-HD in December 2024, its liquid argon was transferred to the ProtoDUNE-VD demonstrator. ProtoDUNE-VD serves as the final large-scale demonstrator for DUNE's VD far detector design, which uses the same charge readout electronics described in this paper for the bottom half of its TPC~\cite{VD_TDR}. ProtoDUNE-VD is currently still under operation at the CERN Neutrino Platform, and its performance will be discussed in a separate future article. In the meantime, the data collected by ProtoDUNE-HD will also enable studies of its other subsystems and the detector as a whole, and the beam data it collected will enable further new studies of interactions in argon that will be important for understanding interactions in DUNE. The results of these studies will be reported in future work.

\acknowledgments

The ProtoDUNE detectors were constructed and operated on the CERN
Neutrino Platform. We gratefully acknowledge the support of the CERN management and the CERN EP, BE, TE, EN, and IT Departments for the ProtoDUNE program, including NP04 and NP02. This document was prepared by
DUNE collaboration using the resources of the Fermi National Accelerator Laboratory (Fermilab), a U.S. Department of Energy, Office of Science, Office of High Energy Physics HEP User Facility. Fermilab is managed by Fermi Forward Discovery Group, LLC, acting under Contract No. 89243024CSC000002. This work was supported by CNPq, FAPERJ, FAPEG, FAPESP and, Funda\c c\~ao Arauc\'aria, Brazil; CFI, IPP and NSERC, Canada; CERN; ANID-FONDECYT, Chile; M\v SMT, Czech Republic; ERDF, FSE+, Horizon Europe, MSCA and NextGenerationEU, European Union; CNRS/IN2P3 and CEA, France; PRISMA+, Germany; INFN, Italy; FCT, Portugal; CERN-RO/CDI, Romania; NRF, South Korea; Generalitat Valenciana, Junta de Andaluc\'ia, MICINN, and Xunta de
Galicia, Spain; SERI and SNSF, Switzerland; T\"UB\.ITAK, Turkey; The Royal
Society and UKRI/STFC, United Kingdom; DOE and NSF, United States of
America.

\bibliographystyle{JHEP}
\bibliography{biblio.bib}

@ARTICLE{ColdADC,
  author={Grace, C. and Braga, D. and Chen, H. and Christian, D. and Dabrowski, M. and Deptuch, G. and Dwyer, D. and Fried, J. and Gao, S. and Hoff, J. and Holm, S. and Lin, C.-J. and Lopriore, E. and Lu, Y. and Luk, K.-B. and Miryala, S. and Prakash, T. and Richardson, H. and Shenai, A. and Tarpara, E. and Vernon, E. and Zhang, J.},
  journal={IEEE Transactions on Nuclear Science}, 
  title={{ColdADC\_P2: A 16-Channel Cryogenic ADC ASIC for the Deep Underground Neutrino Experiment}}, 
  year={2022},
  volume={69},
  number={1},
  pages={105-112},
  doi={10.1109/TNS.2021.3136404}
}

@ARTICLE{KaranicolasCalibration,
  author={Karanicolas, A.N. and Hae-Seung Lee and Barcrania, K.L.},
  journal={{IEEE Journal of Solid-State Circuits}}, 
  title={{A 15-b 1-Msample/s digitally self-calibrated pipeline ADC}}, 
  year={1993},
  volume={28},
  number={12},
  pages={1207-1215},
  doi={10.1109/4.261994}}

@article{Pietropaolo_2017,
doi = {10.1088/1742-6596/888/1/012038},
url = {https://doi.org/10.1088/1742-6596/888/1/012038},
year = {2017},
month = {sep},
publisher = {IOP Publishing},
volume = {888},
number = {1},
pages = {012038},
author = {Pietropaolo, F},
title = {{Review of Liquid-Argon Detectors Development at the CERN Neutrino Platform}},
journal = {J. Phys. Conf. Ser.},
}

@article{Adams_2020,
doi = {10.1088/1748-0221/15/06/P06017},
url = {https://dx.doi.org/10.1088/1748-0221/15/06/P06017},
year = {2020},
month = {jun},
publisher = {},
volume = {15},
number = {06},
pages = {P06017},
author = {Adams, D. and Bass, M. and Bishai, M. and Bromberg, C. and Calcutt, J. and Chen, H. and Fried, J. and Furic, I. and Gao, S. and Gastler, D. and Hugon, J. and Joshi, J. and Kirby, B. and Liu, F. and Mahn, K. and Mooney, M. and Morris, C. and Pereyra, C. and Pons, X. and Radeka, V. and Raguzin, E. and Shooltz, D. and Spanu, M. and Timilsina, A. and Tufanli, S. and Tzanov, M. and Viren, B. and Gu, W. and Williams, Z. and Wood, K. and Worcester, E. and Worcester, M. and Yang, G. and Zhang, J.},
title = {{The ProtoDUNE-SP LArTPC electronics production, commissioning, and performance}},
journal = {JINST}
}

@misc{I2C,
title = {{Two-Wire Bus-System Comprising A Clock Wire And A Data Wire For Interconnecting A Number Of Stations}},
author       = {{Philips N.V.}},
  nationality  = {Netherlands},
  number       = {NL8005976A},
  year         = {1982},
  month        = may,
  note         = {Filed: 1980-10-31; Published: 1982-05-17},
  howpublished = {Patent NL8005976A},
  url          = {https://patents.google.com/patent/NL8005976A/en}
}

@techreport{ethercat,
  type        = {International Standard},
  author      = {{International Electrotechnical Commission}},
  title       = {{Industrial communication networks - Fieldbus specifications - Part 3-12: Data-link layer service definition - Type 12 elements}},
  number      = {IEC 61158-3-12},
  year        = {2019},
}

@misc{8b10b,
title = {{Byte oriented DC balanced (0,4) 8B/10B partitioned block transmission code}},
author       = {{P.A. Franaszek, A.X. Widmer}},
  nationality  = {USA},
  number       = {US4486739A},
  year         = {1984},
  note         = {Filed: 1982-06-30; Published: 1984-12-04},
  howpublished = {Patent US4486739A},
  url          = {https://patents.google.com/patent/US4486739A/en}
}

@article{PDSP_kaon,
  title = {{First measurement of the total inelastic cross section of positively charged kaons on argon at energies between 5.0 and 7.5 GeV}},
  author = {Abud, A. Abed and others},
  collaboration = {DUNE},
  journal = {Phys. Rev. D},
  volume = {110},
  issue = {9},
  pages = {092011},
  numpages = {22},
  year = {2024},
  month = {Nov},
  publisher = {American Physical Society},
  doi = {10.1103/PhysRevD.110.092011},
  url = {https://link.aps.org/doi/10.1103/PhysRevD.110.092011}
}

@article{PDSP_lowkaon,
title = {{Identification of low-energy kaons in the ProtoDUNE-SP detector
}},
author = {Abbaslu, S. and others},
collaboration = {{DUNE}},
journal = {Phys. Rev. D},
volume = {113},
issue = {5},
pages = {052004},
numpages = {21},
year = {2026},
month = {Mar},
publisher = {American Physical Society},
doi = {10.1103/q21l-pl7s},
url = {https://link.aps.org/doi/10.1103/q21l-pl7s}
}

@article{PDSP_pionproton,
year = {2025},
title = {{First Measurement of $\pi^+$-Ar and $p$-Ar Total Inelastic Cross Sections in the Sub-GeV Energy Regime with ProtoDUNE-SP Data
}},
eprint = {2511.11925},
archiveprefix = {arXiv},
doi = {https://doi.org/10.48550/arXiv.2511.11925},
collaboration={DUNE}
}

@article{PDSP_pion,
year = {2025},
title = {{Measurement of Exclusive $\pi^+$--argon Interactions Using ProtoDUNE-SP
}},
eprint = {2511.13462},
archiveprefix = {arXiv},
doi = {https://doi.org/10.48550/arXiv.2511.13462},
collaboration={DUNE}
}

@article{Barcock_2023,
doi = {10.1088/1748-0221/18/01/C01067},
url = {https://dx.doi.org/10.1088/1748-0221/18/01/C01067},
year = {2023},
month = {jan},
publisher = {IOP Publishing},
volume = {18},
number = {01},
pages = {C01067},
author = {Barcock, A. and Cussans, D. and Lindebaum, D. and Newbold, D. and Paramesvaran, S. and Trilov, S.},
title = {{Timing and synchronization of the DUNE neutrino detector}},
journal = {JINST},
}

@article{Acciarri_2017,
doi = {10.1088/1748-0221/12/08/P08003},
url = {https://doi.org/10.1088/1748-0221/12/08/P08003},
year = {2017},
month = {aug},
publisher = {},
volume = {12},
number = {08},
pages = {P08003},
author = {Acciarri, R. and others},
collaboration = {{MicroBooNE}},
title = {{Noise Characterization and Filtering in the MicroBooNE Liquid Argon TPC}},
journal = {JINST},
}

@article{LArASIC_CHARMS,
  author={Mukim, Prashansa and Carini, Gabriella A. and Chen, Hucheng and Deptuch, Grzegorz W. and Gao, Shanshan and De Geronimo, Gianluigi and Mandal, Soumyajit and Narasimha Manyam, Venkata and Radeka, Veljko and Rescia, Sergio},
  journal={IEEE Transactions on Circuits and Systems I: Regular Papers}, 
  title={{Cryogenic Front-End ASICs for Low-Noise Readout of Charge Signals}}, 
  year={2025},
  volume={72},
  number={4},
  pages={1496-1509},
  doi={10.1109/TCSI.2024.3506828}
}

@article{Radeka_2011,
doi = {10.1088/1742-6596/308/1/012021},
url = {https://dx.doi.org/10.1088/1742-6596/308/1/012021},
year = {2011},
month = {jul},
publisher = {},
volume = {308},
number = {1},
pages = {012021},
author = {Radeka, Veljko and Chen, Hucheng and Deptuch, Grzegorz and Geronimo, Gianluigi De and Lanni, Francesco and Li, Shaorui and Nambiar, Neena and Rescia, Sergio and Thorn, Craig and Yarema, Ray and Yu, Bo},
title = {{Cold electronics for "Giant" Liquid Argon Time Projection Chambers}},
journal = {Journal of Physics: Conference Series},
}

@article{CHEN2023167571,
title = {{Cryogenic electronics for noble liquid neutrino detectors}},
journal = {Nucl. Instrum. Methods Phys. Res. A},
volume = {1045},
pages = {167571},
year = {2023},
issn = {0168-9002},
doi = {https://doi.org/10.1016/j.nima.2022.167571},
url = {https://www.sciencedirect.com/science/article/pii/S0168900222008634},
author = {Hucheng Chen and Veljko Radeka},
}

@article{CHEN20121287,
title = {{Front End Readout Electronics of the MicroBooNE Experiment}},
journal = {Physics Procedia},
volume = {37},
pages = {1287-1294},
year = {2012},
note = {Proceedings of the 2nd International Conference on Technology and Instrumentation in Particle Physics (TIPP 2011)},
issn = {1875-3892},
doi = {https://doi.org/10.1016/j.phpro.2012.02.471},
url = {https://www.sciencedirect.com/science/article/pii/S1875389212018366},
author = {H. Chen and G. {De Geronimo} and F. Lanni and D. Lissauer and D. Makowiecki and V. Radeka and S. Rescia and C. Thorn and B. Yu},
}

@INPROCEEDINGS{9875569,
  author={Manyam, V. N. and Deptuch, G. and Gao, S. and Tarpara, E. and Khan, N. and Chen, H. and Carini, G.},
  booktitle={2021 IEEE Nuclear Science Symposium and Medical Imaging Conference (NSS/MIC)}, 
  title={{Optimization of Charge Amplifier Reset Quiescent Current in LArASIC}}, 
  year={2021},
  volume={},
  number={},
  pages={1-2},
  doi={10.1109/NSS/MIC44867.2021.9875569}
}

@ARTICLE{COLDATA_LD,
  author={Wang, Xiaoran and Gui, Ping},
  journal={{IEEE Transactions on Nuclear Science}}, 
  title={{A Hybrid Transmitter With Voltage-Mode SST Preemphasis and Current-Mode Transmitter Equalization Capable of Operating at 77 K in DUNE}}, 
  year={2023},
  volume={70},
  number={3},
  pages={262-271},
  doi={10.1109/TNS.2023.3239055}
}

@misc{WIB_github,
  howpublished = {\url{https://github.com/DUNE-DAQ/dune-wib-firmware/tree/master/sw}}
}

@ARTICLE{Hermes,
  author={Sipos, Roland},
  journal={IEEE Transactions on Nuclear Science}, 
  title={{The Ethernet Readout of the DUNE DAQ System}}, 
  year={2025},
  volume={72},
  number={3},
  pages={317-324},
  doi={10.1109/TNS.2024.3486059}
}

@INPROCEEDINGS{8533137,
  author={Liu, F. and Bishai, M. and Chen, H. and D'Andragora, A. and Fried, J. and Gao, S. and Hou, W. and Li, S. and Radeka, V. and Vernon, E. and Worcester, E. and Worcester, M. and Yethiraj, K. and Yu, B. and Zhang, J. and Deng, Z. and Liu, Y.},
  booktitle={2017 IEEE Nuclear Science Symposium and Medical Imaging Conference (NSS/MIC)}, 
  title={{Cold Electronics System Development for ProtoDUNE-SP and SBND LAr TPC}}, 
  year={2017},
  volume={},
  number={},
  pages={1-4},
  doi={10.1109/NSSMIC.2017.8533137}
}

@INPROCEEDINGS{5874057,
  author={De Geronimo, Gianluigi and D'Andragora, Alessio and Li, Shaorui and Nambiar, Neena and Rescia, Sergio and Vernon, Emerson and Chen, Hucheng and Lanni, Francesco and Makowiecki, Don and Radeka, Veljko and Thorn, Craig and Yu, Bo},
  booktitle={IEEE Nuclear Science Symposium \& Medical Imaging Conference}, 
  title={{Front-end ASIC for a liquid argon TPC}}, 
  year={2010},
  volume={},
  number={},
  pages={1658-1666},
  doi={10.1109/NSSMIC.2010.5874057}
}

@article{HD_TDR,
doi = {10.1088/1748-0221/15/08/T08010},
url = {https://dx.doi.org/10.1088/1748-0221/15/08/T08010},
year = {2020},
month = {aug},
publisher = {},
volume = {15},
number = {08},
pages = {T08010},
author = {Abi, B. and others},
title = {{DUNE Far Detector Technical Design Report Volume IV. The DUNE far detector single-phase technology}},
journal = {JINST},
collaboration = {DUNE}
}

@article{RevModPhys.96.045001,
  title = {The science and technology of liquid argon detectors},
  author = {Bonivento, W. M. and Terranova, F.},
  journal = {Rev. Mod. Phys.},
  volume = {96},
  issue = {4},
  pages = {045001},
  numpages = {40},
  year = {2024},
  month = {Oct},
  publisher = {American Physical Society},
  doi = {10.1103/RevModPhys.96.045001},
  url = {https://link.aps.org/doi/10.1103/RevModPhys.96.045001}
}

@ARTICLE{CryoCMOS,
  author={Clark, W.F. and El-Kareh, B. and Pires, R.G. and Titcomb, S.L. and Anderson, R.L.},
  journal={IEEE Transactions on Components, Hybrids, and Manufacturing Technology}, 
  title={{Low temperature CMOS-a brief review}}, 
  year={1992},
  volume={15},
  number={3},
  pages={397-404},
  doi={10.1109/33.148509}
}

@article{DUNE_Intro,
doi = {10.1088/1748-0221/15/08/T08008},
url = {https://dx.doi.org/10.1088/1748-0221/15/08/T08008},
year = {2020},
month = {aug},
publisher = {},
volume = {15},
number = {08},
pages = {T08008},
author = {Abi, B. and others},
title = {{DUNE Far Detector Technical Design Report Volume I. Introduction to DUNE}},
journal = {JINST},
collaboration = {DUNE}
}

@article{DUNE_physics,
year = {2020},
title = {{DUNE Far Detector Technical Design Report Volume II. DUNE Physics}},
author = {Abi, B. and others},
eprint       = {2002.03005},
archiveprefix = {arXiv},
collaboration = {DUNE}
}

@article{ND_CDR,
year = {2021},
title = {{Deep Underground Neutrino Experiment (DUNE) Near Detector Conceptual Design Report}},
author = {Abed Abud, A. and others},
eprint       = {2103.13910},
archiveprefix = {arXiv},
collaboration = {DUNE}
}

@article{VD_TDR,
year = {2023},
title = {{The DUNE Far Detector Vertical Drift Technology, Technical Design Report
}},
author = {Abed Abud, A. and others},
eprint       = {2312.03130},
archiveprefix = {arXiv},
collaboration = {DUNE}
}

@article{SURF_Muon,
title = {{Muon Flux Measurements at the Davis Campus of the Sanford Underground Research Facility with the Majorana Demonstrator Veto System}},
journal = {Astroparticle Physics},
volume = {93},
pages = {70-75},
year = {2017},
issn = {0927-6505},
doi = {https://doi.org/10.1016/j.astropartphys.2017.01.013},
url = {https://www.sciencedirect.com/science/article/pii/S0927650517300038},
author = {N. Abgrall and others},
collaboration = {MAJORANA}
}

@article{PDSP_TDR,
year = {2017},
title = {{The Single-Phase ProtoDUNE Technical Design Report}},
author = {Abi, B. and others},
collaboration = {DUNE},
eprint       = {1706.07081},
archiveprefix = {arXiv},
}

@article{PDSP_design,
doi = {10.1088/1748-0221/17/01/P01005},
url = {https://dx.doi.org/10.1088/1748-0221/17/01/P01005},
year = {2022},
month = {jan},
publisher = {IOP Publishing},
volume = {17},
number = {01},
pages = {P01005},
author = {Abed Abud, A. and others},
title = {{Design, construction and operation of the ProtoDUNE-SP Liquid Argon TPC}},
collaboration = {DUNE},
journal = {JINST},
}

@Article{app11062455,
AUTHOR = {Majumdar, Krishanu and Mavrokoridis, Konstantinos},
TITLE = {{Review of Liquid Argon Detector Technologies in the Neutrino Sector}},
JOURNAL = {Applied Sciences},
VOLUME = {11},
YEAR = {2021},
NUMBER = {6},
ARTICLE-NUMBER = {2455},
URL = {https://www.mdpi.com/2076-3417/11/6/2455},
ISSN = {2076-3417},
DOI = {10.3390/app11062455}
}

@article{DUNE_LongBaseline,
doi = {10.1140/epjc/s10052-020-08456-z},
url = {https://doi.org/10.1140/epjc/s10052-020-08456-z},
year = {2020},
month = {oct},
publisher = {},
volume = {80},
number = {978},
author = {Abi, B. and others},
title = {{Long-baseline neutrino oscillation physics potential of the DUNE experiment}},
journal = {Eur. Phys. J. C},
collaboration = {DUNE}

}

@article{DUNE_BSM,
doi = {10.1140/epjc/s10052-020-08456-z},
url = {https://doi.org/10.1140/epjc/s10052-020-08456-z},
year = {2021},
month = {April},
volume = {81},
number = {322},
author = {Abi, B. and others},
title = {{Prospects for beyond the Standard Model physics searches at the Deep Underground Neutrino Experiment}},
journal = {Eur. Phys. J. C},
collaboration = {DUNE}

}

@article{DUNE_Supernova,
  title = {{Supernova pointing capabilities of DUNE}},
  author = {Abed Abud, A. and others},
  collaboration = {DUNE},
  journal = {Phys. Rev. D},
  volume = {111},
  issue = {9},
  pages = {092006},
  numpages = {24},
  year = {2025},
  month = {May},
  publisher = {American Physical Society},
  doi = {10.1103/PhysRevD.111.092006},
  url = {https://link.aps.org/doi/10.1103/PhysRevD.111.092006},
}

@article{PDSP_results,
doi = {10.1088/1748-0221/15/12/P12004},
url = {https://dx.doi.org/10.1088/1748-0221/15/12/P12004},
year = {2020},
month = {dec},
publisher = {},
volume = {15},
number = {12},
pages = {P12004},
author = {Abi, B. and others},
title = {{First results on ProtoDUNE-SP liquid argon time projection chamber performance from a beam test at the CERN Neutrino Platform}},
journal = {JINST},
collaboration = {DUNE}
}

@article{DUNEPhaseII,
doi = {10.1088/1748-0221/19/12/P12005},
url = {https://doi.org/10.1088/1748-0221/19/12/P12005},
year = {2024},
month = {dec},
publisher = {IOP Publishing},
volume = {19},
number = {12},
pages = {P12005},
author = {Abed Abud, A. and others},
title = {{DUNE Phase II: scientific opportunities, detector concepts, technological solutions}},
collaboration = {DUNE},
journal = {JINST},
}

@article{PhysRevAccelBeams.20.111001,
  title = {{Low energy tertiary beam line design for the CERN neutrino platform project}},
  author = {Charitonidis, N. and Efthymiopoulos, I.},
  journal = {Phys. Rev. Accel. Beams},
  volume = {20},
  issue = {11},
  pages = {111001},
  numpages = {11},
  year = {2017},
  month = {Nov},
  publisher = {American Physical Society},
  doi = {10.1103/PhysRevAccelBeams.20.111001},
  url = {https://link.aps.org/doi/10.1103/PhysRevAccelBeams.20.111001}
}

@article{PhysRevAccelBeams.22.061003,
  title = {Particle production, transport, and identification in the regime of $1\ensuremath{-}7\text{ }\text{ }\mathrm{GeV}/c$},
  author = {Booth, A. C. and Charitonidis, N. and Chatzidaki, P. and Karyotakis, Y. and Nowak, E. and Ortega-Ruiz, I. and Rosenthal, M. and Sala, P.},
  journal = {Phys. Rev. Accel. Beams},
  volume = {22},
  issue = {6},
  pages = {061003},
  numpages = {12},
  year = {2019},
  month = {Jun},
  publisher = {American Physical Society},
  doi = {10.1103/PhysRevAccelBeams.22.061003},
  url = {https://link.aps.org/doi/10.1103/PhysRevAccelBeams.22.061003}
}

@article{PDSP_Purity,
doi = {10.1088/1748-0221/20/09/P09008},
url = {https://doi.org/10.1088/1748-0221/20/09/P09008},
year = {2025},
month = {sep},
publisher = {IOP Publishing},
volume = {20},
number = {09},
pages = {P09008},
author = {Abbaslu, S. and others},
collaboration = {DUNE},
title = {{Spatial and temporal evaluations of the liquid argon purity in ProtoDUNE-SP}},
journal = {JINST},
}

@article{Adams_2018,
doi = {10.1088/1748-0221/13/07/P07007},
url = {https://dx.doi.org/10.1088/1748-0221/13/07/P07007},
year = {2018},
month = {jul},
publisher = {},
volume = {13},
number = {07},
pages = {P07007},
author = {Adams, C. and others},
title = {{Ionization electron signal processing in single phase LArTPCs. Part II. Data/simulation comparison and performance in MicroBooNE}},
journal = {JINST},
collaboration = {MicroBooNE}
}

\end{document}